\documentclass[aps,pra,twocolumn,showpacs,amsmath,amssymb,superscriptaddress]{revtex4-2}
\usepackage{graphicx}
\usepackage{xcolor}
\definecolor{cerulean}{rgb}{0.0, 0.48, 0.65}
\definecolor{regalia}{rgb}{0.32, 0.18, 0.5}
\usepackage{float}
\usepackage{accents}
\usepackage{hyperref}
\usepackage{bbold}
\usepackage{soul}
\usepackage{multirow}
\hypersetup{
    colorlinks=true,
    linkcolor=regalia,
    filecolor=regalia,      
    urlcolor=regalia,
    citecolor=regalia
}

\newcommand{\re}[1]{{\color{blue}{#1}}}
\newcommand{\bl}[1]{{\color{blue}{#1}}}
\newcommand{\ma}[1]{{\color{blue}{#1}}}

\usepackage[normalem]{ulem}
\usepackage{subfigure}
 \usepackage{tikz} 
 \usetikzlibrary{decorations.pathreplacing,calligraphy}

\def\approxprop{%
  \def\p{%
    \setbox0=\vbox{\hbox{$\propto$}}%
    \ht0=0.6ex \box0 }%
  \def\s{%
    \vbox{\hbox{$\sim$}}%
  }%
  \mathrel{\raisebox{0.7ex}{%
      \mbox{$\underset{\s}{\p}$}%
    }}%
}
\def\be{\begin{equation}}
\def\ee{\end{equation}}
\def\ba{\begin{eqnarray}}
\def\ea{\end{eqnarray}}
\def\bc{\begin{center}}
\def\ec{\end{center}}
\def\p{\partial}
\def\sgn{{\rm sgn}}
\def\Sp{{\rm Sp}}
\def\d{\,\mathrm{d}}
\def\rperp{r_{\!{_\perp}}}
\newcommand{\cam}[1]{\textcolor{blue}{ #1~}}
\newcommand{\mage}[1]{\textcolor{blue}{ #1~}}
\newcommand{\comn}[1]{{ #1~}}
\newcommand{\gsim}{\raisebox{-0.13cm}{~\shortstack{$>$ \\[-0.07cm]
      $\sim$}}~}
\newcommand{\dbtilde}[1]{\widetilde{\raisebox{0pt}[0.85\height]{$\widetilde{#1}$}}}
\newcommand{\chk}[1]{\textcolor{blue}{ #1}}
\newcommand{\crss}[1]{\textcolor{blue}{\sout{ #1}}}

\renewcommand{\vec}[1]{\mathbf{#1}}

\newcommand{\normord}[1]{%
{:\mathrel{\mspace{1mu}#1\mspace{1mu}}:}
}

\usepackage[colorinlistoftodos,prependcaption,textsize=footnotesize]{todonotes}
\newcommand{\noteF}[1]{\todo[inline,,linecolor=blue,backgroundcolor=blue!15,bordercolor=blue!80,]{\textbf{Francesca:} #1}}
\newcommand{\noteM}[1]{\todo[inline,,linecolor=red,backgroundcolor=red!15,bordercolor=red!80,]{\textbf{Markus:} #1}}
\newcommand{\noteG}[1]{\todo[inline,,linecolor=purple,backgroundcolor=purple!15,bordercolor=purple!80,]{\textbf{Guillermo:} #1}}
\newcommand{\noteT}[1]{\todo[inline,,linecolor=green,backgroundcolor=green!15,bordercolor=green!80,]{\textbf{Thiago:} #1}}

\begin{document}
\title{Taming Thiemann's Hamiltonian constraint in canonical loop quantum gravity: reversibility, eigenstates and graph-change analysis}

\author{T. L. M. Guedes}
\email{t.guedes@fz-juelich.de}
\affiliation{Institute for Quantum Information, RWTH Aachen University, D-52056 Aachen, Germany}
\affiliation{Peter Gr{\"u}nberg Institute, Theoretical Nanoelectronics, Forschungszentrum J{\"u}lich, D-52425 J{\"u}lich, Germany}

\author{G. A. Mena Marug{\'a}n}
\email{mena@iem.cfmac.csic.es}

\affiliation{Instituto de Estructura de la Materia, IEM-CSIC, C/ Serrano 121, 28006 Madrid, Spain}

\author{M. M{\"u}ller}
\email{markus.mueller@fz-juelich.de}
\affiliation{Institute for Quantum Information, RWTH Aachen University, D-52056 Aachen, Germany}
\affiliation{Peter Gr{\"u}nberg Institute, Theoretical Nanoelectronics, Forschungszentrum J{\"u}lich, D-52425 J{\"u}lich, Germany}

\author{F. Vidotto}
\email{fvidotto@uwo.ca}
\affiliation{Instituto de Estructura de la Materia, IEM-CSIC, C/ Serrano 121, 28006 Madrid, Spain}
\affiliation{Department of Physics and Astronomy, Department of Philosophy, and Rotman Institute, Western University,
N6A5B7 London, Ontario, Canada.}

\begin{abstract}

One of the key concepts in loop quantum gravity is the quantization of spacetime geometry, with discrete observables such as the quantum area and volume. The quantum state of the gravitational field is encoded in so-called spin networks, and the conventional quantum-mechanical dynamics is substituted by a description in terms of constrained quantum states, in which several constraints define the physical subspace of the Hilbert space. One of these constraints, commonly called Hamiltonian constraint, remains an elusive object in loop quantum gravity because its action on spin networks leads to changes in their corresponding graphs. As a result, calculations in loop quantum gravity are often considered unpractical, and neither the eigenstates of the Hamiltonian constraint, which form the physical space of states, nor the concrete effect of this graph-changing character on observables are entirely known. Much worse, there is no reference value to judge whether the commonly adopted graph-preserving approximations lead to results anywhere close to the non-approximated dynamics. Our work sheds light on several of these issues, by devising a new numerical tool that allows us to implement the action of the Hamiltonian constraint without the need for approximations and to calculate expectation values for the geometric observables. To achieve that, we fill the theoretical gap left in the derivations of the action of the Hamiltonian constraint on spin networks: we provide the first complete derivation of such action for the case of 4-valent spin networks, while updating the corresponding derivation for 3-valent spin networks. Our derivations also include the action of the volume operator. By proposing a new approach to encode spin networks into functions of lists and the derived formulas into functionals, we implement both the Hamiltonian constraint and the volume operator numerically. We are able to transform spin networks with graph-changing dynamics perturbatively and verify that the expectation values for the volume have rather different behaviour from the approximated, graph-preserving results. Furthermore, using our tool we find a family of potentially relevant solutions of the Hamiltonian constraint. Our work paves the way to a new generation of calculations in loop quantum gravity, in which graph-changing results and their phenomenology can finally be accounted for and understood.
\end{abstract}

\maketitle

\section{Introduction}
Although current experiments are still far from observing any traces of quantum behavior in gravity~\cite{search_qgrav1, search_qgrav2, search_qgrav3}, the necessity of a convergence between quantum physics and general relativity has been conceptually established since the pioneering works of Bronstein~\cite{bronstein}, Dirac \cite{Dirac1, Dirac2}  and Hawking~\cite{hawking1}, among others~\cite{Wheeler, Dewitt}. The search for a quantum theory of gravity led to several proposals, one of which, loop quantum gravity (LQG), has at its core the idea of quantized spacetime  geometry. The theory is based on a recasting of the Einstein equation in terms of holonomies in a compact gauge group and fluxes of canonically conjugate densitized triads, constructed with the so-called Ashtekar-Barbero variables~\cite{ashtekar1, Barbero, Lee_knots}.  These new fields allowed for a derivation of Hamiltonian constraints for the gravitational field~\cite{thiemann} and a quantization protocol in the molds of Dirac's quantization~\cite{ashtekar_review, Dirac2}. 

One of the bases commonly used in LQG is spanned by eigenstates of certain geometric operators, the so-called spin networks (these are closely related, for instance, to ribbon graphs and string nets~\cite{koenig, stringnet_simulation, Fibonacci}). These are graphs with spins/colours [or more formally representations of the SU(2) group] assigned to their links and nodes that form singlets out of the spins of incoming and outgoing links (in other words, this enforces the decomposition of the input irreducible representations into the output ones). Spin networks provide a powerful graphical tool to perform and represent otherwise cumbersome calculations~\cite{Ilkka}, and have been in use since the advent of the quantum mechanics of angular momenta~\cite{temperley, Brink, Messiah}. A major difficulty in canonical LQG calculations, however, is the graph-changing effect of the Hamiltonian constraint on spin networks, which generates superpositions of spin networks with different graphs from each input spin network, and therefore can exponentially increase the number of intervening states in computations. This work aims at contributing to fill this gap by providing a complete derivation of the action of the Hamiltonian constraint on 3- and 4-valent nodes. A related goal is to numerically implement the corresponding formulas through a novel spin-network-encoding approach in order to understand the effect of graph changes on geometric observables, like the volume. In this way we can also elucidate the validity of the commonly employed graph-preserving approximations, as well as search for new solutions of the Hamiltonian constraint. 

This article is the companion paper of a letter, where we summarize and highlight the most important results of our investigation without details about technical aspects \cite{letter}. It
is structured as follows. In Sec.~\ref{main} we summarize the main findings of our work, before delving into them. In Sec.~\ref{overview}, we introduce the Hamiltonian constraint and its basic building blocks. In Sec.~\ref{recoupling_theory} and Sec. V, we introduce the mathematical machinery used throughout the calculations, namely, recoupling theory and intertwiners. Sections~\ref{sec_3_val} and \ref{sec_4_val} use recoupling theory to derive the action of the Hamiltonian constraint on 3- and 4-valent node-like spin networks, respectively. In Sec.~\ref{Q_volume} we consider the action of the volume operator.  In Sec.~\ref{numerics}, we introduce our encoding of spin networks and operators, showing how we are able to apply the Hamiltonian constraint on spin networks and evaluate the volume expectation values. In  Sec.~\ref{results}, we present and discuss our results for the volume of spin networks perturbatively transformed by the unitary generated by the constraint. Finally, Sec.~\ref{conclusions} contains our closing remarks, briefly discussing the consequences of our findings. 

\section{Main results}\label{main}

We start with a brief overview of the theory and a derivation of the action of the (scalar) Euclidean Hamiltonian constraint (referred to simply as Hamiltonian whenever there is no risk of confusion) on 3-valent and 4-valent spin networks, the simplest duals to triangulations of bi- and tri-dimensional hypersurfaces. For the former case, as in Ref.~\cite{Ma}, our derivations can be considered an update of those presented in Refs.~\cite{borissov, gaul}, in which an approach based on the much less manageable Temperley-Lieb algebra was employed~\cite{temperley}. Moreover, we show that working with modern conventions leads to somewhat different results (cf. Appendix~\ref{app2}). In the case of 4-valent nodes, our derivations are an extension, as well as a correction, of those presented in Ref.~\cite{Alesci}. This is the first in-depth derivation of the action of the Euclidean Hamiltonian constraint on 4-valent nodes using the modern graphical-calculus machinery~\cite{Ilkka, Messiah}. It serves as a guide for both experts and beginners in LQG, as well as for those generally interested in spin network calculus for other purposes, like studies of non-Abelian topological error-correction codes~\cite{koenig, stringnet_simulation, Fibonacci}. 

\begin{figure*}
    \centering
    \includegraphics[width=0.75\linewidth]{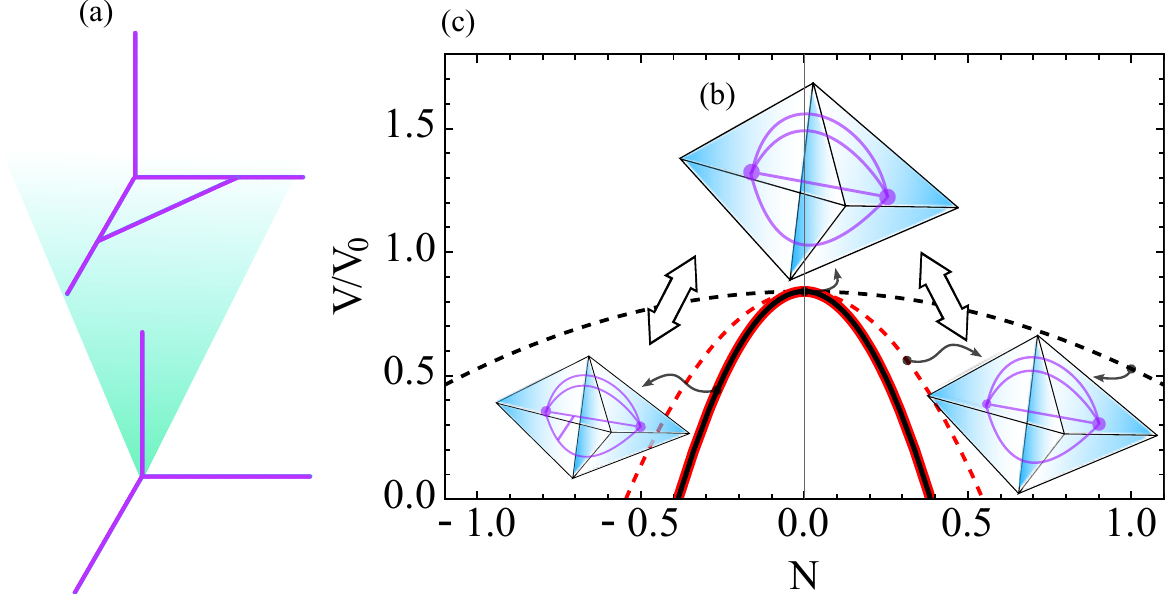}
 \caption{Schematic representations of spin networks. (a) A $3$-valent spin network node (bottom) is transformed under the action of the Hamiltonian to give a modified structure containing an inner loop (top), exemplifying the graph-changing character of the non-approximated Hamiltonian constraint. (b) A minimal example of spin network: the dipole model. Two $4$-valent nodes are connected through their links pairwise, so that the dual to such graph is formed by two tetrahedra with faces glued pairwise (what cannot be visualized in 3D). These tetrahedra represent the quanta of volume. (c) Under the action of a unitary generated by the Hamiltonian, with a perturbative evolution parameter $N$ called lapse (which mathematically behaves similarly to the time in standard quantum mechanics), a transformed spin network node behaves differently when either graph-changing or graph-preserving dynamics are considered. The volume, with contributions up to 2nd order in $N$ displayed, decreases much slower with $N$ when the approximation of nonchanging graphs is adopted (dotted curve, right dipole), while the correct dynamics produces a steeper volume reduction with $N$ at leading perturbative order (solid curve, left dipole). The data for the volume dependence on $N$ for these two cases considers a single spin network node with spins $1/2$ on its four links and spin $0$ (red) or $1$ (black) in the central link (not displayed in the dipole model) and therefore does not entirely represent the dynamics of the dipole model, which was included for illustrative purposes.}
  \label{fig1}
\end{figure*}

In addition, we introduce a new computational tool, concretely as a Mathematica code, that implements the action of the Hamiltonian constraint on spin network nodes through a newly devised numerical approach. A key feature of this approach is a map between spin networks and functions of lists, on which the Hamiltonian acts as a functional. The (symbolic) calculations, performed in a computer based on our analytical formulas, involve no approximations in the Hamiltonian and therefore represent the first complete, graph-changing, application of the Euclidean Hamiltonian constraint on spin networks with low-valence nodes in vacuo (and one of the first numerical works in canonical LQG~\cite{Hanno, Hanno2, Ilkka_GC}), as well as of the unitary transformation it generates, expanded perturbatively. We generate numerical data for the volume expectation values of two perturbatively transformed fiducial spin networks of valence 4 using the lapse as a perturbation parameter and the Hamiltonian constraint as a unitary-transformation generator. Our perturbative expansion of the unitary goes up to the 3rd order when we consider nodes with the same link orientation, up to diffeomorphisms, as the dual graph of a tetrahedron, and up to 4th order when we prevent the Hamiltonian from acting on one of the links. We compare the results with the corresponding data generated with a graph-preserving Hamiltonian and present the first concrete indication that such Hamiltonians fail to capture the proper dynamics of LQG spin networks. As we anticipate in Fig.~\ref{fig1}(c), the expectation values of the quantum volume are rather different between graph-changing and graph-preserving scenarios. 

Furthermore, we use our code to look for low-spin eigenstates of the Euclidean Hamiltonian, showing that simple eigenstates do exist: states with only vanishing spins meeting at the intertwiner. Note that these eigenstates also include spin networks with large numbers of inner loops, as long as the innermost links meeting at the intertwiner are in the trivial representation.  Lastly, we propose a more complex family of solutions built from any desired spin network, the properties of which should be investigated in follow-up works.  

\section{Overview}\label{overview}

Using Ashtekar-Barbero variables, the Einstein-Hilbert action can be recast in terms of smearings over three sets of constraints corresponding to gauge invariance, diffeomorphism invariance and (Euclidean) time reparametrization~\cite{thiemann}. In the quantum theory, gauge invariance is well understood and consideration solely of spin networks with spin singlets at every node (also called intertwiners) suffices to satisfy the corresponding constraints. In precise mathematical terms, gauge invariance enforces that the Clebsch-Gordan inequalities are fulfilled at every node. Spatial diffeomorphism invariance is a key symmetry in general relativity and topological field theories~\cite{koenig}, both of which participate in the construction of LQG. In terms of spin networks embedded in manifolds, spatial diffeomorphisms can be well understood as (invertible) smooth \footnote{In priniciple our focus is on analytic deformations, but see nonetheless the technical subtleties commented in the last paragraph of this section.}
deformations of the spin network graphs. In a very simplified description, to satisfy the diffeomorphism constraint, one needs to consider equivalence classes of (dual) spin networks with respect to diffeomorphisms~\cite{rovelli2004book, ashtekar_review}: all graphs related to each other by smooth deformations should be superposed to compose states that satisfy the diffeomorphism constraints. 

The last constraint, commonly known as scalar or Hamiltonian constraint, dictates the dynamics of spin networks, and we will sometimes refer to it simply as the Hamiltonian, for the sake of analogy with standard quantum mechanics. When neither matter nor a cosmological constant are considered, the eigenstates of all three constraints with null eigenvalue are the physical states of the theory (strictly speaking, those normalizable with respect to a suitable inner product). On the other hand, for the scalar constraint, nonzero eigenvalues might represent, for example, physical states of the geometry in the presence of classical matter or a nonzero cosmological constant, both of which are common in the formulation of loop quantum cosmology~\cite{Ashtekar_cosmo, LQC_Guillermo}.

In the absence of matter or a cosmological constant, we can construct the scalar constraint from the volume operator $\hat{V}$ and the holonomies $\hat{h}[p]$ (the link-related parallel-transport operators commonly encountered in lattice gauge theories) along a path $p$, 
\begin{equation}
\begin{aligned}
    & \hat{C}_s =\\ &\lim_{\boxtimes \to 0} \sum_{\boxtimes} \frac{iN_{\boxtimes}\epsilon_{ijk}}{3 l^2_0} \text{tr}  \left\{ \hat{h}[\alpha_{ji}]-\hat{h}[\alpha_{ij}], \hat{h}[p_k] \hat{V} \hat{h}^{-1}[p_k] \right\} .
    \label{scalar_constr}
    \end{aligned}
\end{equation}
The above definition follows the proposal introduced by Thiemann in Refs.~\cite{thiemann, QSD, QSD2} and further investigated in Ref.~\cite{borissov}. In this equation, the large curly brackets stand for the anticommutator, the symbol $\text{tr}$ is the trace and $\epsilon_{ijk}$ is the totally antisymmetric symbol. The symbol $\boxtimes$ represents a partition of the manifold into tetrahedra and the limit $\boxtimes\to 0$ means that the size of those tetrahedra gets infinitesimally small (yet still nonzero~\cite{QSD2}), while their number diverges. It must be noted, however, that different partition schemes using, e.g., prisms of choice (which can still be broken down into tetrahedra), are possible, all of which lead to the same equations up to some prefactors \cite{Thiemann2007book, QSD}. As a result of this regularization procedure, only the tetrahedra based at the nodes of the spin networks will contribute to Eq.~\eqref{scalar_constr} and no tetrahedron will ever contain more than one node~\cite{rovelli2004book} (in fact, shrinking the tetrahedra to a size at which they contain either one or no nodes suffices to describe the effect of this limit). The prefactor $N_\boxtimes$ is called lapse and results from the Riemannian discretization of a distribution that provides a Lagrange multiplier to integrate  the constraint. In Eq.~\eqref{scalar_constr}, the lapse serves as an amplitude modulator for the action of the constraint in each tetrahedron, which effectively translates a 3D foliation of spacetime (or a triangulation thereof) along a timelike vector proportional in absolute value to $ N_\boxtimes$. The paths $\alpha_{ij}$ and $\alpha_{ji}$ are ``triangular'' loops of opposite orientations (i.e., $\alpha_{ij}=\alpha^{-1}_{ji}$),  with segments tangent to two linearly independent links (labelled by $i$ and $j$) from a (physical) node and span one of the faces of a regularization tetrahedron. The path $p_k$ is a line segment tangent to yet another link from the node (labelled by $k$), linearly independent from the two links $i$ and $j$, and spanning one of the edges of a tetrahedron. The holonomies couple additional spins (in the sense of a Clebsch-Gordan spin addition) to the respective links of the spin network on which $\hat{C}_s $ acts. Moreover, because $\alpha_{ij}$ and $\alpha_{ji}$ are (closed) loops, the term $\hat{h}[\alpha_{ij}]-\hat{h}[\alpha_{ji}]$ can add one additional link (between $i$ and $j$) to the spin network [see Fig.~\ref{fig1}(a)].

Finally, the volume is a key geometric operator in LQG~\cite{Ashtekar_area, ashtekar_volume, Thiemann_volume, Giesel_volume1, Giesel_volume2, volume, Ma_volume_arxiv}, extracting information about the (quantum) geometry of spacetime from quantum states. Discrete eigenvalues of the volume are based on the spins of the links connected to a certain node of valence 4 or higher, with $l_0$ being the Planck length. This provides an interpretation of spin networks as the dual to a triangulation of a manifold, associating a link to each face (2-simplex) of the triangulation and a node to each tetrahedron (3-simplex), or polyhedron in the most general case. In this sense, two nodes connected by four links can be seen as two tetrahedra with pairwise connected faces [see Fig.~\ref{fig1}(b)]. The matrix elements of $\hat{V}$ will be introduced later in the calculations. Since the volume depends on the relative arrangement of links at each node~\cite{ashtekar_volume}, we consider spin networks with linearly independent triplets of links, each oriented along a face of a tetrahedron. Diffeomorphisms (or averaging by them) should not influence the effect of the volume or the constraint on these spin networks~\cite{QSD}. One possible exception, however, is given by spin networks that had one or more of their links removed by the constraint, in which case two diffeomorphically nonequivalent spin networks can generate equivalent ones after removing a link by the action of the constraint~\cite{QSD2}.

At this point, it is worth noting that Thiemann presents two possibilities for the implementation of a symmetric constraint operator in Ref.~\cite{QSD2} . In his first proposal, a repeated action of the constraint~\eqref{scalar_constr} introduces inner loops progressively deeper, without ever removing them. Then, in order to obtain a symmetric operator on spin networks, the author suggests to add by ``brute force" the Hermitian-conjugate term to every matrix element of the operator, i.e.,  $\langle \psi | \hat{C}_s|\phi\rangle \to  \langle \psi | \hat{C}_s|\phi\rangle + \langle \phi | \hat{C}_s|\psi\rangle^* $. On the other hand, the second proposal considers that the added links belonging to the inner loops are smooth rather than analytic, and that they intersect a pair of analytic spin network links at nodes such that all the links have collinear tangents. The collinearity of the links allows these nodes of the inner loops to have arbitrary valence, yet not be changed by the Hamiltonian because nodes with collinear links (as well as 3-valent nodes with coplanar links) are volume eigenstates with zero eigenvalue~\cite{QSD2}, granting an anomaly-free action for this symmetric constraint. Although this would allow two inner loops with a single common link to intersect at the same node of this link, we will neglect these subtleties while adhering to the second proposal for a symmetric constraint. It is worth noting that this symmetrization additionally requires a modification of the triangulation scheme: when the Hamiltonian acts on a node introducing a loop in the location where the deepest inner loop is, these loops get coupled (i.e., the regularization tetrahedra there should match the deepest inner loop and therefore cannot be shrunk to arbitrarily small sizes). In this way, a loop link can decrease its spin and be removed. We also note that both symmetrization approaches involve changes caused exclusively in the vicinity of the spin network nodes, preventing ``long-range'' couplings between different nodes by the constraint and implying that the nodes created by loop couplings cannot have additional loops coupled to them (so that the constraint commutes with itself, rendering it anomaly free~\cite{QSD}), differently from what happens in covariant loop quantum gravity.

\section{SU(2) recoupling theory}\label{recoupling_theory}

Before deriving the action of Eq.~\eqref{scalar_constr} on spin network nodes, we introduce the main working tools from recoupling theory, i.e., the graphical calculus involving elements and representations of the SU(2) group. Early papers in LQG~\cite{borissov, gaul, volume, Smolin} made use of the now ``old fashioned", yet more graphically intuitive description of such systems in terms of Temperley-Lieb tangles~\cite{temperley}, which are closely related to knots~\cite{rovelli_knots}. Tangles are usually proportional to spin networks~\cite{volume}, but the complicated conversion factors between them, which lead to the need for normalization not only in the states, but also in commonly used functions like the Wigner 3j, 6j and 9j symbols, make the Temperley-Lieb approach less attractive when one aims at robust calculations that can be performed numerically. For completeness, we report the derivations using the Temperley-Lieb algebra in Appendix~\ref{Temperley-Lieb}, while here instead we focus on the modern convention for recoupling theory, mainly following the notation of Ref.~\cite{Ilkka}, as well as some identities from Ref.~\cite{Messiah}. In this convention, the Wigner 3j, 6j and 9j symbols are the same that are implemented in Mathematica. 

The key idea of recoupling theory is to represent the SU(2) group elements $g$ in a given representation $j \in \mathbb{N}/2$ (where we consider the naturals $\mathbb{N}$ to include zero), as well as its coupling to other elements of the same group in possibly different representations, in graphical form. Starting from the simplest element in any representation, the identity, we write a single straight line with ends carrying the two indices of the identity matrix (e.g., for $j=1/2$, the $2\times 2$ matrix has two indices commonly associated with spin magnetic numbers $\pm 1/2$),
\begin{equation}
\begin{tikzpicture}[baseline=(current  bounding  box.center)];
\draw (0.7,0) --  node[above] {$j$} (2.2,0);
\node at (-0.4,0.05) {$\delta^{(j){n}}_{m}=  $};
\node at (0.4,0) {$m$};
\node at (2.5,0) {$n$} ;
\node at (2.7,0) {.};
\end{tikzpicture}   
\label{identity}
\end{equation}
and with the representation indicated above the corresponding link. For a given $j$, there are $d_j = 2j+1$ possible choices of indices, and once two matrices are contracted, summation over the indices at the corresponding connected ends of the graphical representation is implied. If one therefore connects the two opposite ends of the identity, forming a closed loop, one ends up with its trace, which is simply $d_j$. 

Another SU(2) element that deserves its own graphical representation is given by the $j$-representation tensor $\epsilon^{(j) mn}=\epsilon^{(j)}_{mn}=(-1)^{j-m}\delta^{(j)}_{m,-n}=(-1)^{2j}\epsilon^{(j)}_{nm}$, for which  $\epsilon^{(j)}_{mn}\epsilon^{(j)nk}=(-1)^{2j}\delta^{(j)k}_{m}$. Graphically, this tensor is represented by a small solid arrow pointing from $m$ to $n$,
\begin{equation}
\begin{tikzpicture}[baseline=(current  bounding  box.center)];
\draw (0.7,0) --  node[above] {$j$} (2.2,0);
\draw[-stealth] (0.7,0) --  (1.45,0);
\node at (-0.4,0.05) {$\epsilon^{(j)}_{mn}=  $};
\node at (0.4,0) {$m$};
\node at (2.4,0) {$n$};
\node at (2.7,0) {.};
\end{tikzpicture} 
\label{levi}
\end{equation}
Consequently, one has the graphical relations
\begin{equation}
\begin{tikzpicture}[baseline=(current  bounding  box.center)];
\draw (0.7,0) --  node[above] {$j$} (2.7,0);
\draw[-stealth] (0.7,0) --  (1.2,0);
\draw[-stealth] (1.2,0) --  (2.2,0);
\node at (-0.8,0.05) {$\epsilon^{(j)}_{mn}\epsilon^{(j) nk}=  $};
\node at (4.3,0.05) {$=(-1)^{2j}\delta^{(j)k}_{m}  $};
\node at (0.4,0) {$m$};
\node at (3,0) {$k$};
\node at (5.6,-0.05) {,};
\end{tikzpicture} 
\label{levi2}
\end{equation}
\begin{equation}
\begin{tikzpicture}[baseline=(current  bounding  box.center)];
\draw (0.7,0) --  node[above] {$j$} (2.7,0);
\draw[-stealth] (0.7,0) --  (1.2,0);
\draw[stealth-] (2.2,0) --  (2.7,0);
\node at (-0.8,0.05) {$\epsilon^{(j)}_{mn}\epsilon^{(j) kn}=  $};
\node at (3.8,0.05) {$=\delta^{(j)k}_{m}  $};
\node at (0.4,0) {$m$};
\node at (3,0) {$k$};
\node at (4.6,-0.05) {.};
\end{tikzpicture} 
\label{levi3}
\end{equation}
It should be noted that flipping the arrow in Eq.~\eqref{levi} corresponds to swapping the order of the indices, which leads to a prefactor of $(-1)^{2j}$. Since $(-1)^{4j}=1$, Eq.~\eqref{levi3} can be derived from Eq.~\eqref{levi2} through an arrow flip. The tensor $\epsilon^{(j) mn}$ is invariant under SU(2) transformations: given a Wigner matrix $D^{(j)m}_n (g)$ for the SU(2) element $g$, $D^{(j)m}_n (g)\epsilon^{(j)}_{mp} D^{(j)p}_q (g)=\epsilon^{(j)}_{ nq}=D^{(j)n}_m (g)\epsilon^{(j)}_{ mp} D^{(j)q}_p (g)$. Graphically, the Wigner matrix is represented by a triangle with the group element $g$ within,
\begin{equation}
 
\label{element}
\end{equation}
The invariance of the tensor $\epsilon^{(j)}_{ nq}$ is therefore graphically represented as
\begin{equation}
%
 
\label{levi_invar}
\end{equation}
A similar relation holds when the triangles point toward the ``free" ends of the links. It is worth noting that the invariance of $\epsilon^{(j)}_{ nq}$ holds for any $g$ as long as it is present on both Wigner matrices. Using Eq.~\eqref{levi_invar}, it is possible to show that the Wigner matrices can be inverted through contraction with $\epsilon^{(j)}_{mn}$ on both of their indices,
\begin{equation}
%
 
\label{wigner_inverse}
\end{equation}

Equations \eqref{identity}-\eqref{wigner_inverse} span the basic relations for representing single SU(2) elements graphically as links. Consideration of graphs, however, requires the coupling of several such links at nodes according to the SU(2) decomposition rule into irreducible representations. This enforces the Clebsch-Gordan (also known as triangularity) conditions on the spins meeting at a certain node. Mathematically, the (nontrivial) minimal-valence coupling is enforced by a Wigner 3j symbol, an object proportional to the Clebsch-Gordan coefficients. The Wigner 3j symbol is graphically represented as a 3-valent node with a certain cyclicity that describes whether the columns of the 3j symbol are ordered clockwise ($-$) or counter-clockwise ($+$) at the node,
\begin{equation}
\begin{tikzpicture}[baseline=(current  bounding  box.center)];
\node at (1.5,0) {$ \begin{pmatrix}
j_1 & j_2 & j_3\\
m_1 & m_2 & m_3
\end{pmatrix} =$};
\draw (3.1,-0.7)  --(3.7,-0.1)--(3.7,0.75);
\draw (3.7,-0.1)--(4.3,-0.7);
\node at (3,-0.4) {$j_2$};
\node at (3.5,0.6) {$j_1$};
\node at (4.4,-0.4) {$j_3$};
\node at (3.7,-0.35) {$+$};
\draw (5.3,-0.7)  --(5.9,-0.1)--(5.9,0.75);
\draw (5.9,-0.1)--(6.5,-0.7);
\node at (5.2,-0.4) {$j_3$};
\node at (5.7,0.6) {$j_1$};
\node at (6.6,-0.4) {$j_2$};
\node at (5.9,-0.35) {$-$};
\node at (4.8,0) {$=$};
\end{tikzpicture}.
\label{wigner3j}
\end{equation}
Swapping the cyclicity of the node, which is also known as braiding, leads to a phase factor of $(-1)^{j_1+j_2+j_3}$. Since $j_1, j_2, j_3$ must fulfill the Clebsch-Gordan conditions, such that $j_1+j_2+j_3 \in \mathbb{N}$, a double braid leads to a prefactor of $(-1)^{2(j_1+j_2+j_3)}=1$.
On the leftmost side of Eq.~\eqref{wigner3j}, the Wigner 3j symbol contains an upper row of spins (irreducible representations) and a lower row of associated spin projections/magnetic numbers (tensor indices in a given irreducible representation). The Wigner matrices $D^{(j)m}_n (g)$, represented as in Eq.~\eqref{element} [which includes both {Eqs.} \eqref{identity} and \eqref{levi} as special cases] are then contracted with the node legs of the same representation in Eq. \eqref{wigner3j}, which implies a summation over the lower entries in the latter. The Wigner 3j symbol is also invariant with respect to SU(2) transformations, i.e., it returns the node when this is contracted with three inwards oriented or three outwards oriented Wigner matrices representing the same SU(2) element $g$,
\begin{equation}
\begin{tikzpicture}[baseline=(current  bounding  box.center)];
\draw (3.3,-0.5) --(3.5,-0.7)--(3,-0.8)--(3.1,-0.3)--(3.3,-0.5)  --(3.7,-0.1)--(3.7,0.35)--(4,0.35)--(3.7,0.75)--(3.4,0.35)--(3.7,0.35);
\draw (3.7,-0.1)--(4.1,-0.5)--(3.9,-0.7)--(4.4,-0.8)-- (4.3,-0.3)--(4.1,-0.5);
\draw  (4.4,-0.8)--(4.5,-0.9);
\draw  (3.7,0.75)--(3.7,0.9);
\draw  (3,-0.8)--(2.9,-0.9);
\node at (3.2,-0.6) {$g$};
\node at (3.7,0.5) {$g$};
\node at (4.2,-0.6) {$g$};
\node at (2.85,-0.25) {$j_2$};
\node at (3.4,0.75) {$j_1$};
\node at (4.55,-0.25) {$j_3$};
\node at (3.7,-0.35) {$+$};
\draw (5.4,-0.7)  --(6,-0.1)--(6,0.75);
\draw (6,-0.1)--(6.6,-0.7);
\node at (5.3,-0.4) {$j_2$};
\node at (5.8,0.6) {$j_1$};
\node at (6.7,-0.4) {$j_3$};
\node at (6,-0.35) {$+$};
\node at (4.95,0) {$=$};
\end{tikzpicture}.
\label{wigner3j_invar}
\end{equation}
The SU(2) invariance of the Wigner 3j symbol means that, regardless of their representations, the Wigner matrices can be partially or even entirely absorbed into the nodes in graphical notation, depending on which elements $g$ are connected to each node. As an example, if all three legs of a given node are contracted with $\epsilon^{(j)}_{ nq}$ symbols, therefore displaying three outwards or inwards oriented solid arrows in graphic form, they can be all simultaneously incorporated into the node. Lastly, there is a correspondence between $\epsilon^{(j)}_{ nq}$ and the Wigner 3j symbols, which graphically takes the form
\begin{equation}

\label{wigner3j_to_levi}
\end{equation}

Now that the coupling of three links at a node has been introduced, we present a few graphical relations involving nodes and links. The first two relations allow for major simplifications during calculations,
\begin{equation}
%
 
\label{bubble}
\end{equation}
Note that these graphical relations are of topological nature, in the sense that rotating or deforming them without crossing links does not affect the outcomes. 

Two (identity) links of arbitrary spins can be coupled by introducing two nodes connecting them according to
\begin{equation}
%
 
\label{merge}
\end{equation}
The sum on the right-hand side of Eq.~\eqref{merge} runs, in principle, over $j \in \mathbb{N}/2$, but considering that the two nodes enforce that the triangle inequality has to be fulfilled by the three spins meeting at them, the sum over $j$ runs effectively from $|j_1-j_2|$ to $j_1+j_2$.
Although Eq.~\eqref{merge} can also be used on links with arbitrary group elements assigned to them, this requires extending the corresponding Wigner-matrix links by contracting one of their ends with identities, so that the identity segments of the links can be merged through \eqref{merge} and the Wigner matrices, in the form of Eq.~\eqref{element}, sit at the external legs of one of the nodes on the right-hand side of Eq.~\eqref{merge}. When the same group element is assigned to both of the links merged through this relation, however, the node invariance [Eq.~\eqref{wigner3j_invar}] can be used to give an expression corresponding to the direct coupling of two Wigner matrices of the same $g$,
\begin{equation}
 
\label{element_merge_inv}
\end{equation}
The manipulation of structures like the ones on the right-hand side of Eqs. \eqref{merge}-\eqref{element_merge_inv} is facilitated by
\begin{equation}
%
 
\label{prepachner}
\end{equation}
which accounts for the swapping of the legs of spins $j_2$ and $j_3$. The symbol within curly brackets in Eq.~\eqref{prepachner} is the Wigner 6j symbol, defined through the contraction of four Wigner 3j symbols (note that, for contraction, one needs $\epsilon^{(j)}_{ nq}$ to convert upper into lower indices and vice versa). In graphical form, this contraction is represented as
\begin{equation}
%
 
\label{tetrahedron}
\end{equation}
The three arguments in the upper row of the Wigner 6j symbol [right-hand side of Eq.~\eqref{tetrahedron}] are the spins of the three legs of any chosen node of the tetrahedron; the remaining arguments are the corresponding spins of opposite links of the tetrahedron (e.g., $j_i$ is opposite to $k_i$) organized column-wise. The  Wigner 6j symbol has a high degree of symmetry: permuting its columns gives the same outcome, as well as swapping the upper and lower arguments in any two chosen columns. On top of that, owing to the presence of four Wigner 3j symbols, the Wigner 6j symbol is nonzero only if the triangle inequalities are simultaneously satisfied in all of the nodes of the tetrahedron, i.e., for the sets $\{j_1, j_2, j_3\}$, $\{j_1, k_2, k_3\}$, $\{k_1, j_2, k_3\}$ and $\{k_1, k_2, j_3\}$.
Similarly, contracting six Wigner 3j symbols gives the Wigner 9j symbol,
\begin{equation}
 
\label{hexagon}
\end{equation}
Under permutations of any two of its rows or columns, the Wigner 9j symbol picks up a phase $(-1)^{j_1 + j_2 + j_3 + k_1 + k_2 + k_3 + l_1 + l_2 + l_3}$. If two of its columns or rows are identical and the phase factor is negative, the Wigner 9j symbol is therefore zero.

Combining Eq.~\eqref{prepachner} with Eqs. \eqref{levi} and \eqref{wigner3j}, one can derive the so-called 2-2 Pachner move (see Appendix~\ref{appendix_recoup}),
\begin{equation}
%
 
\label{pachner}
\end{equation}
This transformation represents a change of intertwiner basis, which will be further discussed in the next section. The 2-2 Pachner move is also known as F-move in the field of non-Abelian anyonic quantum error correction and plays an important role in implementing lattice surgeries, Dehn-twists and braid-moves~\cite{koenig, stringnet_simulation}.

Considering the large number of sums over spins, as well as of Wigner symbols summed over, it is advantageous to employ a few identities involving such quantities (cf. Ref.~\cite{Messiah}, Eqs. C.35a-e and C.37):
\begin{equation}
 \, .
\label{messiah5}
\end{align}
\end{widetext}

The SU(2) generators, being elements of the su(2) algebra, can also be represented similarly to Wigner matrices. However, the presence of an extra index, say $i$, means that they require one additional leg with respect to Eq.~\eqref{element}, with spin 1. These elements are sometimes called grasps and are a key component of the quantum volume operator. Their graphical representation is
\begin{equation}
%
 
\label{grasp}
\end{equation}

\section{Intertwiners and spin networks}\label{intertwiners_SN}

Intertwiners are equivariant multilinear maps between tensor products of SU(2) representations. In other words, they are invariant tensors of the SU(2) group. Given an arbitrary number of representations $j_i$ acting on Hilbert spaces $\mathcal{H}_{j_i}$, the intertwiners are the elements of the space (of spin singlets) $\text{Inv}_{SU(2)}(\otimes_i \mathcal{H}_{j_i} )$. The simplest (which we will refer as trivial) intertwiner is $\epsilon^{(j)}_{ nq}$, which is the basis element of the 1-dimensional space $\text{Inv}_{SU(2)}(\mathcal{H}_{j}\otimes \mathcal{H}_{j})$ for a given $j$. The next (nontrivial) intertwiner, corresponding to the Wigner 3j symbol, is the sole basis element of the space $\text{Inv}_{SU(2)}(\mathcal{H}_{j_1}\otimes \mathcal{H}_{j_2} \otimes \mathcal{H}_{j_3})$ for a given choice of $j_1$, $j_2$ and $j_3$. As can be seen from Eq.~\eqref{3valcirc}, which can be recast as the contraction of two nodes with the same three spins but different cyclicities [cf. Eq.~\eqref{wigner3j}], i.e., as an inner product of the basis element of $\text{Inv}(\mathcal{H}_{j_1}\otimes \mathcal{H}_{j_2} \otimes \mathcal{H}_{j_3})$ with itself, the norm of the Wigner 3j symbol is 1. The identity of this space can therefore be resolved simply as two ({noncontracted}) copies of the Wigner 3j symbols, a ket followed by a bra (in Dirac's notation). 

In general, $n$-valent intertwiners can be built from 2- and 3-valent ones through contraction. The protocol for construction of $n$-valent intertwiners requires the use of $n-2$ Wigner 3j symbols, each of which has one or two of its indices contracted with one out of $n-3$ $\epsilon^{(j)}_{ nq}$, which bridge the 3j symbols pairwise. The chain of 3-valent nodes so constructed has therefore one ``free" 3-valent-intertwiner leg per inner node and two such legs at the nodes at the ends of the chain. These constructions, however, are in general not unique, as the corresponding space $\text{Inv}_{SU(2)}(\otimes_i \mathcal{H}_{j_i} )$ might have several basis elements corresponding to all possible choices of inner-link spins. The identity $\mathbb{1}$ of such spaces can then be resolved as the weighted sum over all inner spins of the (noncontracted) doubled $n$-valent intertwiners,
\begin{equation}
 .
\label{general_identity}
\end{equation}
For simplicity, we will not use Dirac's bra-ket notation explicitly, and instead we will merely present the spin networks (or intertwiners at their nodes) as quantum states, with ``free" legs pointing in opposite directions for bras and kets [right and left spin networks in Eq.~\eqref{general_identity}, respectively], so that inner products tie legs of corresponding spins [the Wigner 3j symbols are real, so in the absence of Wigner matrices
there is no need for complex conjugation when converting kets into bras]. The sum in Eq.~\eqref{general_identity} runs over all (Clebsch-Gordan-)allowed values for the set of $n-3$ internal spins $\{i\}$. Using Eq.~\eqref{bubble}, it is possible to show that the squared norm of each of the $n$-valent intertwiners in Eq.~\eqref{general_identity} is $ (\prod_l d_{i_l})^{-1}$, what explains the weighting factors in the sum. Eq.~\eqref{general_identity} can greatly simplify calculations, since introducing the intertwiner-space identity graphically accounts for ``breaking" any number of links of a spin network and introducing~\eqref{general_identity} for the corresponding number of external legs, which are contracted with the broken-link ends of the spin network. The concept is similar to lattice surgery~\cite{koenig,surgery1,surgery2}. By introducing the resolution of the identity in spin network calculations, one can therefore ``surgically remove" portions of the spin network which, once contracted with suitable intertwiners, form closed secondary spin networks that can be converted into functions through, e.g.,  Eqs.~\eqref{tetrahedron} and \eqref{hexagon}. As an example that will be useful for later calculations, we consider the following use of the resolution of the identity:
\begin{equation}
 
\label{theorem}
\end{equation}
In Eq.~\eqref{theorem}, the three (magenta) dots represent the specific locations chosen for ``breaking" the links and contracting with the legs of the 3-valent intertwiners that resolve the identity of the space $\text{Inv}(\mathcal{H}_{j_1}\otimes \mathcal{H}_{j_2} \otimes \mathcal{H}_{j_3})$. Note that this contraction requires the braiding of the links of spins $k_1$, {$j_2$} and $k_3$ on the right-hand side of Eq. \eqref{theorem}, explaining the flip in cyclicity. We note that, after accounting for some ``arrow-flipping" and ``braiding-related" phase factors, a 2-2 Pachner move applied on the spin-$j_1$ link of the spin network on the left-hand side of Eq.~\eqref{theorem} would lead to a 6J symbol as the coefficient of a spin network of the form given in Eq.~\eqref{bubble}, with an additional 3-valent intertwiner at one of its ends. After resolving the ``bubble", one equivalently obtains the right-hand side of Eq.~\eqref{theorem}. 

As a special case, which will be of great importance in the study of the action of the LQG Hamiltonian constraint on spin networks, we look at the 4-valent intertwiners. They are composed by two Wigner 3j symbols contracted by means of one $\epsilon^{(j)}_{ nq}$. The choice of the legs that are paired leads to different bases of $\text{Inv}_{SU(2)}(\mathcal{H}_{j_1} \otimes \mathcal{H}_{j_2} \otimes \mathcal{H}_{j_3} \otimes \mathcal{H}_{j_4} )$, which can be mapped into each other by a 2-2 Pachner move {[cf. Eq.~\eqref{pachner}]}. The number of elements in these bases is determined solely by the number of allowed inner-spin values connecting the two nodes: if the central link of spin $i$ pairs the external links of spins $j_1$, $j_2$, $j_3$ and $j_4$, forming triangularity-fulfilling sets $\{j_1, j_3, i\}$ and $\{j_2, j_4, i\}$, then $i$ runs from $\text{max}\{|j_1-j_3|,|j_2-j_4|\}$ to $\text{min}\{|j_1+j_3|,|j_2+j_4|\}$. The 4-valent intertwiners have squared norm $d^{-1}_i$, as can be derived with the aid of Eq.~\eqref{bubble} (which removes the ``inner bubble") followed by Eq.~\eqref{levi3} (to remove the arrows) and Eq.~\eqref{3valcirc},
\begin{equation}
\begin{tikzpicture}[baseline=(current  bounding  box.center)];
\draw (1.4,-0.8) ellipse (1.2 and 0.9);
\draw (1.4,-0.8) ellipse (0.6 and 0.45);
\draw (2.6,-0.8)-- (2,-0.8);
\draw[stealth-] (2.25,-0.8)--(2.6,-0.8);
\draw (0.8,-0.8)--(0.2,-0.8);
\draw[-stealth] (0.2,-0.8)--(0.55,-0.8);
\node at (1.4,-1.03) {$j_2$};
\node at (1.4,0.3) {$j_1$};
\node at (1.4,-0.56) {$j_4$};
\node at (1.4,-1.9) {$j_3$};
\node at (3.7,-0.8) {$-\,=d^{-1}_i\delta_{i,i'} \, .$};
\node at (0,-0.8) {$+$};
\node at (1,-0.8) {$-$};
\node at (1.8,-0.8) {$+$};
\node at (0.55,-0.6) {$i$};
\node at (2.25,-0.6) {$i'$};
\draw[red,fill=red] (1.4,-1.25) circle (0.02);
\draw[red,fill=red] (1.4,-1.7) circle (0.02);
\draw[red,fill=red] (1.4,-0.35) circle (0.02);
\draw[red,fill=red] (1.4,0.1) circle (0.02);
\end{tikzpicture} 
\label{4valnorm1}
\end{equation}
Note that the 4-valent intertwiner on the right part of the graph on the left-hand side of the equation (corresponding to Dirac's ket) has both its 3-valent nodes braided, so that the cyclicities are inverted. If the inner spins $i$ and $i'$ do not match, the inner product gives zero. Similarly, if the external spins of any of the contracted legs do not match, the inner product is also zero, but we will omit the corresponding Kronecker deltas whenever possible. In Eq.~\eqref{4valnorm1}, in order to make its interpretation clearer, we have displayed the connection points between external links of the two 4-valent intertwiners participating in the inner product (represented in red). It is worth noting that, by choosing the intertwiners in a different basis, such as the one obtained after applying a 2-2 Pachner move, the same result can be obtained in a different way. First Eq.~\eqref{bubble} is used to get rid of the upper and lower ``bubbles" in the graph, then Eq.~\eqref{levi3} merges the arrows, and finally the remaining loop, which represents the trace over the identity in the inner-spin representation, gives a factor of $d_i$. This calculation is represented as
\begin{equation}
\begin{tikzpicture}[baseline=(current  bounding  box.center)];
\draw[rounded corners] (0.6, -1.8) rectangle  (2.2, 0.2);
\draw (0.6,-1.4)-- (2.2,-1.4);
\draw[stealth-] (0.6,-0.78)--(0.6,-0.81);
\draw (0.6,-0.2)--(2.2,-0.2);
\draw[-stealth] (2.2,-0.79)--(2.2,-0.82);
\node at (1.4,-1.15) {$j_2$};
\node at (1.4,0.45) {$j_1$};
\node at (1.4,-0.4) {$j_4$};
\node at (1.4,-2.0) {$j_3$};
\node at (3.5,-0.8) {$=d^{-2}_i\delta_{i,i'}$};
\node at (0.4,-0.2) {$+$};
\node at (2.4,-0.2) {$-$};
\node at (0.4,-1.4) {$-$};
\node at (2.4,-1.4) {$+$};
\node at (0.45,-0.8) {$i$};
\node at (2.425,-0.8) {$i'$};
\draw[red,fill=red] (1.4,-1.4) circle (0.02);
\draw[red,fill=red] (1.4,-1.8) circle (0.02);
\draw[red,fill=red] (1.4,-0.2) circle (0.02);
\draw[red,fill=red] (1.4,0.2) circle (0.02);
\draw[rounded corners] (0.6+3.75, -1.6) rectangle  (2+3.75, -0);
\node at (6.85,-0.8) {$=d^{-1}_i\delta_{i,i'}$ \, .};
\node at (1.35+3.75,0.25) {$i$};
\draw[stealth-] (0.6+3.75,-0.78)--(0.6+3.75,-0.81);
\draw[-stealth] (2+3.75,-0.79)--(2+3.75,-0.82);
\end{tikzpicture} 
\label{4valnorm2}
\end{equation}
Note that the deformation of the lines in the graphs, e.g., rounded in {Eq.}~\eqref{4valnorm1} and blocky in {Eq.}~\eqref{4valnorm2}, is irrelevant. All that matters is the adjacency between graphical elements. For the sake of visual simplicity, we will henceforth omit these (red) contraction points.

It might seem that once {we have completed} the study of bases of the $\text{Inv}_{SU(2)}(\otimes_i \mathcal{H}_{j_i} )$ spaces, {we are} ready to proceed with the action of the Euclidean Hamiltonian {constraint on} general spin networks. This perspective, however, fails to acknowledge the importance of a fundamental component of spin networks, namely the Wigner matrices at the links. The holonomies in Eq.~\eqref{scalar_constr} act on spin networks according to Eqs.~\eqref{element_merge} and \eqref{element_merge_inv}, so that the spin networks at both the domain and image of the map \eqref{scalar_constr} contain SU(2) group elements assigned to their links. Once the quantum states are described by spin networks, the inner product requires taking integrals with Haar measure over all SU(2) elements at the links of the spin network. These integrals couple two links with the same assigned SU(2) element, one from the bra and another one from the ket, according to 
\begin{equation}
\begin{tikzpicture}[baseline=(current  bounding  box.center)];
\draw (0.7,0) --  (1,0) -- (1,-0.3) -- (1.5,0) -- (1,0.3) -- (1,0);
\draw (1.5,0) --  (1.8,0);
\node at (1.15,0) {$g$};
\node at (1.2,-0.45) {$j$};
\draw (0.7,0.9) --  (1,0.9) -- (1,0.6) -- (1.5,0.9) -- (1,1.2) -- (1,0.9);
\draw (1.5,0.9) --  (1.8,0.9);
\node at (1.15,0.9) {$g$};
\node at (1.2,1.45) {$j'$};
\node at (2.85,0.45) {$=d^{-1}_j \delta_{j,j'}$};
\node at (0.1,0.45) {$\int \mathrm{d}g$};
\draw (-0.05+4,0.7+0.45)--(-0.05+4,-0.7+0.45);
\draw (0.2+4,0.7+0.45)--  (0.2+4,-0.7+0.45);
\draw[-stealth] (-0.05+4,-0.7+0.45)--(-0.05+4,0.05+0.45);
\draw[-stealth] (0.2+4,-0.7+0.45)--  (0.2+4,0.05+0.45);
\node at (4.4,1.0) {$j$};
\node at (3.75,1.0) {$j$};
\node at (5,0) {.};
\end{tikzpicture} 
\label{haar}
\end{equation}
Note that the orientation of the arrows in Eq.~\eqref{haar} is irrelevant: as long as they are parallel, flipping both of them gives a phase $(-1)^{4j}=1$.
We show below that the Kronecker delta in Eq.~\eqref{haar} is critical to ensure orthogonality of states in the image of the map~\eqref{scalar_constr}.

The operator defined in Eq.~\eqref{scalar_constr} acts on 3-valent node-like spin networks (i.e., spin networks that include the intertwiner and its ``neighborhood", potentially representing a local component of a larger spin network), henceforth denoted NLSNs,
of the form
\begin{equation}

\label{3val_spinnet_total}
\end{equation}
Using Eqs.~\eqref{levi_invar} and \eqref{wigner3j_invar}, several of the SU(2) elements in Eq.~\eqref{3val_spinnet_total} can be absorbed into the ones effectively contributing to the action of the Hamiltonian, namely the set of elements $\{t, l, r, u, e, d\} \to \{1, lt^{-1}, rt^{-1}, u, e, d\} \to \{1, 1, 1, tl^{-1}urt^{-1}, elt^{-1}, drt^{-1}\}$. Renaming $tl^{-1}urt^{-1}=g$ and pushing the elements $elt^{-1}$ and $drt^{-1}$ further down through the spin network with the aid of Eq.~\eqref{levi_invar}, leaves us with the following NLSN:
\begin{equation}
\begin{tikzpicture}[baseline=(current  bounding  box.center)];
\draw (3.7+0.4,0.45) --  (4+0.5,0.45) -- (4+0.5,0.15) -- (4.5+0.5,0.45) -- (4+0.5,0.75) -- (4+0.5,0.45);
\draw (4.5+0.5,0.45) --  (4.8+0.6,0.45);
\draw[-stealth] (5.4,0.45) --  (5.1,0.45);
\draw (3.814,0.05) --  (4.1,0.45)-- (4.75,1.55);
\draw (5.686,0.05) --  (5.4,0.45)-- (4.75,1.55)--(4.75,2.5);
\draw[-stealth] (4.1,0.45)-- (4.425,1);
\draw[-stealth] (5.4,0.45)-- (5.075,1);
\node at (4.65,0.45) {$g$};
\node at (4.75,0) {$\varepsilon$};
\node at (4.55,2.3) {$j_1$};
\node at (5.65,0.45) {$+$};
\node at (3.85,0.45) {$+$};
\node at (4.75,1.25) {$+$};
\node at (5.35,1.05) {$b$};
\node at (3.6,0) {$j_2$};
\node at (5.9,-0) {$j_3$};
\node at (4.15,1.05) {$a$};
\node at (6.6,0) {.};
\end{tikzpicture} 
\label{3val_spinnet}
\end{equation}
In other words, it is possible to assign SU(2) group elements to the links in the representations $j_1$, $a$ and $b$ in Eq.~\eqref{3val_spinnet}, although this would only further complicate calculations. Similarly, the same can be done in the other two nodes connected with the Wigner matrix of the element $g$ in the representation $\varepsilon$, leading to different elements assigned to each of the links in Eq. \eqref{3val_spinnet}. We note, however, that a temporary conversion back to Eq.~\eqref{3val_spinnet_total} is implicit prior to the application of the volume operator, since the grasp operators actually act as derivatives on the SU(2) elements. By comparison with Refs.~\cite{borissov, gaul, Alesci}, it might seem surprising that we are considering a 3-valent NLSN with an additional bridging link between two of its legs, even more when one considers the presence of a Wigner matrix there. As we will show in the next section, however, this is precisely the general form of the spin network nodes generated by the scalar Euclidean Hamiltonian constraint~\eqref{scalar_constr}. In fact, it is also the general form of the input nodes acted upon by Eq.~\eqref{scalar_constr}. If one wants to consider a spin network node with no added loops, applying Eq.~\eqref{wigner3j_to_levi} on both right and left lower nodes of Eq. \eqref{3val_spinnet} when $\varepsilon=0$ gives, up to a normalization factor, the ``naked" 3-valent spin network node (i.e, the intertwiner). In order to investigate the effects of the symmetry/Hermitianity of Eq.~\eqref{scalar_constr} (and therefore also the reversibility of its action) on spin networks, one must by any means start with a spin network of the form~\eqref{3val_spinnet} to be able to observe that the Hamiltonian both increases and decreases $\varepsilon$, therefore removing the Wigner matrix when $\varepsilon \to 0$. Note that setting the SU(2) element in the bridging link equal to $g=1$ in the output spin network precludes the use of Eq.~\eqref{haar} when taking the inner product between spin networks, which basically destroys the orthogonality between NLSNs of the form \eqref{3val_spinnet}. In order to enforce orthogonality between 3-valent NLSNs with different values of any of the spins $j_1$, $j_2$, $j_3$, $a$, $b$ or $\varepsilon$, one applies Eq.~\eqref{haar} on the contraction of two spin networks of the form \eqref{3val_spinnet} according to
\begin{equation}
 
\label{3val_spinnet_norm}
\end{equation}
Note that, since the bra spin network is Hermitian conjugated, its Wigner matrix in Eq.~\eqref{3val_spinnet_norm} (left of the graph) is inverted.
In the first row of Eq.~\eqref{3val_spinnet_norm}, we use Eq.~\eqref{wigner_inverse}, then perform the integral over $g$ according to Eq.~\eqref{haar} and finally use Eq.~\eqref{levi3} to achieve the form in the second row. After application of Eq.~\eqref{bubble}, the graph in the second row of Eq.~\eqref{3val_spinnet_norm} can be converted into the graph in the third row. The bottom-most graph is then converted into a function with the aid of Eqs.~\eqref{levi3} and \eqref{3valcirc}. For spin networks of the form \eqref{3val_spinnet} with different spins at the external legs (say, $\{j_1, j_2, j_3\}$ on the bra and $\{j'_1, j'_2, j'_3\}$ on the ket), the inner product then gives $\delta_{\varepsilon ,\gamma} \delta_{a,\alpha} \delta_{b ,\beta}\delta_{j_1,j'_1} \delta_{j_2,j'_2} \delta_{j_3,j'_3}d^{-1}_{a}d^{-1}_{b}d^{-1}_{\varepsilon}$. 

Similarly, [after properly absorbing SU(2) elements that effectively do not contribute to the action of the Hamiltonian] the 4-valent NLSNs acted upon and generated by Eq.~\eqref{scalar_constr} have the form 
\begin{equation}
 
\label{4val_spinnet}
\end{equation}
Orthogonality between spin networks of the form \eqref{4val_spinnet} can be shown through
\begin{equation}
%
 
\label{4val_spinnet_norm}
\end{equation}
The derivation of Eq.~\eqref{4val_spinnet_norm} follows the same steps as in Eq.~\eqref{3val_spinnet_norm}, with one additional step in the last row, where Eqs.~\eqref{bubble} and~\eqref{levi3} have to be used before Eq.~\eqref{3valcirc}. For spin networks of the form \eqref{4val_spinnet} with different spins at the external legs (say $\{j_1, j_2, j_3, j_4\}$ on the bra and $\{j'_1, j'_2, j'_3, j'_4\}$ on the ket), the inner product then gives $\delta_{\varepsilon ,\gamma} \delta_{a,\alpha} \delta_{b ,\beta}\delta_{i,i'}\delta_{j_1,j'_1} \delta_{j_2,j'_2} \delta_{j_3,j'_3}\delta_{j_4,j'_4}d^{-1}_{a}d^{-1}_{b}d^{-1}_{\varepsilon}d^{-1}_{i}$. It is easy to show that changing the basis of 4-valent intertwiners in both bra and ket does not affect the inner product. 

When applying Eq.~\eqref{scalar_constr} several times on spin networks, patterns with deeper and deeper inner loops connecting pairs of legs emerge. Equation~\eqref{wigner3j_to_levi} shows that, by considering the inner product between two such spin networks with different inner-loop structures, their graphs can be made the same through inclusion of links in the trivial representation [e.g., $\varepsilon=0$ in Eq.~\eqref{4val_spinnet}]. The group elements assigned (pairwise) to the corresponding links on each of the spin networks participating in the inner product are the same, and through Eq.~\eqref{haar} it is possible to see that, when the representations do not coincide, as is the case when any inner loop is missing in one of the spin networks, the inner product is zero [owing to the Kronecker delta in Eq.~\eqref{haar}]. We emphasize that such orthogonality is a direct result of the integration ~\eqref{haar}, and neglecting the presence of a Wigner matrix on any of the inner loops makes sets of non-equivalent spin networks non-orthogonal. One is therefore left only with the task of calculating the norm of such complicated spin networks. This can be performed by induction. Let us assume that we know the squared norm $N_{n-1}$ of a spin network $s_{n-1}$ with $n-1$ inner loops. Adding one extra inner loop as close as possible to the intertwiner (i.e., deeper than any of the $n-1$ loops), gives us the spin network $s_n$. On the other hand, $N_{n-1}$ can be cast as $N^{>1}_{n-1} d^{-1}_i$, where $N^{>1}_{n-1}$ is the coefficient obtained from all inner loops and $d^{-1}_i$ is the norm of the intertwiner, which remains after the contributions of all inner loops are factored out of the graph. If we now consider $s_n$, its squared norm will be $N^{>1}_{n-1} N_{1}$, where $N_1$ is the squared norm of the spin network \eqref{4val_spinnet}, $s_1$, which remains after converting the $n-1$ outermost loops into coefficients. If the innermost loop in $s_1$ contains links of spins $i$, $a_n$, $b_n$ and $\varepsilon_n$, then, from Eq.~\eqref{4val_spinnet_norm}, we see that $N_1/d^{-1}_i = d^{-1}_{a_n}d^{-1}_{b_n}d^{-1}_{\varepsilon_n}$, which means that adding a new loop as close as possible to the intertwiner will lead to a multiplicative factor of $N_1/d^{-1}_i$ in the squared norm. By induction we then see that, if the ``naked" intertwiner has squared norm $d^{-1}_i$, adding loops with external links {with} spins $a_k$, $b_k$ and $\varepsilon_k$ gives the total squared norm $d^{-1}_i \prod_k d^{-1}_{a_k} d^{-1}_{b_k} d^{-1}_{\varepsilon_k} $, to which the only spins that do not contribute are those of the outermost links of the NLSN. 

It will be useful to work with normalized spin networks for some calculations, such as the derivation of the quantum volume operator, which requires diagonalization. Normalization of spin networks can be achieved by multiplying them with the inverse of their norm, rendering {orthonormal} the set of spin networks of the form \eqref{3val_spinnet} or {\eqref{4val_spinnet}.} 

\section{Action of the scalar constraint on 3-valent spin networks}\label{sec_3_val}

\subsection{Overview}

In this section, we present a detailed derivation of the action of the constraint~\eqref{scalar_constr} on 3-valent NLSNs of the form~\eqref{3val_spinnet}. The derivation consists in a sequential implementation, via recoupling theory, of the action on the considered NLSN of each of the operators appearing in Eq.~\eqref{scalar_constr}. For readers that are not interested in these technical details, we recommend skipping directly to the final formula of this section, Eq.~\eqref{3val_Hstep13}, which explicitly shows the form of the Hamiltonian matrix elements.

\subsection{Action on 3-valent NLSNs}

Let us start with a spin network of the form \eqref{3val_spinnet}.  Following Refs.~\cite{borissov, gaul}, both the links (i.e., the Wigner matrices) and the paths of the holonomies in Eq.~\eqref{scalar_constr} will be oriented towards the node, so that inverse holonomies are associated with segments oriented away from the node. The orientation is important, since the consecutive application of the holonomies in Eq.~\eqref{scalar_constr} should follow a chain of contractions closed by the trace. 

We proceed with the application of the first holonomy of the directly ordered term on the right-hand side of Eq.~\eqref{scalar_constr} to the fiducial spin network. At first, we will consider the action of $\hat{h}^{-1}[p_k]$ only along the path $p_1$ co-curvilinear to the link of spin $j_1$, i.e.,
\begin{equation}
 
\label{3val_Hstep1}
\end{equation}
As previously explained, the Wigner matrices of inverse SU(2) elements corresponding to inverse holonomies are directed outward from the node. The holonomies are in the fundamental representation, i.e., spin $1/2$, as shown in Eq.~\eqref{3val_Hstep1}. We extend the lower end of the spin-$1/2$ Wigner matrix on the right-hand side of Eq.~\eqref{3val_Hstep1} by contracting it with the $1/2$-representation identity, Eq.~\eqref{identity}, so that the latter can be coupled to the spin network link of spin $j_1$ with the aid of Eq.~\eqref{merge}. The spin resulting from the coupling assumes all values $m$ allowed by the Clebsch-Gordan conditions, namely $m=j_1 \pm 1/2$. Note that the lower ``free" link of spin $1/2$ in Eq.~\eqref{3val_Hstep1} is technically one of the legs of the node, the intertwiner of which temporarily becomes 4-valent, having an inner link with spin $j_1$. The action of the holonomies $\hat{h}^{-1}[p_2]$ and $\hat{h}^{-1}[p_3]$ along the links of spins $j_2$ and $j_3$, which we will not explicitly show, follow analogous relations with permuted labels. 

Spin networks like the one in the lower row of Eq.~\eqref{3val_Hstep1} are eigenstates of the volume operator. The latter acts on the (physical) nodes of the graph, giving zero contribution from nodes with valence below 4, while the central node contributes with a volume defined by the spins of the links attached to it. We will provide a detailed derivation of the action of the volume operator on such spin networks in Sec.~\ref{Q_volume}, but for now we merely represent the eigenvalues of the operator as $V^{(3.5)}_{m,j_1,a,b}$, where the superscript denotes that the intertwiner has valence 4, but one of its links (the holonomy one) is merely temporary. The next operator on the right-hand side of Eq.~\eqref{scalar_constr} is the holonomy $\hat{h}[p_k]$. For the specific case of the path $p_1$ along the $j_1$-representation link, $\hat{h}[p_1]$ is graphically represented by a Wigner matrix directed toward the node, with its lower index contracted with the upper index of $\hat{h}^{-1}[p_1]$. It is worth noting that, since $\hat{h}[p_1]\hat{h}^{-1}[p_1]=\mathbb{1}$ (the SU(2) identity in the fundamental representation), the contraction of these holonomies, graphically shown as a contraction between two Wigner matrices, one for $g$ and one for $g^{-1}$, leads to temporary spin-1/2 links corresponding only to identities (straight lines):
\begin{widetext}
\begin{equation}
 
\label{3val_Hstep2}
\end{equation}
Note that the deformation of the links on the last two rows of Eq.~\eqref{3val_Hstep2} is irrelevant, i.e., only the arrangement of the ends of the links matters, as we will show below. The choice of the SU(2) element $g$ for the Wigner matrices corresponding to the holonomies applied in Eqs.~\eqref{3val_Hstep1} and \eqref{3val_Hstep2} results from another absorption of noncontributing group elements. In fact, the SU(2) element associated with the holonomies in Eq.~\eqref{scalar_constr} could be chosen at random, yet expressing it as $k=y^{-1}gy$ allows one to implement the coupling shown in Eq.~\eqref{3val_Hstep1} with $g^{-1}$ replaced with $g^{-1}y$ and leave an additional Wigner matrix for the element $y^{-1}$ on the lower spin-$1/2$ link. Similarly, the coupling between $g^{-1}y$ and $k$ in Eq.~\eqref{3val_Hstep2} leads to a left Wigner matrix for the element $y$ on the upper spin-$1/2$ link. The Wigner matrix for the loop holonomies $\hat{h}[\alpha_{ij}]$, once contracted to the Wigner matrices of $y$ and $y^{-1}$, perfectly matches the element $g$ of the inner-loop link of the NLSN. Thus, one can in fact directly consider $g$ as the SU(2) element associated with the holonomies in the Hamiltonian.

The holonomies over the triangular loop, $\hat{h}[\alpha_{ij}]-\hat{h}[\alpha_{ji}]$, should be applied on the final NLSN obtained in Eq.~\eqref{3val_Hstep2} in such a way that the sequence of contractions in Eq.~\eqref{scalar_constr} is properly ordered and closed. Since $\alpha_{ij}$ and $\alpha_{ij}$ have opposite orientations, they will be attached to the loose ends of the two spin-$1/2$ temporary links (which are physically at the same point of the manifold) in different ways. The presence of a trace in Eq.~\eqref{scalar_constr} enforces that all virtual links should be tied together, such that no loose virtual links remain. As a result, for $\hat{h}[\alpha_{23}]$ we get
\begin{equation}
 
\label{3val_Hstep3}
\end{equation}
While transitioning from the left-hand to the right-hand side in the upper row of the above equation, Eq.~\eqref{wigner3j_invar} has been used on the node joining the links of spins $\{j_3, b, \varepsilon\}$. In the (upper) right-hand side of Eq.~\eqref{3val_Hstep3}, the holonomy along the loop $\alpha_{ij}$ has been graphically represented as a Wigner matrix in the fundamental representation with ends that have been extended by identities, one of which, on the right side, is converted into two $\epsilon^{(1/2)}$ symbols with the aid of Eq.~\eqref{levi3}. The trace in Eq.~\eqref{scalar_constr}, which we do not write explicitly in Eq.~\eqref{3val_Hstep3}, enforces the contraction of each of the free indices of this Wigner matrix (or of its identity-extended version) to the temporary links introduced by the holonomies along the link $p_1$ (``vertical" direction in the graphs), leading to the formation of ``square" loops in the bottom row of Eq.~\eqref{3val_Hstep3} after a braiding operation on the uppermost node according to Eq.~\eqref{wigner3j} (notice that the kinks have no physical meaning and serve merely for graphical convenience). Additionally, at the node containing spins $\{a, \alpha, 1/2\}$, Eq.~\eqref{wigner3j_invar} has been used to introduce three solid arrows. The effect of coupling the identity extensions of the Wigner matrix of $g$ to the links of spins $a$ and $b$ can be accounted for by applying Eq.~\eqref{merge} to each of these links. In particular, for the right fundamental-representation identity, the coupling to the spin-$b$ link involves only the segment between the two $\epsilon^{(1/2)}$ symbols, which then become integrated into loops. Similarly, the $\epsilon^{(a)}$ symbol is moved down in order for the identities in the $a$ and $1/2$ representations to be coupled. The Wigner matrix of the holonomy is coupled through Eq.~\eqref{element_merge} to the Wigner matrix of $g$ in the spin network. Each of the sums resulting from couplings between the spin network and the holonomy runs over the original link spin plus or minus $1/2$. These three couplings leave triangular loops at the legs with spins $j_2$ and $j_3$, which, alongside with the small ``square" loop, can be factored out of the spin network with the help of the resolution of the identity [cf. Eq.~\eqref{theorem}]: 
\begin{equation}
 
\label{3val_Hstep4}
\end{equation}

Note that the tetrahedra in the first row of Eq.~\eqref{3val_Hstep4} contain a different number and arrangement of $\epsilon$ symbols when compared to Eqs.~\eqref{tetrahedron} and \eqref{theorem}. Yet, through Eqs.~\eqref{wigner3j_invar} and \eqref{levi3}, it is possible to see that these are equivalent up to phase terms [cf. Eq.~\eqref{app3}]. The latter can originate either from changes in the cyclicity of nodes [cf. Eq.~\eqref{wigner3j}] or changes in the direction of the solid arrows [see text preceding Eq.~\eqref{levi}]. After the three loops are removed in the first row of Eq.~\eqref{3val_Hstep4}, a fourth loop can be removed, as shown in the second row of the same equation. The four tetrahedral structures factored out of the spin network with Eq.~\eqref{theorem} can be converted into Wigner 6j symbols according to Eq.~\eqref{tetrahedron}, and for some of the tetrahedra braiding at the nodes or flips in the direction of solid arrows are required, leading to the phase factor in the third row of Eq.~\eqref{3val_Hstep4}.

We now proceed with the application of $\hat{h}[\alpha_{32}]$ on the outcome of Eq.~\eqref{3val_Hstep2}:
\begin{equation}
 
\label{3val_Hstep5}
\end{equation}
\end{widetext}
Similar to the procedure in Eq.~\eqref{3val_Hstep3}, Eq.~\eqref{wigner3j_invar} has been used at the node joining the links of spins $\{j_3, b, \varepsilon\}$ between the right- and left-hand sides in the upper row of the above equation. Furthermore, the holonomy along the loop $\alpha_{32}$ {has been} graphically represented as the inverse Wigner matrix to the one representing $\hat{h}[\alpha_{23}]$, with ends also extended by identities. The contraction of each of the free indices of this Wigner matrix to the temporary links introduced by the holonomies along the link $p_1$ therefore happens in opposite order, leading to the formation of new ``square" loops in the bottom row of Eq.~\eqref{3val_Hstep5} after a braiding operation on the $\{m, j_1, 1/2\}$ node [cf. Eq.~\eqref{wigner3j}]. The effect of coupling the identity extensions of the Wigner matrix of $g$ to the links of spins $a$ and $b$ is described by Eq.~\eqref{merge}. The $\epsilon^{(a)}$ symbol has been moved up in order for the identities in the $a$ and $1/2$ representations to be coupled. The Wigner matrix itself is coupled through Eq.~\eqref{element_merge_inv} to the Wigner matrix of the NLSN, and the intertwiners contracted to the resulting Wigner matrix are braided to properly contract with the free legs from adjacent coupled links. After introducing pairs of arrows through Eq.~\eqref{levi3} at the links of spins $j_2, \beta$ and $m$, the resulting triangular loops at the legs with spins $j_2$ and $j_3$, alongside with the small ``square" loop, can be factored out of the spin network with the help of the resolution of the identity [cf. Eq.~\eqref{theorem}]: 
\begin{widetext}
\begin{equation}
 
\label{3val_Hstep6}
\end{equation}
\end{widetext}
Between the second and third rows of Eq.~\eqref{3val_Hstep6} we have used Eq.~\eqref{wigner3j_invar} to introduce $\epsilon$ tensors at the node connecting the spins $\{j_2, \alpha, \gamma\}$, leading to a double arrow on the spin-$\gamma$ link or, through Eq.~\eqref{levi3}, a $(-1)^{2\gamma}$ phase factor instead. As in Eq.~\eqref{3val_Hstep4}, the conversion between the tetrahedra in the first row of Eq.~\eqref{3val_Hstep6} and the one in Eq.~\eqref{tetrahedron} requires braiding operations and flips in the directions of the solid arrows, resulting in the appearance of phases. 
\newpage 

The contributions from terms inversely ordered in the anticommutator in Eq.~\eqref{scalar_constr} require the application of holonomies in a different order. First, the holonomies along the loops $\alpha_{ij}$ or $\alpha_{ji}$ are applied, leaving virtual spin-$1/2$ links connected to two different links of the spin network. And then the holonomies along the path $p_k$, interleaved by the volume operator, are applied in such a way that all virtual links form closed loops to be factored out. The application of $\hat{h}[\alpha_{23}]$ on the spin network~\eqref{3val_spinnet} gives
\pagebreak \\

\begin{equation}
 
\label{3val_Hstep7}
\end{equation}

In Eq.~\eqref{3val_Hstep7}, just as in Eq.~\eqref{3val_Hstep3}, a pair of solid arrows has been created on one of the identity extensions of the Wigner matrix to be coupled to the spin network. The coupling between the spin-$1/2$ identity segment connecting the two arrows and the spin-$b$ link of the spin network leads to the splitting of the $\epsilon$ tensors as seen in the bottom row of  Eq.~\eqref{3val_Hstep7}. We then proceed with the application of the inverse holonomy $\hat{h}^{-1}[p_1]$, 

\vspace{5 cm}

\newpage 

\begin{equation}
%
 
\label{3val_Hstep8}
\end{equation}
Note that, to pass from the second to the third row of Eq.~\eqref{3val_Hstep8}, a braiding [cf. Eq.~\eqref{wigner3j}] {has been} performed in the uppermost node of the spin network after applying Eq.~\eqref{merge}. Before applying the volume operator in the last row of Eq.~\eqref{3val_Hstep8}, we can factor out the loops introduced on the spin network by the coupling 
to the holonomies:
\begin{widetext}
\begin{equation}
%
 
\label{3val_Hstep9}
\end{equation}
In the second row of Eq.~\eqref{3val_Hstep9} we have used the 2-2 Pachner move \eqref{pachner} to convert the spin network into a form on which the action of the volume operator is well known. Since this NLSN is one of the eigenstates of the volume operator, one merely gets a coefficient from $\hat{V}$. The action of the holonomy $\hat{h}[p_1]$ on this spin network then simply ties the temporary spin-$1/2$ links together,
\begin{equation}
%
 
\label{3val_Hstep10}
\end{equation}
\end{widetext}
In the first equality of this equation, we have made use of Eq.~\eqref{wigner3j_invar} in the central intertwiner of the spin network to recover the desired arrangement of $\epsilon$ tensors. In the resulting term, the two contracted Wigner matrices leave an identity behind, forming a loop that can be removed with Eq.~\eqref{bubble}. For this loop to be factored out, however, a braiding according to Eq.~\eqref{wigner3j} has to be implemented in one of the nodes forming the loop, leading to the phase factor $(-1)^{\frac{1}{2}+m+j_1}$. 

Finally, we look at the action of the holonomy $\hat{h}[\alpha_{32}]$ on the spin network~\eqref{3val_spinnet},
\begin{equation}
 
\label{3val_Hstep11}
\end{equation}
As in the case of Eq.~\eqref{3val_Hstep4}, the couplings between links in Eq.~\eqref{3val_Hstep11} are dictated by Eqs.~\eqref{merge} and \eqref{element_merge_inv}. We then proceed with the application of $\hat{h}^{-1}[p_1]$,
\begin{equation}
%
 
\label{3val_Hstep12}
\end{equation}
The loops in the spin network in the third row of Eq.~\eqref{3val_Hstep12} can then be factored out with the aid of the resolution of the identity [cf. Eq.~\eqref{general_identity}]. We thus obtain
\begin{widetext}
\begin{equation}
%
 
\label{3val_Hstep13}
\end{equation}
\end{widetext}
Note that on the right-hand side of the upper row of Eq.~\eqref{3val_Hstep13} we have used the invariance of intertwiners, Eq.~\eqref{wigner3j_invar}, on the node of spins $\{\alpha, b, m\}$. The same property has also been used later on the node of spins $\{\alpha, j_2, \gamma\}$, and the resulting double $\epsilon^{(\gamma)}$ has directly been converted into a phase factor $(-1)^{2\gamma}$ at the last row of Eq.~\eqref{3val_Hstep13}.

The last step in this calculation is precisely the application of $\hat{h}[p_1]\hat{V}$, as explained in Eq.~\eqref{3val_Hstep10}.

To summarize Eqs.~\eqref{3val_Hstep1}-\eqref{3val_Hstep13}, the action of Eq.~\eqref{scalar_constr} on a spin network of the form~\eqref{3val_spinnet} for a fixed choice of directions for the holonomies reads
\begin{widetext}

\begin{equation}
 
\label{3val_total}
\end{equation}

\end{widetext}

In {this equation,} we have used $(-1)^{2a +2m+2\beta}=1$ in the first term within square brackets (namely in the prefactor of $V^{(3.5)}_{m,j_1, \alpha, \beta} $) and $(-1)^{2b +2m+2\beta+2j_1}=(-1)^{4\alpha}=1$ in the second one (in the prefactor of $V^{(3.5)}_{m,j_1, a, b} $). Note that taking $a\leftrightarrow \alpha$, $b\leftrightarrow\beta$ and $\epsilon \leftrightarrow \gamma$ in Eq.~\eqref{3val_total}, what swaps input and output states, only affects the coefficient on the right-hand side of the equation by interchanging the two terms within square brackets, which effectively corresponds to a $-1$ prefactor [to {help in} the conversion of prefactors, note that  $(-1)^{2b +2j_1+2\alpha}=(-1)^{2a+2\alpha}=-1$, {$(-1)^{\varepsilon +\gamma}=(-1)^{1-\varepsilon -\gamma}$ and} $(-1)^{m+j_1}=(-1)^{m+j_1+2(1/2+m+j_1)}=(-1)^{1-j_1-m}$]. The negative prefactor is in consonance with the definition of symmetry/Hermitianity, i.e., for any two spin networks $|s\rangle$ and $|s'\rangle$, $\langle s' | \hat{C}_s|s\rangle =\langle s | \hat{C}_s|s'\rangle^*$ [note that Eq.~\eqref{3val_total} contains an imaginary prefactor $i$ as well]. The complete action of $\hat{C}_s$ on the spin network requires considering all possible rotations of $p_k$ and $\alpha_{jk}$, with the caveat that new loops $\alpha_{jk}$ should be applied deeper (i.e., closer to the intertwiner) if other loops $\alpha'_{j'k'}$ with $j'\neq j$ or $k'\neq k$ are already present in the input spin network.

\section{Action of the scalar constraint on 4-valent spin networks}\label{sec_4_val}

\subsection{Overview}

In this section, we present a detailed derivation of the action of the constraint~\eqref{scalar_constr} on 4-valent NLSNs of the form~\eqref{4val_spinnet}. The derivation consists again in the sequential implementation of the action on the considered NLSN of each of the operators in Eq.~\eqref{scalar_constr}, via recoupling theory. Since, for a 4-valent NLSN, choosing the links to which the loop holonomy couples (say, by fixing the loop $\alpha_{23}$) still leaves two possible links with which one can span the regularization tetrahedra (e.g., $p_1$ or $p_4$), we arrive at two inequivalent contributions to the matrix elements of the Hamiltonian acting on 4-valent spin networks, namely Eqs.~\eqref{4val_total1} and \eqref{4val_total2}. These equations correspond, respectively and for the aforementioned example, to the choice of $p_1$ or $p_4$ as the directions of the support links when a loop is introduced between directions $p_2$ and $p_3$. It is worth noting that these expressions are not mere extensions of Eq.~\eqref{3val_total}, derived for 3-valent NLSNs. For readers that are not interested in the lengthy details of the derivation, we recommend skipping it directly and going to the final formulas of this section and their subsequent discussion.\\

\subsection{Action on 4-valent NLSNs}

Let us thus investigate the action of the scalar constraint~\eqref{scalar_constr} on spin networks of the form~\eqref{4val_spinnet}. As in the previous section, we start with the action of the first holonomy of the directly ordered term on the right-hand side of Eq.~\eqref{scalar_constr} on a NLSN. It is important to notice that, for each possible inclusion of an inner loop (say, connecting links along the directions $p_2$ and $p_3$), there are up to two possible choices of a third direction along which the holonomies $\hat{h}[p_k]$ and $\hat{h}^{-1}[p_k]$ can be applied. Since the directions of $p_k$ and of the support links of $\alpha_{ij}$ are linearly independent, whether $p_k$ can be chosen to be directed along one or two legs of the spin network for each choice of $\alpha_{ij}$ depends on the geometric arrangement of the spin network links relative to each other. In the canonical approach, it is usually assumed that the spin networks are embedded in 3D (as a consequence of the foliation adopted for the definition of the Ashtekar-Barbero variables).  We will therefore derive contributions to the action of the scalar constraint arising from both choices of $p_k$ for each choice of inner loop, assuming that the four links are arranged as the dual graph of a tetrahedron, up to diffeomorphisms. Our choice provides the most general possible result, so that consideration of a simpler Hamiltonian action that excludes one of the links of the spin network can be attained by selectively removing certain terms from our expressions. Let us first consider the application of the holonomy $\hat{h}^{-1}[p_1]$ along the path $p_1$ co-curvilinear to the link of spin $j_1$,
\begin{widetext}
\begin{equation}

\label{eq:1.1}
\end{equation}

We reiterate that, after coupling the identity extension of the introduced Wigner matrix to the spin-$j_1$ link of the spin network through Eq.~\eqref{merge}, the new coupled spin takes values $m=j_1\pm 1/2$. Note that the direction of the temporary link containing the Wigner matrix of $g^{-1}$ is drawn off-tangent relative to the direction the holonomy is applied in, since the temporary link has actually no spatial extension and its direction is arbitrary as long as one remembers the correct order of contraction of its ends. In order for the spin network structure to match the ones we investigate the action of the volume operator on [cf. Eq.~\eqref{eq:5.1}], we employ Eq.~\eqref{prepachner} to swap the upper legs of the spin network at the end of Eq.~\eqref{eq:1.1} [note that the $-n-l$ terms in the phase factor come from flipping $\epsilon^{(l)}$ and $\epsilon^{(n)}$ before {and} after applying Eq.~\eqref{prepachner}]. When acting on the spin network in the last row of Eq.~\eqref{eq:1.1}, the volume operator then gives
\begin{equation}

\label{eq:1.2}
\end{equation}
Spin networks of this form are not eigenstates of the volume operator. As will be shown in Sec.~\ref{Q_volume}, the volume operator maps such (normalized)  spin networks to linear combinations of spin networks with the same structure. Luckily, each such spin network belongs to an equivalence class, and the action of the volume operator never maps spin networks from a certain equivalence class into another. For now, the coefficients of the linear combination resulting from the application of the volume operator on these spin networks will be denoted $\big(V^{(4.5)p,q}_{l,j_1;j_4,b,a,l}\sqrt{d_p d_q d_l^{-1} d_{j_1}^{-1}}\big)$, where the square-root contribution comes from normalizing and denormalizing the spin network prior and posterior to the action of the volume operator, respectively, and the superscript $(4.5)$ denotes that the intertwiner has valence 5, but one of its links (the holonomy one) is merely temporary, while $p$ and $q$, on the one hand, and $n$ and $j_4$, on the other hand, are the two-entry indices of its matrix elements. In the last row of Eq.~\eqref{eq:1.2}, we employed again Eq.~\eqref{prepachner} to swap the upper legs of the spin network.

We now apply the holonomy $\hat{h}[p_1]$ on these spin networks, resulting in an NLSN with two virtual spin-$1/2$ links located at the 4-valent node (note that the spin-$k$ link has no physical extension). As in previous equations, the Wigner matrices of $g$ and $g^{-1}$ contract to give the identity. Accordingly,
\begin{equation}

\label{eq:1.3.1}
\end{equation}

Finally, we apply the holonomy $\hat{h}[\alpha_{32}]$, which is represented by a Wigner matrix with ends extended by identities, one of which, on the left side, includes two arrows introduced through Eq.~\eqref{levi3}. The trace is also applied to contract all the holonomy indices in the end, but we will not explicitly write it in the following equations: 
\begin{equation}
%
\label{eq:1.3.2}
\end{equation}

The effect of coupling the identity extensions of the Wigner matrix of $g$ to the links of spins $a$ and $b$ can be accounted for by applying Eq.~\eqref{merge} to each of these links. In particular, for the left fundamental-representation identity, the coupling to the spin-$a$ link involves only the segment between the two $\epsilon^{(1/2)}$ symbols, which then become integrated into loops. The $\epsilon^{(a)}$ and $\epsilon^{(b)}$ symbols have both been moved up to get the identities in the $a$ and $b$ representations coupled to the fundamental-representation ones. The Wigner matrix of the holonomy is coupled through Eq.~\eqref{element_merge} to the Wigner matrix of the spin network. Each of the sums resulting from couplings between the spin network and the Wigner matrix corresponding to the holonomy runs over the original spin of the spin network link plus or minus $1/2$. These three couplings leave triangular loops at the links with spins $j_2$ and $j_3$. On top of that, contracting all the holonomies converts the temporary spin-$1/2$ links into two loops, one of which connects the links of the spin network diagonally (with a ``tilde" shape). This loop has to be removed before the other loop on the left upper side of the spin network could be factored out through the resolution of the identity. We show in Appendix \ref{appendix_recoup}, Eq.~\eqref{app1}, that the diagonal link can also be factored out to give
\begin{equation}

\label{eq:1.4}
\end{equation}
where we have made use of the invariance of intertwiners, Eq.~\eqref{wigner3j_invar}, to create $\epsilon$ tensors on the nodes of spins $\{m, q, 1/2\}$, $\{q, a, k\}$ and $\{\gamma, \varepsilon, 1/2\}$ on the left-hand side of Eq.~\eqref{eq:1.4}, as well as on the intertwiners of spins $\{m, a, u\}$ and $\{u, \beta, j_4\}$ on the right-hand side of the same equation. Note the factor $(-1)^{2\varepsilon}$ originating from the contraction of two $\epsilon^{(\varepsilon)}$ tensors and the use of the simplification $(-1)^{2u+2a+2m}=1$.

The remaining loops can be also factored out with the aid of Eq.~\eqref{theorem},
\begin{equation}

\label{eq:1.5}
\end{equation}
For this specific contribution to the constraint, we use Eq.~\eqref{messiah5} to simplify the final expression. 

Starting from the final spin network in Eq.~\eqref{eq:1.3.1}, we can also apply the inverse holonomy  $\hat{h}[\alpha_{23}]$, with ends extended by identities. After reorganizing the arrows around the node joining the spins $\{j_2, \varepsilon, b\}$ using Eq.~\eqref{wigner3j_invar}, the spin-$a$ and spin-$b$ links of the spin network can be coupled to the identity extensions of the holonomy with the help of Eq.~\eqref{merge}, while the Wigner matrices can be merged through Eq.~\eqref{element_merge_inv}. We obtain 
\begin{equation}

\label{eq:1.6}
\end{equation}
In this case, however, Eq.~\eqref{theorem} can be used right away to factor out {three} inner loops from the resulting spin network, while the remaining diagonal link can be factored out afterwards (see Appendix~\ref{appendix_recoup}). To facilitate this factorization, we include pairs of arrows in the links of spins $m$, $j_3$ and $\beta$ with the aid of Eq.~\eqref{levi3} [and use $(-1)^{2u+2\alpha+2j_1}$ to simplify the final expression]. This leads to
\begin{equation}
%
\label{eq:1.7}
\end{equation}

While still considering  $\hat{h}[p_1]$ to be directed along the link of spin $j_1$, we now look at the {other term in the anticommutator of} Eq.~\eqref{scalar_constr}, namely, the contribution that starts with the application of  $\hat{h}[\alpha_{32}]$ on the spin network~\eqref{4val_spinnet}. For that, we include two $\epsilon^{(1/2)}$ tensors on the right identity extension of the applied Wigner matrix. Furthermore, we use the invariance of the nodes {[cf. Eq.~\eqref{wigner3j_invar}]} to reorganize the arrows at the node joining the spins $\{\epsilon, j_2, b\} $. Using Eq.~\eqref{merge}, we couple the spin-$b$ link of the spin network to the spin-$1/2$ identity between the two $\epsilon^{(1/2)}$ tensors of the right extension of the Wigner matrix. Likewise, the section of the spin-$a$ link of the spin network right above the $\epsilon^{(a)}$ tensor is coupled to the left extension of the Wigner matrix. In sequence, using Eq.~\eqref{wigner3j_invar}, we can rearrange the arrows in the negative-cyclicity  node joining the spins $\{\varepsilon, 1/2, \gamma \} $ [effectively creating an $\epsilon^{(\gamma)}$ out of $\epsilon^{(\varepsilon)}$ and $\epsilon^{(1/2)}$] before applying Eq.~\eqref{theorem} to factor out the two inner loops. This gives

\begin{equation}

\label{eq:2.1}
\end{equation}

We then proceed with the application of $\hat{h}^{-1}[p_1]$. {Its} Wigner matrix contracts with one of the temporary spin-$1/2$ links created by $\hat{h}[\alpha_{32}]$. We apply a braid [cf. Eq.~\eqref{wigner3j}] at the node where this link is connected, so that the latter can be moved to the ``interior" of the spin network. The resulting diagonal link can be factored out, as shown in Eq.~\eqref{app1}. Thus, we obtain  
\begin{equation}
%
\end{equation}
For the spin network in the last row of this equation, we then use Eq.~\eqref{pachner} between the nodes of spins $\{\alpha, a, 1/2 \} $ and $\{a, u, m \} $. As before, we use Eq.~\eqref{prepachner} to swap the upper legs of the spin network and braid the temporary links, getting
\begin{equation}
%
\label{eq:2.2}
\end{equation}
We can then use the invariance of intertwiners to rearrange the arrows around the nodes of spins $\{j_4, \alpha, k\}$ and $\{k, \beta, l\}$, followed by the application of the volume operator according to Eq.~\eqref{eq:1.2}. Note that the rearrangement of arrows includes a flip in the tensor $\epsilon^{(k)}$, leading to a phase factor $(-1)^{2k}$.

Finally, we apply the holonomy $\hat{h}[p_1]$, which ties the two temporary spin-$1/2$ links of the spin network while contracting the Wigner matrices of $g$ and $g^{-1}$ to give the identity. The formed ``bubble" can be factored out with the help of Eq.~\eqref{bubble}, resulting in the desired NLSN:
\begin{equation}

\label{eq:2.4}
\end{equation}

The last contribution that considers holonomies applied along $p_1$ is the one starting with the application of $\hat{h}[\alpha_{23}]$. Both indices of the corresponding Wigner matrix are contracted with identities, which extend themselves to embrace the NLSN as in the previous cases. Before coupling the holonomy to the spin network, we use the invariance of the Wigner 3j symbols, Eq.~\eqref{wigner3j_invar}, to reorganize the arrows associated with the node connecting the links of spins $\{ j_2, \varepsilon, b\}$. After employing Eqs.~\eqref{merge} (coupling identity segments below the arrows) and \eqref{element_merge_inv}, we apply Eq.~\eqref{levi3} to the spin-$j_3$ and spin-$\beta$ links to create a pair of arrows on each of them. One arrow from each pair can be extracted alongside the inner loops with the aid of Eq.~\eqref{theorem}, and we can then apply Eq.~\eqref{wigner3j_invar} to the nodes joining the spins $\{a, \alpha, 1/2\}$, $\{ b, \beta, 1/2\}$ and $\{ j_2, \gamma, \beta\}$ to obtain the final spin network [note that the doubled $\epsilon^{(\alpha)}$ symbols are canceled via Eq.~\eqref{levi3}]. This gives
\begin{equation}

\label{eq:2.5}
\end{equation}

Acting with $\hat{h}^{-1}[p_{1}]$ on the spin network resulting from Eq.~\eqref{eq:2.5} allows for coupling the identity extension of $\hat{h}[\alpha_{23}]$ to the spin-$j_1$ link of the NLSN. Before and after applying Eq.~\eqref{merge}, we braid the fundamental-spin links around the nodes joining the spins $\{a, \alpha, 1/2\}$ and $\{m, j_1, 1/2\}$, respectively. The Wigner matrix itself remains in one of the free ends of the temporary fundamental-spin links, while the other end is contracted with the corresponding temporary link resulting from the application of the holonomy on the loop $\alpha_{23}$. The inner loop formed by this contraction can then be factored out employing Eq.~\eqref{theorem}. We can next use Eq.~\eqref{wigner3j_invar} to create a trio of arrows around the node where the links of spins $\{n, \alpha, m\}$ meet [see the third row of Eq.~\eqref{eq:2.6}]. One of these arrows, representing the $\epsilon^{(n)}$ symbol, allows us to use Eq.~\eqref{prepachner} to swap the upper legs of the spin network. Similarly, creating a trio of arrows around the node with spins $\{b, \beta, 1/2\}$ allows us to use Eq.~\eqref{pachner} in the connecting spin-$b$ link. Note that, before applying the volume operator, one needs to reduce the tensors $\epsilon^{(u)}$, $\epsilon^{(1/2)}$ and $\epsilon^{(m)}$ on the third row of Eq.~\eqref{eq:2.6} to a prefactor $(-1)^{2m}$. As in Eq.~\eqref{eq:1.2}, the action of the volume operator on the resulting spin network [see the last row of Eq.~\eqref{eq:2.6}] gives a linear combination of spin networks with the same structure. We can then act on it with the holonomy $\hat{h}[p_{1}]$ to generate the desired spin networks, as in Eq.~\eqref{eq:2.4}. In total, we obtain

\begin{equation}

\label{eq:2.6}
\end{equation}

Since we are considering the most general case for the application of Eq.~\eqref{scalar_constr} acting on the spin network \eqref{4val_spinnet}, we must also take into account holonomies applied along the direction $p_4$. For the first term in Eq.~\eqref{scalar_constr}, we consider the application of the holonomy $\hat{h}^{-1}[p_4]$ on the spin network,
\begin{equation}
%
\label{eq:3.1}
\end{equation}

The coupling of the identity extension of the Wigner matrix corresponding to $\hat{h}^{-1}[p_4]$ with the spin-$j_4$ link of the spin network is described by Eq.~\eqref{merge}. The resulting spin network is already in the graphical form suitable for application of the volume operator, as described in Eq.~\eqref{eq:5.1}, resulting in the linear combination
\begin{equation}
%
\label{eq:3.2}
\end{equation}

We then apply the holonomy $\hat{h}[p_4]$. {Its} Wigner matrix contracts with the free end of the $g^{-1}$ Wigner matrix, forming a temporary identity link. This gives
\begin{equation}
%
\label{eq:3.3}
\end{equation}

The two temporary spin-$1/2$ links are then contracted by the trace with the identity extensions of the Wigner matrix representing the holonomy $\hat{h}[\alpha_{32}]$ after performing a braiding at the node of spins $\{ m, 1/2, j_4 \}$. We apply Eq.~\eqref{levi3} in both {identity} extensions of the Wigner matrix, as well as Eq.~\eqref{wigner3j_invar} to the node of spins $\{ \varepsilon , j_2, b \}$ in order to reorganize the arrows. The inter-arrow sections of the identity extensions are then coupled to the spin-$a$ (namely the section below $\epsilon^{(a)}$) and spin-$b$ links by using Eq.~\eqref{merge}. Similarly, the Wigner matrices are coupled by means of Eq.~\eqref{element_merge}. We obtain
\begin{equation}

\label{eq:3.4}
\end{equation}

After using Eq.~\eqref{levi2} on the spin-$1/2$ link with two arrows in the spin network obtained in the last row of this equation and introducing new pairs of arrows at the links of spins $\beta$ and $j_3$ via Eq.~\eqref{levi3}, we can use Eq.~\eqref{theorem} to factor out three inner loops from the spin network. We can then use Eq.~\eqref{wigner3j_invar} to redistribute the arrows around the {resulting} node of spins $\{j_2, \beta, \gamma\}$ [see the second row of Eq.~\eqref{eq:3.5}], leading to a double $\epsilon^{(\beta)}$ that can be factored out employing Eq.~\eqref{levi2}. By braiding the node of spins $\{\alpha, 1/2, a\}$ [note the corresponding factor $(-1)^{\alpha+a+1/2}$] and rearranging arrows around that same node, as well as the node below, the remaining diagonal link can be factored out via Eq.~\eqref{app2} (cf. Appendix \ref{appendix_recoup}). In this way, we get
\begin{equation}

\label{eq:3.5}
\end{equation}

Starting from the spin network in the last row of Eq.~\eqref{eq:3.3}, we can apply $\hat{h}[\alpha_{23}]$ (and the trace), using Eqs.~\eqref{merge} and \eqref{element_merge_inv} [as well as Eq.~\eqref{wigner3j_invar} to reorganize arrows around the node $\{j_2, \varepsilon, b\}$] to obtain a spin network with temporary inner loops (note the double braiding on the leg along direction $p_4$ and the resulting phase factor): 
\begin{equation}
%
\label{eq:3.6}
\end{equation}
In the spin network in the last row of Eq.~\eqref{eq:3.6}, we can use Eq.~\eqref{app1} to factor out the diagonal link:
\begin{equation}
%
\label{eq:3.7}
\end{equation}
The remaining inner loops in Eq.~\eqref{eq:3.7} can be factored out with the help of Eq.~\eqref{theorem}. Note that we can use Eq.~\eqref{levi3} on the links of spins $\beta$, $j_4$ and $j_3$ to introduce additional arrows that can be extracted when factoring out the tetrahedral structures. This leads to
\begin{equation}
%
\label{eq:3.8}
\end{equation}
The phase factors in the last row of Eq.~\eqref{eq:3.8} are the result of recombining arrows using Eqs.~\eqref{wigner3j_invar}, \eqref{levi2} and \eqref{levi3}.

In order to consider the contribution arising from the other order of operators in the anticommutator of
Eq.~\eqref{scalar_constr} for $p_k=p_4$, we start from Eq.~\eqref{eq:2.1} and apply $\hat{h}[p_4]$ on its final spin network. The identity extension of the Wigner matrix of $g^{-1}$ is coupled to the spin-$j_4$ link through Eq.~\eqref{merge}, and its lower end is contracted with one of the spin-$1/2$ links produced by $\hat{h}[\alpha_{32}]$, forming an inner loop. We can factor out the inner loop with Eq.~\eqref{theorem}, braid the remaining temporary spin-$1/2$ link and use Eq.~\eqref{wigner3j_invar} to introduce three arrows around the node connecting the links of spins $\{a, n, j_1\}$ [this leads to the second row of Eq.~\eqref{eq:3.8}]. Using Eq.~\eqref{pachner}, we can then move the remaining spin-$1/2$ link to the center of the spin network, 
\begin{equation}

\label{eq:3.8.2}
\end{equation}
After braiding the fundamental-spin link from the interior to the exterior of the NLSN and changing the cyclicity of its node, we apply once again Eq.~\eqref{pachner}, now to the spin-$n$ link. We also use Eq.~\eqref{wigner3j_invar} to rearrange the arrows around the node of spins $\{\alpha, k, j_1\}$ {[this gives the second row of Eq.~\eqref{eq:4.1}]}. An additional braiding at the node of spins $\{1/2, l, m\}$, followed by the introduction of $\epsilon$ tensors on the node of spins $\{k, \beta, l\}$ [cf. Eq.~\eqref{wigner3j_invar}], gives the following spin network, on which the volume operator can be directly applied: 
\begin{equation}

\label{eq:4.1}
\end{equation}

Finally, we apply $\hat{h}[p_4]$, which forms a closed loop that can be factored out with the aid of Eq.~\eqref{bubble} to recover the desired NLSN,
\begin{equation}
%
\label{eq:4.2}
\end{equation}

The last contribution in the scalar constraint~\eqref{scalar_constr} can be derived from the spin network obtained in Eq.~\eqref{eq:2.5}. We apply $\hat{h}^{-1}[p_4]$, using Eq.~\eqref{merge} to couple the identity extension of the Wigner matrix to the spin-$j_4$ link. The contraction of indices forms a diagonal link that can be factored out with the help of Eq.~\eqref{app1}. This gives

\begin{equation}
%
\end{equation}

After using Eq.~\eqref{wigner3j_invar} to reorganize the arrows around the node of spins $\{b, m, l \}$, we can employ Eq.~\eqref{pachner} to move the temporary spin-$1/2$ link upwards. We thus obtain
\begin{equation}
%
\label{eq:4.3}
\end{equation}
A simple rearrangement of arrows in the last spin network in Eq.~\eqref{eq:4.3} allows us to apply the volume operator, Eq.~\eqref{eq:3.2}, followed by the remaining holonomy, according to Eq.~\eqref{eq:4.2}.

Putting equations \eqref{eq:1.1}--\eqref{eq:2.6} together, we obtain all the contributions in Eq.~\eqref{scalar_constr} when a specific choice of loop, $\alpha_{32}$, and holonomies along $p_1$ are considered,
\begin{equation}
%
 
\label{4val_total1}
\end{equation}

Similarly, Eqs.~\eqref{eq:3.1}-\eqref{eq:4.3} give the result corresponding to Eq.~\eqref{scalar_constr} when $\alpha_{32}$ and holonomies along {$p_4$} are considered,

\begin{equation}
%
 
\label{4val_total2}
\end{equation}
\end{widetext}
Note that Eqs.~\eqref{4val_total1} and \eqref{4val_total2}
are seemingly not simply related to each other by Eq.~\eqref{prepachner}. In these equations, we explicitly wrote two separate sums over $p$ [and also over $l$ in the case of Eq.~\eqref{4val_total1}] because the values summed over are different for each sum. Furthermore, there are five other different possible choices for the loop $\alpha_{ij}$. For the loop $\alpha_{14}$, bridging the links of spins $j_1$ and $j_4$ in our assignment of spins, Eqs.~\eqref{eq:1.1}-\eqref{eq:4.3} can be directly used if preceded and followed by a flip in the direction of the central $\epsilon^{(i)}$ [cf. Eq.~\eqref{4val_spinnet}]. For the loops $\alpha_{13}$ and $\alpha_{24}$, however, application of the formulas derived in this section has to be preceded and followed by Eq.~\eqref{pachner}, which allows us to change the basis in the 4-valent intertwiner space so that the formulas hold. The last two loops, $\alpha_{12}$ and $\alpha_{34}${, extend} themselves diagonally through the spin network and require a braid between either legs in the $p_1$ and $p_3$ directions or $p_2$ and $p_4$ directions, both before and after employing Eqs.~\eqref{eq:1.1}-\eqref{eq:4.3}. It can be shown that applying the derived Hamiltonian between braid moves gives the same result as directly employing Eq.~\eqref{scalar_constr} with holonomies applied along the twisted loops $\alpha_{12}$ and $\alpha_{34}$.

A last relevant point concerning the action of the Euclidean scalar constraint on $4$-valent NLSN is how Eqs.~\eqref{4val_total1} and \eqref{4val_total2} can be modified to render their action graph preserving. The general approach in LQG is based on an extension of the loops $\alpha_{ij}\to\tilde{\alpha}_{ij}$, so that the enlarged loops $\tilde{\alpha}_{ij}$ cover an entire spin network ``patch'' with borders defined by (a minimal number of) links and intertwiners~\cite{Giesel_GP, AQG}. The precise form of the loops $\tilde{\alpha}_{ij}$ therefore depends on the spin network of choice, as well as on which ``patch'' the graph-preserving Hamiltonian is acting on. This renders the analysis of graph-preserving dynamics limited to NLSNs impossible and one is forced to extend the fiducial intertwiner in such a way that all its legs are connected to other intertwiners, forming closed loops $\tilde{\alpha}_{ij}$ on which the graph-preserving Hamiltonian can act. We note that a small modification of the NLSN studied here can still cover a small range of (modular segments of) connected spin networks on which a few such loops $\tilde{\alpha}_{ij}$ can be applied. If we include one inner loop at location 2 and one at location 4, promoting their ``additional'' links to virtual central links of two other independent intertwiners, we can apply Eqs.~\eqref{4val_total1} and \eqref{4val_total2} limiting their action solely to these loops at locations 2 and 4. As a result, one considers a ladder-like spin network (or rather a ``chain"-shaped one, when accounting for the choice of link arrangement at each node) with intertwiners connected
by their upper or lower pairs of legs and loop couplings restricted to solely happen above and below the fiducial intertwiner, neglecting large loops coupled from the sides (for which the action of the graph-preserving Hamiltonian depends on the precise number and connectivity of intertwiners through the entire ``side patches'' of the spin network). It is worth noting that these modified NLSNs can cover five other (modular segments of) spin network graphs, namely by setting one or two of the spins of the outermost links to zero (e.g., by having all of them equal to zero, one creates a spin network ``bubble'' in which the two lower legs of the intertwiner are connected to each other, and similarly for the two upper legs). Let us finally comment that other realizations of graph-preserving dynamics are possible~\cite{loopySN, reduced_LQG, phoenix}, such as the coarse graining of spin networks to a fixed graph starting from the action of a graph-changing constraint~\cite{loopySN}. 
Since the challenge of characterizing the properties of the (graph-changing) Hamiltonian constraint \eqref{scalar_constr} is already extremely demanding, we will not discuss such alternatives in the present work, nor explore their physical consequences.\\

\section{Action of the quantum volume operator}\label{Q_volume}

The quantum volume operator is a key observable in LQG, both owing to its presence in the scalar Hamiltonian constraint \eqref{scalar_constr} and to the conceptual implications of its eingenvalues and expectation values (which imply, among other things, that geometric properties of spacetime itself can have quantum features). There are, however, some open questions regarding the most suitable regularization approach to obtain the action of the quantum volume operator on spin networks. Several different regularizations lead nonetheless to the same expression for the quantum volume operator up to a prefactor \cite{rovelli2004book,Thiemann2007book, ashtekar_volume}. We therefore leave an arbitrary prefactor $V_0 $ on the volume operator, proportional to the Planck volume. This approach allows for some freedom of choice of a preferred regularization and renders dimensionless results that can be later properly scaled by the desired prefactor. 

It is important to emphasize that, when acting with the scalar constraint on an $n$-valent NLSN, the volume operator appearing in Eq.~\eqref{scalar_constr} actually acts on an $(n+1)$-valent node, since the holonomy $\hat{h}^{(-1)}[p_{k}]$ temporarily raises the valence of the node. Still, this increased valence does not imply that the volume operator simply acts on an $(n+1)$-valent NLSN, since the volume cannot directly ``grasp" the temporary spin-$1/2$ link introduced by the holonomy, and therefore acts on $n$ out of the $n+1$ legs of the spin network~\cite{borissov, gaul}. When the volume operator is directly applied on an $n$-valent NLSN, however, its action is different, and all triplets of linearly independent links selected out of all the $n$ node legs can be ``grasped"~\cite{volume}. 

In the basis of spin networks, the volume operator is composed of other operators $\hat{W}$ that can be represented as spin networks which attach themselves to the input spin network through so-called ``grasps". The resulting spin network contains additional inner structures that can be factored out (with the help of the expressions obtained with recoupling theory and introduced in Secs.~\ref{recoupling_theory} and \ref{intertwiners_SN}) to recover the input graphical structure, rendering $\hat{W}$ a map between spin networks with the same graph. In fact, when acting on the entire (kinematical) Hilbert space of spin networks, the volume operator forms an infinite block-diagonal matrix, and we can without loss of generality restrict the analysis to the specific block to which each spin network of interest belongs, forming equivalence classes of spin networks that can be mapped into each other by $\hat{W}$. The matrices corresponding to these operators have to be diagonalized in the basis of normalized spin networks in order to build the (desired block of the) volume-operator matrix. Formally, for each possible choice of three linearly independent links from each node, one applies the operator $\hat{W}$ on all spin networks of the same equivalence class to build its matrix. The corresponding matrices for each choice of three links are summed with certain weights, taking the square root of the absolute value of the result. The sum over all nodes of such square-root matrices gives the volume-operator matrix representation in the normalized spin network basis. It is worth noting that this protocol for construction of the quantum-volume matrix follows Ref.~\cite{ashtekar_volume}, yet more complicated protocols, requiring the sum of absolute values of $\hat{W}$ for each triplet of links before calculating the square root (which demands multiple diagonalization steps to derive a single volume matrix) are also found in the literature~\cite{volume}. Our choice of volume operator does not only admit a much more practical numerical implementation, but also renders the Hamiltonian constraint anomaly free (i.e., the commutator of the constraint with itself for different lapses is zero on spin networks)~\cite{QSD}. For details on the derivation of the quantum-volume operator, we refer to Refs.~\cite{volume, ashtekar_volume}.

Smearing densitized triads (the conjugated fields to the holonomies) results in angular-momentum-like operators $\hat{J}_i^{(e,v)}$. The index $i$ refers to a choice of SU(2) generator and $(e,v)$ is an assignment of link $e$ incoming {to} or outgoing {from} a node $v$ of the spin network. The operator $\hat{J}_i^{(e,v)}$ acts on holonomies, given as in Eq.~\eqref{element}, by applying on them the generators of SU(2) [cf. Eq.~\eqref{grasp}]. If $e$ is the link along which the holonomy is applied, $\hat{J}_i^{(e,v)}D^{(j)m}_k(g)=-i(\tau^{(j)}_i)^m_n D^{(j)n}_k(g)$ when $e$ is incoming {to} $v$ and $\hat{J}_i^{(e,v)}D^{(j)m}_k(g)= iD^{(j)m}_n(g)(\tau^{(j)}_i)^n_k$ when $e$ is outgoing from $v$, otherwise $\hat{J}_i^{(e,v)}D^{(j)n}_k(g)=0$. 

We define $\hat{W}^{(v)}_{\{e_\alpha, e_\beta , e_\gamma \}}=\eta^{ijk}\hat{J}^{(e_\alpha,v)}_i\hat{J}^{(e_\beta,v)}_j\hat{J}^{(e_\gamma,v)}_k$, where $\eta^{ijk}$ is the structure constant of su(2) and $\{e_\alpha, e_\beta , e_\gamma \}$ is a set of links meeting at the node $v$. Following Ref.~\cite{ashtekar_volume}, we define the operator $\hat{Q}=(1/48)\sum_{\{e_\alpha, e_\beta , e_\gamma \}} \kappa (\{e_\alpha, e_\beta , e_\gamma \} )\hat{W}^{(v)}_{\{e_\alpha, e_\beta , e_\gamma \} }$, where $\kappa (\{e_\alpha, e_\beta , e_\gamma \} )$ is a factor that usually depends on the regularization scheme. Through a process of averaging, $\kappa (\{e_\alpha, e_\beta , e_\gamma \} )$ can be made independent of the regularization, assuming values $\pm 1$ depending on the relative orientation of the linearly independent links in its argument (with respect to the natural orientation of reference frames on the manifold). For the sake of simplicity, we equivalently assume from here on that for 3-valent nodes $\kappa (\{e_\alpha, e_\beta , e_\gamma \} )=6$ whenever the set $\{e_\alpha, e_\beta , e_\gamma \}$ has an ascending index order and $\kappa (\{e_\alpha, e_\beta , e_\gamma \} )=0$ otherwise (note, e.g., that for $\{e_1, e_2 , e_3 \}$ there are six ordering choices, three for which $\kappa =1$ and three with $\kappa=-1$, but in this last case a rearrangement in the indices of $\eta_{ijk}$ gives an extra factor of $-1$), while for 4-valent nodes $\kappa (\{e_2, e_3 , e_4 \} )=\kappa (\{e_1, e_2 , e_4 \} )=6=-\kappa (\{e_1, e_2 , e_3 \} )=-\kappa (\{e_1, e_3 , e_4 \} )$, and $\kappa (\{e_\alpha, e_\beta , e_\gamma \} )=0$ otherwise. The volume operator for a single node is then $\hat{V}=V'_0\sqrt{|\hat{Q}|}$ (or a sum thereof over different nodes, for general spin networks), where $V'_0$ differs from $V_0$ by {a numerical factor extracted from} $\sqrt{|\hat{Q}|}$. 
  
The structure constant in $\hat{W}^{(v)}_{\{e_\alpha, e_\beta , e_\gamma \}}$, which is given by the Levi-Civita symbol, can be graphically represented by a 3-valent node with spin-1 links up to a {factor} of $i\sqrt{6}$, which we include in $V'_0$ together with the $1/48$ factor arising from $\hat{Q}$. The operator $\hat{W}^{(v)}_{\{e_\alpha, e_\beta , e_\gamma \}}$ can therefore be represented by {three} grasps [cf. Eq.~\eqref{grasp}], connected via spin-1 links to a single node. 

We first consider the action of $\hat{W}^{(v)}_{\{e_\alpha, e_\beta , e_\gamma \}}$ on a 3-valent spin network previously modified by a holonomy, as in the last row of Eq.~\eqref{3val_Hstep1} (with the Wigner matrix omitted). In this case
  \begin{widetext}
\begin{equation}
\begin{tikzpicture}[baseline=(current  bounding  box.center)];

\begin{scope}[shift={(0,0)}]
\draw (1.95,1.1) -- (2.75,1.67) -- (3.55,1.1);
\draw (2.75,1.67)-- (2.75,2.226) -- (2.75,2.798);
\draw (2.75,2.226) -- (3.55,2.798);
\node at (2.75,2.798+0.15) {$m$};
\node at (3.7,2.798+0.15) {$\frac{1}{2}$};
\node at (2.52,2.226) {$+$};
\node at (3,1.948) {$j_1$};
\node at (2.75,1.47) {$+$};
\node at (1.85,0.95) {$a$};
\node at (3.65,0.95) {$b$};
\node at (1.3,2.1) {${6^{-1/2}\hat{W}}$};
\end{scope}

\begin{scope}[shift={(11,0)}]
\draw (1.95,1.1) -- (2.75,1.67) -- (3.55,1.1);
\draw (2.75,1.67)-- (2.75,2.226) -- (2.75,3.226);
\draw (2.75,2.226+0.214) -- (3.848,3.226);

\draw (2.75,3.226-0.3333) arc [start angle=90, end angle=50, radius=0.5cm];
\draw (3.2,2.65) arc [start angle=40, end angle=-25, radius=0.8cm];
\draw (3.31,1.8) arc [start angle=120, end angle=165, radius=0.5cm];
\draw (3.31,1.8) arc [start angle=45, end angle=-28, radius=0.5cm];
\draw (3.08,1.37) arc [start angle=350, end angle=200, radius=0.45cm];

\draw[-stealth] (2.75,2.226+0.214) -- (2.75,3.1);
\draw[-stealth] (1.95,1.1) -- (2.44,1.45);
\draw[-stealth] (3.55,1.1) -- (2.85,1.6);

\node at (2.75,3.226+0.15) {$m$};
\node at (3.998,3.376) {$\frac{1}{2}$};
\node at (2.52,3.226-0.3333) {$+$};
\node at (3.53,1.82) {$+$};
\node at (3.63,1.45) {$1$};
\node at (2.53,2.055) {$j_1$};
\node at (3.51,2.35) {$1$};
\node at (2.75,1.47) {$+$};
\node at (2.1,1.38) {$+$};
\node at (2.6,0.83) {$1$};
\node at (3.31,1.1) {$+$};
\node at (1.85,0.95) {$a$};
\node at (3.65,0.95) {$b$};
\node at (-2.75,2.1) {$=-i\left[m(m+1)(2m+1)a(a+1)(2a+1)b(b+1)(2b+1)\right]^{\frac{1}{2}}$};
\node at (4.8,1.3) {$.$};
\end{scope}

\end{tikzpicture}
\label{eq:5.6}
\end{equation}
In Eq.~\eqref{eq:5.6}, the three grasps are contracted with the spin network links of spins $a$, $b$ and $m$, while the temporary spin-$1/2$ link is ``ignored" by the grasps. The orientation chosen for attaching the grasps follows the order of contraction of the indices assuming that holonomies for the spin-$a$, spin-$b$ and spin-$j_1$ links are directed towards the node [cf. Eq.~\eqref{3val_spinnet_total}]. Using Eq.~\eqref{wigner3j_invar}, applied at the node with spins $\{ m, 1/2, j_1\}$, the holonomy along the spin-$j_1$ link can be shifted to the links of spins $m$ and $1/2$, both of which will be directed outwards from the node, resulting in an opposite order for the grasp with the spin-$m$ link compared to the other two grasps. The prefactor in Eq.~\eqref{eq:5.6} comes from the three grasps [cf. Eq.~\eqref{grasp}]. We can then factor out the spin-1 structure of the spin network on the right-hand side of Eq.~\eqref{eq:5.6} as follows:

\begin{equation}

\label{eq:5.7}
\end{equation}
\end{widetext}
In the first equality of Eq.~\eqref{eq:5.7}, the arrows around the node joining the spins $\{m, 1/2, m \}$ have been rearranged using Eq.~\eqref{wigner3j} and a braiding has been performed at the same node, moving the spin-$1$ link to the left. After changing the direction of $\epsilon^{(m)}$, giving a phase factor of $(-1)^{2m}$, a Pachner move, Eq.~\eqref{pachner}, gives the spin network in the second equality of~\eqref{eq:5.7}, in which the spin-$1$ has been braided back to the right side. Applying then Eq.~\eqref{wigner3j_invar} simultaneously to the nodes of spins $\{ a, b, j_1\}$ and $\{k, 1, j_1\}$, as well as Eq.~\eqref{levi3} to the spin-$k$ link, allows us to get rid of all the arrows, obtaining the expression in the second row of Eq.~\eqref{wigner3j}. Finally, using relation \eqref{app2} (cf. Appendix \ref{app2}), we can extract a hexagonal spin network of the form \eqref{hexagon}, which is represented in the last row of Eq.~\eqref{wigner3j} as a Wigner 9j symbol. It is worth noting that summation over $k$ only covers the values $j_1$ or $j_1\pm 1$, but the choice of sign is constrained by the value of $m$: if $m=j_1+1/2$, the Clebsch-Gordan conditions only allow $k=j_1$ or $k=j_1+1$. The Wigner 9j symbol has the interesting property that swapping any two of its columns or rows gives the unswapped symbol up to a phase factor of $(-1)^s$, with $s$ being the sum of all entries of the symbol. For the 9j symbol in the last row of Eq.~\eqref{wigner3j}, $s=3+2a+2b+j_1+k$, while $a+b+j_1\in \mathbb{N}$ by the gauge invariance of spin networks. This implies that,  for $k=j_1$, swapping the rows or columns in the Wigner 9j symbol gives the same symbol multiplied by $-1$. Therefore the symbol has to be zero. As a result, the action of the operator $\hat{W}$ on the considered spin network leads to two possible anti-symmetric $2\times 2$ matrices. One of them couples the spin networks with $k=j_1$ and $k=j_1+1$ for $m=j_1+1/2$, whereas the other couples the spin networks with $k=j_1-1$ and $k=j_1$ for $m=j_1-1/2$. It is easy to show that such matrices have two real eigenvalues with the same absolute value. Therefore, taking the square root of the absolute value of $\hat{Q}$ gives a matrix proportional to the identity. The volume operator in this case acts diagonally.  

When we consider the volume operator acting on $4$-valent nodes modified by a holonomy, we must account for the four possible ways in which $\{e_\alpha, e_\beta , e_\gamma \}$ can be chosen while neglecting the spin-$1/2$ link. Calling $W_{\alpha \beta \gamma }$ the action of $\hat{W}_{\{e_\alpha, e_\beta , e_\gamma \}}$ on the input spin network up to prefactors, the action of $\hat{Q}$ reads
  \begin{widetext}
 \begin{equation}
\begin{tikzpicture}[baseline=(current  bounding  box.center)];

\begin{scope}[shift={(-7,0)}]
\draw (1.1,0.45) -- (1.672,1.25)--(1.1,2.05);
\draw (2.4+0.4,0.45) -- (1.828+0.4,1.25)--  (2.4+0.4,2.05);
\draw (2.4+0.4,2.05) -- (2.8,3.0);
\draw (2.4+0.4,2.05) -- (2.4+0.4+0.95,2.05);
\draw (1.672,1.25)-- (1.828+0.4,1.25);
\draw[dotted,rotate=28] (3.2,0.3) ellipse (1.8cm and 0.85cm);
\draw[-stealth] (1.672,1.25)-- (1.78+0.2,1.25);
\node at (1.75+0.2,1.5) {$l$};
\node at (1.1,2.3) {$j_1$};
\node at (1.828+1,1.65) {$j_4$};
\node at (3.05,2.3) {$+$};
\node at (3,2.65) {$c$};
\node at (3.95,2.05) {$\frac{1}{2}$};
\node at (2.1+0.4,1.25) {$+$};
\node at (1.4,1.25) {$+$};
\node at (1.1,0.2) {$j_3$};
\node at (2.8,0.2) {$j_2$};
\node at (0.4,1.25) {$\frac{8}{\sqrt{6}}\hat{Q}$};
\node at (10.5,1.25) {$=\ i\kappa_{123}L_{j_1}L_{j_2}L_{j_3}W_{123}-i\kappa_{134}L_{j_1}L_{j_3}L_cW_{134}-i\kappa_{124}L_{j_1}L_{j_2}L_cW_{124}-i\kappa_{234}L_{j_2}L_{j_3}L_cW_{234}$ \, .};
\end{scope}

\end{tikzpicture}
\label{eq:5.1}
\end{equation}
In this equation, the dotted ellipsis serves a mere illustrative purpose, crossing the links that can be acted upon by the grasps. We have used the notation $L_j = \sqrt{j d_{j/2} d_{j}}$ and $\kappa_{\alpha \beta \gamma}=\kappa(\{e_\alpha, e_\beta, e_\gamma\})/6$ (with $\kappa_{123}=\kappa_{134}=-1$ and $\kappa_{234}=\kappa_{124}=1$). Note that the sign of the first term on the right-hand side of Eq.~\eqref{eq:5.1} is different because it does not include a ``grasp" on the outward-oriented spin-$c$ link. We consider any trio of links to be linearly independent. As shown in Ref.~\cite{ashtekar_volume}, our choice of $\kappa_{\alpha \beta \gamma}$ corresponds to an arrangement of links with tangents diffeomorphic to the vectors $\{(1,0,0), (0,1,0),(0,0,1),(-1,-1,-1)\}$. Other geometric arrangements [for which the last vector is diffeomorphic to $(1,1,1), (1,1,0),(-1,-1,0),(0,0,1)$ or $(0,0,-1)$] imply different choices of $\kappa_{\alpha \beta \gamma}$, which will not be addressed here. 

The first contribution to the right-hand side of Eq.~\eqref{eq:5.1} is 
\begin{equation}

\label{eq:5.2}
\end{equation}
Note that, given our choice of labels for the links of the spin network, the grasps for the links of spins $j_2$ and $j_3$ have to be braided, causing the node that joins the grasps to have a negative or clockwise cyclicity. Braiding the spin-1 link around the spin-$j_2$ one (around the node joining the spins $\{1, j_2, j_2\}$, where the second grasp has been attached) gives a phase factor of $(-1)^{1+2j_2}$ and allows us to use Eq.~\eqref{pachner} along the link containing $\epsilon^{(j_2)}$ to obtain the spin network in the second row of Eq.~\eqref{eq:5.2}. Simultaneously, we can use Eq.~\eqref{wigner3j_invar} at the node of spins $\{j_1, j_3, l\}$ to transfer the arrows to the spin-$l$ link, and then remove its doubled $\epsilon^{(l)}$ via Eq.~\eqref{levi3}. We also change the cyclic orientation of the other two nodes where the other grasps were attached. Using the resolution of the identity [Eq.~\eqref{app1} in Appendix~\ref{appendix_recoup}], we can factor out the additional structures in the penultimate row of Eq.~\eqref{eq:5.2}, obtaining the result in the last row, where the input spin network graph is recovered.

The second contribution, $W_{134}$, contains a grasp in the spin-$c$ link, which, as previously explained, has its Wigner matrix oriented outwards from its node, since the spin-$j_4$ link has a Wigner matrix oriented toward the node of spins $\{l, j_2, j_4\}$. The corresponding contribution reads
\begin{equation}

\label{eq:5.3}
\end{equation}
Between the first and second rows of Eq.~\eqref{eq:5.3}, we have braided the spin-1 link around the spin-$c$ one [cf. Eq.~\eqref{wigner3j}], while also employing Eqs.~\eqref{wigner3j_invar} and \eqref{levi3} to move $\epsilon^{(c)}$ to the links connected with it through the node of spins $\{ c, c ,1\}$, picking up a phase [namely, $(-1)^{1+c+c}$ from the braiding and $(-1)^{2c}$ from the cancellation of a doubled arrow]. Equation~\eqref{pachner} then allows us to move leftward the attachment point of the spin-$1$ link related to the rightmost grasp. A similar combination of Eqs.~\eqref{wigner3j_invar} and \eqref{levi3} applied to the node joining the spins $\{l, j_1, j_3\}$ converts $\epsilon^{(j_1)}$, $\epsilon^{(j_3)}$ and $\epsilon^{(l)}$ into a phase factor {of} $(-1)^{2l}$. These operations lead to the spin network in the second row of Eq.~\eqref{eq:5.3}. Using Eq.~\eqref{wigner3j_invar} to convert $\epsilon^{(k)}$ and $\epsilon^{(1)}$ into $\epsilon^{(j_4)}$, a second application of Eq.~\eqref{pachner} on the spin-$j_4$ link, followed by a swap of cyclic order at the node of spins $\{j_1, 1 , j_1\}$ and braids on the other two spin-1 links, gives the third row of the equation. Finally, Eq.~\eqref{app2} allows us to factor out a term of the {form}~\eqref{hexagon}, and flipping $\epsilon^{(m)}$ results in the last row of Eq.~\eqref{eq:5.3}.

The next term is given by
\begin{equation}

\label{eq:5.4}
\end{equation}
As in Eq.~\eqref{eq:5.3}, the passage from the first to the second row of Eq.~\eqref{eq:5.4} involves a braiding of the spin-1 link around the spin-$c$ one [cf. Eq.~\eqref{wigner3j}], picking up a phase $(-1)^{1+c+c}$, followed by the application of Eqs.~\eqref{wigner3j_invar} and \eqref{levi2} to move $\epsilon^{(c)}$ to the links connected with it through the node of spins $\{ c, c ,1\}$. Equation~\eqref{pachner} permits us to reallocate the corresponding spin-$1$ link. Analogously, braiding the spin-$1$ link around the spin-$j_1$ one gives a phase $(-1)^{1+j_1+j_1}$. The resulting spin network is given in the second row of Eq.~\eqref{eq:5.4}. We {can} then use Eq.~\eqref{pachner} on the link containing $\epsilon^{(j_1)}$ and combine $\epsilon^{(j_2)}$ and $\epsilon^{(l)}$ into $\epsilon^{(j_4)}$ employing Eq.~\eqref{wigner3j_invar} (this gathers the arrows on a single node, allowing them to be effectively removed). The spin network in the third row is then obtained after performing a braiding at the node joining the spins $\{j_4 , k, 1\}$. The final expression {is reached} by factoring out a term of the form~\eqref{hexagon} by employing Eq.~\eqref{app2}, recovering in this way the input spin network graph. 

The last contribution to the action of $\hat{Q}$ on the modified 4-valent nodes is
\begin{equation}

\label{eq:5.5}
\end{equation}
  \end{widetext}

Between the first and second rows of Eq.~\eqref{eq:5.5}, we have performed a braiding of the spin-1 link around the spin-$c$ one [cf. Eq.~\eqref{wigner3j}], followed by the application of Eqs.~\eqref{wigner3j_invar} and \eqref{levi2} to move $\epsilon^{(c)}$ to the links connected with it through the node of spins $\{ c, c ,1\}$, picking up a phase $(-1)^{1+c+c}$.  We have used Eq.~\eqref{pachner} on the link containing $\epsilon^{(j_3)}$ and combined $\epsilon^{(j_2)}$ and $\epsilon^{(l)}$ into $\epsilon^{(j_4)}$ by means of Eq.~\eqref{wigner3j_invar}. Inverting the cyclic order around the node connecting the grasps gives a $-1$ prefactor. This has led to the expression in the second row of Eq.~\eqref{eq:5.5}. We have then applied Eq.~\eqref{pachner} to the link containing $\epsilon^{(c)}$ and performed braidings at the nodes joining the spins $\{j_4 , k, 1\}$ and $\{m , l, 1\}$, flipping also $\epsilon^{(m)}$. The final expression has been derived by factoring out a term of the form~\eqref{hexagon} by means of Eq.~\eqref{app2}, a procedure that allows us to recover the input spin network graph.

Normalization of spin networks results in the change $d_m d_k \to \sqrt{d_l d_{j_4} d_m d_k}$ in Eqs.~\eqref{eq:5.2}-\eqref{eq:5.5}. Once these equations have been introduced into Eq.~\eqref{eq:5.1}, the matrix elements of $\hat{Q}$ can be calculated between spin networks with spins $\{ l, j_4 \}$ at the input and $\{m, k\}$ at the output, for every choice of these spins that fulfills triangularity with respect to the fixed ``external" spins {$j_1$, $j_2$, $j_3$} and $c$. Since $c=j_4\pm 1/2$, its value also determines the admissible values for $k$ when the Wigner 6j symbols are taken into account. In more detail, since both $\{k, 1, j_4\}$ and $\{c, 1/2, k\}$ must fulfill triangularity, $c=j_4\pm 1/2$ implies that $k\in \{j_4, j_4 \pm 1\}$ (the choice of plus or minus is fixed by the value of $c$). There are, therefore, two choices of $k$ for each $c$. The values that $m$ can assume depend on the spins $\{j_1, j_2, j_3, j_4, k\}$, but since $k$ can take different values on its own, we consider $|j_1 - j_3|\leq m\leq  j_1 + j_3$, what gives $2\min \{j_1,j_3\}+1$ possible values. The matrix representing $\hat{Q}$ has therefore dimension $2(2\min \{j_1,j_3\}+1)$. Once the $i$ factor is included, this matrix can be diagonalized to give a matrix of purely real eigenvalues. The square root of their absolute values gives the volume matrix after the application of the inverse diagonalizing unitaries. 

Since we are also interested in the expectation values of the volume operator, we still need to calculate its action on spin networks that have not been modified by holonomies. For the $3$-valent case [as in Eq.~\eqref{3val_spinnet}], it is easy to show that the volume operator always gives zero~\cite{volume, Ilkka}. When the volume operator is applied on spin networks of the form~\eqref{4val_spinnet}, however, it acts in a nontrivial manner that differs both from Eqs.~\eqref{eq:5.6} and \eqref{eq:5.1}, because now all of the four links connected with the intertwiner can be grasped (as long as they are linearly independent). In terms of the spin network~\eqref{4val_spinnet}, only the links of spins $j_1$, $a$, $b$ and $j_4$ are effectively grasped, because the two $3$-valent nodes give zero contributions to the volume. The action of the volume operator can be derived in a similar manner to what was done in Eqs.~\eqref{eq:5.1}-\eqref{eq:5.5}, having four grasp arrangements when 4-valent nodes are considered. A much simpler derivation, however, can be obtained by setting the spin of the temporary link in the spin network on the left-hand side of Eq.~\eqref{eq:5.1} to zero and using Eq.~\eqref{wigner3j_to_levi} \footnote{Note that the way the grasp is attached to the link along the direction $p_4$ is different when the temporary spin-$1/2$ link is present. Nonetheless, after applying Eq.~\eqref{wigner3j_to_levi} when we take the limit in which the spin $1/2$ tends to $0$, the difference boils down to a braid operation and an arrow flip, giving an overall phase of $(-1)^{1+j_4+j_4}(-1)^{2j_4}=-1$, which cancels out the minus in the prefactors in Eq.~\eqref{eq:5.1}. This shows graphically how the volume for 4-valent intertwiners can be obtained from Eq.~\eqref{eq:5.1}.}. Therefore, we omit the re-derivation of the action of the volume operator for a 4-valent intertwiner. A general derivation of the action of the volume operator on nodes of arbitrary valence can be found in Ref.~\cite{Ma}, and its spectral analysis when acting on nodes of valence up to 7 can be found in Ref.~\cite{Brunnemann}.   
  
\section{Numerical implementation of the scalar constraint and the volume operator}\label{numerics}

The analytical discussion presented in Secs.~\ref{sec_3_val} and \ref{sec_4_val} shows the complexity of the action of the scalar constraint on spin networks. Even though this action is confined to the vicinity of the nodes of the spin networks, the changes it induces forces us to consider a rapidly growing set of spin networks with different graph structures and different spin attributions to their links. It is well known that these complications render the study of the scalar constraint in LQG almost unfeasible with currently available analytical and numerical tools~\cite{Hanno, Hanno2}. As a consequence, many questions still remain open in the field. As a remedy, approximations like the graph-preserving ones have been proposed, yet the regime of validity of most of these approximations (or even their validity overall) remains obscure.

As an effort to understand the graph-changing properties of Eq.~\eqref{scalar_constr} and to overcome some of the problems imposed by its action on spin networks, we develop here a new numerical approach that allows us to apply Eq.~\eqref{scalar_constr} on 3-valent and 4-valent spin networks without resorting to approximations. With this aim, we have implemented this new numerical framework as a code in Mathematica, but we believe that it can similarly be implemented (and potentially further optimized) in other programming languages and computational software.  The interested reader that wants to know the code in full detail, check how it works in practical cases or even develop it for other  applications can access it in Ref.~\cite{link}.

A key idea of our approach is encoding spin networks in a way that numerical tools can easily understand and process. Graphical input and output are rather unpractical and resource consuming, yet we must store information not only about spins assigned to links, but also about the (constantly changing) arrangement of these links on a substrate manifold (or, more generally, their adjacency relations). This varying complexity renders adjacency matrices unpractical. We therefore fix a certain valency, and consider either three or four external legs (the outermost links) such that their internal structures can accommodate, in principle, an arbitrarily large number of inner loops ordered in terms of proximity to the central virtual link. This spin and location information is stored as ordered lists, each being in one-to-one relation (up to a padding of zeros) with a given spin network (see further discussion for details). Spanning a vector space out of these lists is, however, not possible in a direct manner, and for this reason we instead adopt a vector space of functions for which the arguments are these lists. The functions are never truly defined (i.e., at no point during the computation they are assigned a functional form), and for this reason we call them ghost functions. All needed information for the computation is stored in the arguments of ghost functions. These arguments can have arbitrary sizes and the orthogonality relations of ghost functions are only based on whether their arguments coincide. If $s_i = \{s_{i,1}, s_{i,2}, \ldots\}$ denotes lists encoding the spin and structural information of distinct (normalized) spin networks, the inner product $I$ on functions $f(s_i)$ is defined such that $I[ f(s_i)| f(s_j)] = \delta_{i,j} $. The scalar constraint is then included in our code as a linear functional ${C}_s$ that acts on the ghost functions by reading and manipulating their arguments, i.e., $C_s [f(s_i)]=\sum_j c_j(s_i) f(s_j)$ for coefficients $c_j$ taken from Eqs. \eqref{3val_total} or \eqref{4val_total1} and \eqref{4val_total2} for 3-valent or 4-valent spin networks, respectively. Linearity then implies that $C_s [\sum_i c_i f(s_i)]=\sum_i c_i C_s[f(s_i)]$, so that the constraint functional can be used recursively, e.g., to generate perturbative outputs. 

Let us start with the discussion of the functional for 4-valent spin networks, since this is the most relevant and intricate case. We do not constrain ourselves to the consideration of structures of the form~\eqref{4val_spinnet}, but instead assume that we start with an NLSN with four external legs, an inner virtual link and an arbitrary number of inner loops. The inner loops can be arranged in six different ways, by connecting links belonging to each possible pair of directions (say, $p_1$, $p_2$, $p_3$ or $p_4$ according to our previous notation). We label the locations of such inner loops from 1 to 6, corresponding to inner loops connecting the links along the pairs of directions $\{p_1,p_3\}$, $\{p_2,p_3\}$, $\{p_2,p_4\}$, $\{p_1,p_4\}$, $\{p_1,p_2\}$ {and} $\{p_3,p_4\}$, respectively [cf. Fig.~\ref{pseudocode}(a)]. One important thing is that the presence of certain inner loops affects the ways in which Eq.~\eqref{scalar_constr} can attach new inner loops. If a loop is present in location 1 (placed between directions $p_1$ and $p_3$), for example, the scalar constraint attaches a new loop in the same location by coupling its holonomies with the already existing loop links, leading to no change in the graph structure, but changing the spins attributed to these links (unless the spin of the connecting link is reduced to zero, which effectively changes the graph by removing the loop). On top of that, the Hamiltonian also applies holonomies to form inner loops in the locations 2, 3, 4, 5 and 6, but the presence of a loop in location 1 means that loops in locations 2, 4, 5 and 6 (which share a common link with loop 1) would have to be introduced further inwards (or closer to the node) relative to the location-1 loop, while the introduction of a loop in location 3 is completely unaffected by this subtlety. A diagrammatic representation of these loop-attachment relations is shown in Fig.~\ref{pseudocode}(b) in the form of a pseudocode, which also summarizes our implementation of $C_s$. As a consequence of these relations, recursive application of Eq.~\eqref{scalar_constr} leads to structures with increasingly deeper inner loops, with depths that depend on the positions of outer loops. 

\begin{figure*}
    \centering
    \includegraphics[width=\linewidth]{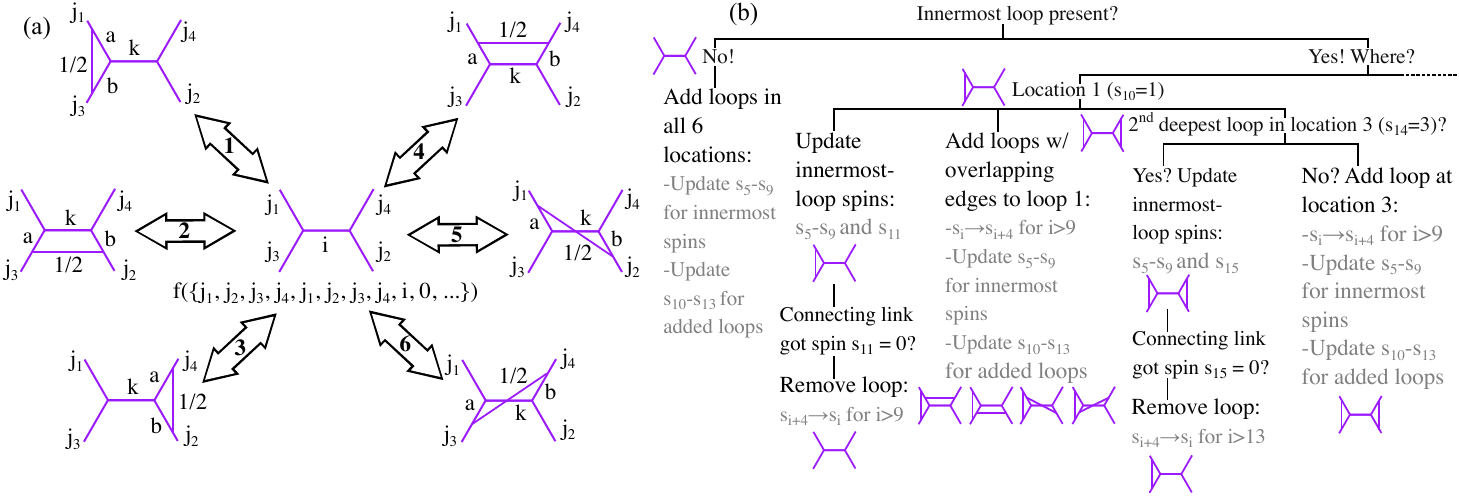}
 \caption{(a) Schematic representation of the graphical changes introduced by the Hamiltonian constraint on 4-valent node-like spin networks. The central spin network contains its encoding function represented below. The rest of spin networks, differing from the central one by the addition of an inner loop at the location indicated within the arrows, are represented by the functions $f(\{j_1, j_2, j_3, j_4, a, j_2, b, j_4, k, 1, 1/2, j_1, j_3, 0, \ldots , 0 \})$, $f(\{j_1, j_2, j_3, j_4, j_1, b, a, j_4, k, 2, 1/2, j_2, j_3, 0, \ldots , 0 \})$, $f(\{j_1, j_2, j_3, j_4, j_1, b, j_3, a, k, 3, 1/2, j_2, j_4, 0, \ldots , 0 \})$, $f(\{j_1, j_2, j_3, j_4, a, j_2, j_3, b, k, 4, 1/2, j_1, j_4, 0, \ldots , 0 \})$, $f(\{j_1, j_2, j_3, j_4, a, b, j_3, \newline j_4, k, 5, 1/2, j_1, j_2, 0, \ldots , 0 \})$ and $f(\{j_1, j_2, j_3, j_4, j_1, j_2, a, b, k, 6, 1/2, j_3, j_4, 0, \ldots , 0 \})$, respectively for loop insertions at positions 1, 2, 3, 4, 5 and {6.} (b) Pseudocode for the Hamiltonian implementation. The code checks whether an inner loop is present. If absent, it introduces inner loops in all six locations, with spin $1/2$ on the newly created link. If present, for each possible location, a series of steps are followed. The case for location 1 is shown, while for other locations the dashed-line continuation of the diagram implies the presence of similar rules not shown. The corresponding innermost loop has its spins shifted by all allowed values without graph changes, and if the connecting link reaches spin 0, it is removed, and the inner-loop data in the corresponding list is shifted to the left by four entries. Additionally, inner loops are added to all other positions, but if a loop was added at position $3$ right before adding one at position $1$ (these loops share no links), it is again possible to remove its extra link or simply change its spin. The diagram contains examples for the simplest spin networks for which the rules apply.}
  \label{pseudocode}
\end{figure*}

We choose our spin
network encoding lists to have the first four entries representing the spins of the four outermost links: $j_1$, $j_2$, $j_3$ and $j_4$, following the convention of Eq.~\eqref{4val_spinnet}. These values are fixed and unaffected by the functional $C_s$, but should be stored in the list for the purpose of normalization after Eq.~\eqref{scalar_constr} is applied a desired number of times. The next four entries in the list are the four innermost spins adjacent to the central virtual link along directions $p_1$, $p_2$, $p_3$ and $p_4$, which will be affected by the scalar constraint [e.g., $j_1$, $b$, $a$ and $j_4$ in Eq.~\eqref{4val_spinnet}]. The 9th entry is the spin of the central virtual link. If the spin network has no inner loops, all remaining entries in the list are zero [cf. ghost function under central NLSN in Fig.~\ref{pseudocode}(a)]. For spin networks containing inner loops, the innermost loop will occupy the next four entries of the list, and every following loop, in decreasing order of depth, will be described by four additional entries. From each such quadruple of entries, the first two store the location of the loop and the spin of the connecting link of the loop, while the other two store the spins adjacent to the loop along the directions that the loop connects. For the sake of an example, let us consider the spin network ~\eqref{4val_spinnet}. According to our convention, the aforementioned quadruple is 2, $\varepsilon$, $j_2$ and $j_3$, and the full encoding list corresponding to the spin network is $\{j_1, j_2, j_3, j_4, j_1, b, a, j_4, i, 2,\varepsilon, j_2, j_3, 0, \ldots, 0\} $, where the zero padding should be chosen in such a way as to accommodate as many inner-loop entries as one intends to recursively apply the Hamiltonian constraint. The size of the lists should be fixed prior to any calculations, so that orthogonality relations can be properly applied. 

The action of the functional $C_s$ reorganizes the lists contained as arguments in the ghost functions according to all $p_i$-direction permutations of Eqs.~\eqref{4val_total1} and \eqref{4val_total2} (which are actually not simply obtained by permuting arguments, as explained below). When $C_s$ creates a new inner loop, it effectively moves all list entries from 10th onward to the right by four entries, so that entries corresponding to inner loops are now moved down in depth order for the data corresponding to a new loop to be included. The new spins immediately adjacent to the central nodes are encoded in entries 5 to 8, and the new central spin becomes the 9th entry. Entries 10 to 13 receive the information about the added innermost loop, according to our loop-description convention. In this manner, the zero padding in the lists is gradually filled from the left with inner-loop information as new loops are included in the spin network by the action of the scalar constraint. Following our example, if we add a new loop in location 1 to the spin network~\eqref{4val_spinnet}, its list will change to $\{j_1, j_2, j_3, j_4, c, b, d, j_4, k, 1,\gamma, j_1, a, 2,\varepsilon, j_2, j_3, 0, \ldots, 0\} $, where $c$ and $d$ are the new innermost spins along directions $p_1$ and $p_3$, respectively, $k$ is the new central spin and $\gamma$ is the connecting spin of the new loop at location 1. Although the spin networks and their encoding lists become increasingly complicated with each application of Eq.~\eqref{scalar_constr}, only the two deepest inner loops of a 4-valent NLSN are acted upon by the Hamiltonian. Since in our encoding the information about these two loops is stored between the 5th and 17th entries of the list, the coefficients $ c_j(s_i)$ in $C_s [f(s_i)]=\sum_j c_j(s_i) f(s_j)$ depend only on these entries of the input ghost-function argument, avoiding the search for entries scattered among large lists (in fact, the size of the list does not affect the Hamiltonian functional).

It turns out that the constraint functional might output the same spin network with different coefficients as independent terms in a linear combination, what slows recursive application of the Hamiltonian by forcing its functional to evolve the same spin network multiple times. To remedy that, consecutive applications of the constraint functional are intercalated by a ``collector" functional (inbuilt in Mathematica as ``Collect[\,]" command) that collects all coefficients of the same spin network into one.

The Hamiltonian acts on and also generates non-normalized spin networks, so normalization takes place after $C_s$ has been recursively applied a number of times, and denormalization is employed prior to any calculations if one decides to start with normalized spin networks. We have developed a ``normalizer" functional, which linearly implements the normalization discussed in Sec.~\ref{intertwiners_SN} according to the formula $f(s_i)\to [d_{j_1}d_{j_2}d_{j_3}d_{j_4}\prod_k d^{-1}_k]f(s_i)$, where $d_{j_i}$ is associated with the outermost legs and $k$ runs over the spins of all links in the spin network, including the outermost ones (so that they effectively cancel out from the normalization factor). To achieve this, the normalizer reads the first six entries of each ghost-function argument list, as well as the $j$-th, $(j-1)$-th and $(j-2)$-th entries for $j = 4n+9$ ($n\in \mathbb{N}$). Note that the zero padding does not contribute since $d_0 = 1$. Similarly, a ``denormalizer'' functional has been implemented to perform the inverse of the normalizer functional.

After normalization has been performed, one can either calculate inner products or act with observables on the output states in order to estimate expectation values. As key observable of our work, we have implemented a quantum-volume functional, which generates matrices depending on the input spin networks. Note that the volume operator only ``sees" the innermost spins in the spin network, i.e., those closest to the inner virtual link. These spins determine the size of the matrix $\hat{Q}$ generated by the volume operator. Since the volume operator maps a 4-valent NLSN with central spin $i$ to a linear combination of 4-valent NLSNs with all possible central spins, the size of the matrix it generates runs from $ |j'_1 - j'_3|$ to $ j'_1 + j'_3$ (for innermost spins $j'_1$, $j'_2$, $j'_3$ and $j'_4$), and the indices are the input and output spin values of the central link. Note that this is equivalent to considering the range of spins from  $\max \{ |j'_1 - j'_3|, |j'_2-j'_4|\}$ to $\min \{ j'_1 + j'_3, j'_2+j'_4\}$, since for $|j'_1 - j'_3| < |j'_2-j'_4|$ and/or $ j'_1 + j'_3 > j'_2+j'_4$ the matrix we generate contains the actual volume matrix as a block, and the remaining elements are all zero. Following the calculations in Sec.~\ref{Q_volume}, the $\hat{Q}$ matrix is generated according to Eq.~\eqref{eq:5.1}, and the resulting matrix is diagonalized, so that the square root of the absolute value of its entries can be taken before applying the inverse of the diagonalizing transformation. The resulting matrix is the volume operator in a basis of 4-valent spin network states, and can be turned into a linear functional $V$ by using its matrix elements as coefficients of the output linear combination, $V[f(s_i)]=\sum_j V_{j, i} f(s_j)$. 

Finally, the inner product is introduced as a functional $I$ that is antilinear in its first argument and linear in its second one, 
\begin{equation}
I[\sum_i c_i f(s_i), \sum_j d_j f(s_j)]=\sum_{i,j} c^*_i d_j I[f(s_i), f(s_j)].
\end{equation}
Once two (linear combinations of) normalized spin networks {are given} in the form of ghost functions with suitable list arguments{, the} inner-product {functional} evaluates {the} orthonormality based on the criterion of whether the lists in the arguments of the functions are the same (i.e., $I[f(s_i), f(s_j)]=\delta_{i,j}$, as defined above). 

For 3-valent NLSNs, we use a similar scheme for encoding graphs and spins as lists. The first {three} entries of the list carry the information about the spins of the outermost links of the spin network following counter-clockwise order starting from the top [e.g., for the spin network~\eqref{3val_spinnet} these would be $j_1$, $j_2$ and $j_3$]. The following three entries correspond to the counter-clockwise ordered innermost spins [once again, for the spin network~\eqref{3val_spinnet} these would be $j_1$, $a$ and $b$]. For each inner loop in descending order of depth we assign groups of four entries, starting at position 7 in the ordered list. In each of these quadruples, the first entry indicates the position of the inner loop (1 for upper left, 2 for bottom and 3 for upper right). The second entry stores the spin of the bridging link in this loop and the remainder entries give the spins of the links adjacent to (but not included in) the loop. The scheme is similar to the one introduced for $4$-valent NLSNs, and allows us to implement the scalar constraint as a functional acting on the arguments of ghost functions. When it creates a new inner loop, it effectively moves all list entries from 7th onward to the right by four entries, so that entries corresponding to inner spins are now moved down in depth for the data corresponding to a new loop to be included. The new spins immediately adjacent to the node are encoded in entries 4 to 6, while entries 7 to 10 receive the information about the new innermost loop.

\section{Results}\label{results}

Our numerical approach allows us to investigate a variety of properties of the scalar constraint. One of the open questions regarding the operator~\eqref{scalar_constr} is finding its zero-eigenvalue eigenstates, since these ultimately span the space of physical states in LQG. A solution to Eq.~\eqref{scalar_constr} was given in Ref.~\cite{Alesci}, although the corresponding derivation was based on an incomplete application of the scalar constraint on 4-valent spin networks. In Ref.~\cite{Lewandowski_state} it was shown that, in the cosmological symmetry-reduced case where a massless scalar field serves as relational clock variable, solutions to the scalar constraint of the joint matter and gravitational fields in a certain region of phase space can be given by cylindrical functions generated from transformed wave functions depending solely on the Ashtekar-Barbero one-form. Furthermore, in Ref.~\cite{Pullin_state} it was shown that when the quantum-deformed Temperley-Lieb algebra is considered, certain states generated through transforms with a Chern-Simons kernel are eigenstates of the (deformed) Thiemann's Hamiltonian constraint. Using our code, which allows to implement the scalar constraint acting on spin networks of the forms~\eqref{3val_spinnet} and \eqref{4val_spinnet} with arbitrary spins assigned to their links, we have searched for zero-eigenvalue solutions of Eq.~\eqref{scalar_constr}. Our protocol is based on ``For" loops (a routine that runs a section of code repeatedly while varying some parameters) covering all possible spin values within a certain range on each of the links besides the one containing a Wigner matrix, for which the spin is fixed at zero. Spin assignments not fulfilling triangularity at the nodes are excluded from the search, since they violate one of the LQG constraints. Whenever $C_s [f(s_i)]=0$ for a certain $s_i$ within the search range, our protocol prints the corresponding spin assignments that led to this result. For spin networks of the form~\eqref{3val_spinnet}, we vary each of the spins $j_1, j_2, j_3 \in \mathbb{N}/2$ from 0 to $7/2$ while keeping $\varepsilon =0$ (therefore $a=j_2$ and $b=j_3$). The only spin values for which the condition $C_s [f(s_i)]=0$ is fulfilled are $j_1= j_2= j_3 =0$. Additional numerical data for spin networks with more inner loops suggest that any spin network with zero innermost spins connected to the $3$-valent intertwiner is also an eigenstate of Eq.~\eqref{scalar_constr} with null eigenvalue. By inspecting Eq.~\eqref{3val_total} we can easily understand and generalize these results. If we set, e.g., $j_1=a=b=0$ for the input \eqref{3val_spinnet}, we see that the two terms within square brackets in that equation turn out to be equal and cancel out [cf. also Eq.~\eqref{messiah1}]. The $\mathcal{C}_3$ rotational symmetry at the innermost-node level assures that the result extends to the three possible ways of inserting an inner loop. 

Running the search protocol on spin networks of the form~\eqref{4val_spinnet} with $\varepsilon=0$ (i.e., allowing for all possible terms derived in Sec.~\ref{sec_4_val}) reveals that the same single family of eigenstates of the Hamiltonian with $j_1=j_2=j_3=j_4=i=0$ can be found. Not surprisingly, when we assume that the link along the direction $p_4$ does not contribute to the action of the constraint, we get the same zero-eigenvalue family of eigenstates of the Hamiltonian. These eigenstates have no volume, since their innermost spins are all zero, yet, due to their possible ``shielding" by inner loops with nontrivial spins, they can still have nonzero areas, and hence these eigenstates might potentially serve as boundaries. Once the spins of the innermost links of a 4-valent NLSN are set to zero, the volume in the Hamiltonian constraint makes the action of the latter vanish on the central NLSN node. Owing to cylindrical consistency, only a graph effectively containing 3-valent nodes that belong to the inner loops is left from the NLSN. These 3-valent intertwiners have coplanar links (therefore the Hamiltonian acts trivially on them, with vanishing result) and possess zero volume, and consequently so does the entire NLSN. On the other hand, these NLSNs still possess (non-zero) lengths, areas (along certain ``cuts" of the manifold), and dihedral angles. One can hence triangulate a 3-dimensional region in four dimensions with a spin network composed of 4-valent nodes in its bulk, but effectively containing only 3-valent nodes at its boundaries.

When acting on any single spin network or on a linear combination of spin networks that cannot be generated from one another by inner-loop couplings, the Hamiltonian generates a linear combination of spin networks that has no overlap with the input state. In fact, starting from a certain $|s_0\rangle$ for which $\hat{C}_s |s_0\rangle =c^*_{1} |s_{1}\rangle$ (e.g., a state from which one cannot remove inner loops), if we denote the (normalized linear combinations of) states generated by $i$ loop insertions as $|s_i\rangle$, with $\langle s_i | s_j \rangle = \delta_{ij}$, we have  $\hat{C}_s |s_i\rangle = c_i |s_{i-1}\rangle + c^*_{i+1} |s_{i+1}\rangle =\langle s_{i-1} |\hat{C}_s |s_i\rangle |s_{i-1}\rangle +\langle s_{i+1}|\hat{C}_s |s_i\rangle|s_{i+1}\rangle$. From this relation and any suitable $|s_0\rangle$, containing intertwiners of any valency, we can generate the following solution of the Hamiltonian constraint:
\begin{widetext}

\begin{equation}
    | E_0\rangle = |s_0\rangle - \frac{\langle s_{1}|\hat{C}_s |s_0\rangle}{\langle s_{1}|\hat{C}_s |s_2\rangle}|s_2\rangle + \frac{\langle s_{1}|\hat{C}_s |s_0\rangle \langle s_{3}|\hat{C}_s |s_2\rangle}{\langle s_{1}|\hat{C}_s |s_2\rangle \langle s_{3}|\hat{C}_s |s_4\rangle}|s_4\rangle + \ldots = 
    \sum_{i \, \text{even}}(-1)^{i/2}\frac{\langle s_{1}|\hat{C}_s |s_0\rangle }{\langle s_{1}|\hat{C}_s |s_2\rangle }\cdots\frac{\langle s_{i-1}|\hat{C}_s |s_{i-2}\rangle }{\langle s_{i-1}|\hat{C}_s |s_i\rangle }|s_i\rangle \, .
    \label{eq_eigenstate}
\end{equation}
Although it is not clear whether this state can be normalized, it is easy to check that it is annihilated by the action of Eq.~\eqref{scalar_constr},
\begin{equation}
    \hat{C}_s| E_0\rangle = 
    \sum_{i \, \text{even}}(\cdots)\bigg[\frac{\langle s_{i-1}|\hat{C}_s |s_{i-2}\rangle }{\langle s_{i-1}|\hat{C}_s |s_i\rangle }\langle s_{i+1}|\hat{C}_s |s_{i}\rangle |s_{i+1}\rangle-\frac{\langle s_{i-1}|\hat{C}_s |s_{i-2}\rangle\langle s_{i+1}|\hat{C}_s |s_{i}\rangle }{\langle s_{i-1}|\hat{C}_s |s_i\rangle\langle s_{i+1}|\hat{C}_s |s_{i+2}\rangle }\langle s_{i+1}|\hat{C}_s |s_{i+2}\rangle |s_{i+1}\rangle\bigg] = 0 \, .
    \label{eq_eigenstate_check}
\end{equation}
\end{widetext}
 The idea underlying the construction of the solution $|E_0\rangle$ is that two consecutive terms on the right-hand side of Eq.~\eqref{eq_eigenstate} differ by two loop insertions, and applying the Hamiltonian constraint converts them into the same linear combination of spin networks (inserting loops on one and removing loops from another) with opposite prefactors. Furthermore, although Eq.~\eqref{eq_eigenstate} holds for entire spin networks if the values of the lapse contained in $\hat{C}_s$ are the same at all nodes, if one allows different values at each node, one needs to build NLSN solutions via Eq.~\eqref{eq_eigenstate} for each ``building block'' of the spin network and then contract these NLSN solutions to create a full spin network solution. This construction assures that $|E_0\rangle$ remains independent from the values of the lapse [note the mutual cancellation of these values between numerators and denominators in the coefficients in Eq.~\eqref{eq_eigenstate}]. Put simply, breaking a large spin network into NLSNs and using Eq.~\eqref{eq_eigenstate} on each of them before reassembling a spin network assures that Eq.~\eqref{eq_eigenstate_check} is fulfilled at each spin network node, since the action of the Hamiltonian on the entire spin network is the sum of its action at each node. Note that the series~\eqref{eq_eigenstate} does not necessarily need to converge. A suitable habitat for solutions of this form is the algebraic dual of the linear span of spin network states. If this dual contains all relevant solutions, endowing it with a convenient inner product (and averaging over diffeomorphims)  should suffice to construct a Hilbert space of physical states. Nonetheless, exclusion of NLSNs with (non-exceptional) links removed by the Hamiltonian might be necessary to implement diffeomorphism invariance on these solutions~\cite{QSD2}, and additional modifications of the Hamiltonian action might be needed if diffeomorphism-invariant spin networks with pairs of nodes sharing two or more links are considered~\cite{Madhavan}.

Another interesting open question in LQG is the regime of validity of some commonly used approximations, such as the assumption that the graph does not change. A hypothesis that is frequently employed to support this assumption is that a coarse-grained triangulation of a manifold might produce a spin network capturing the key features of the quantum geometry and such that its dynamics effectively leaves the graph unaffected~\cite{rovelli_stepping_2008, Giesel_GP, loopySN}. One can indeed see that a first application of Eq.~\eqref{4val_spinnet} on a spin network implements a graphical change, but a consecutive application of the constraint should recover the ``original" spin network. Nonetheless, as our discussion in Secs.~\ref{sec_3_val} and \ref{sec_4_val} shows, while the scalar constraint is formally symmetric and maps output into input once applied a second time, it also maps these ``1st-order" output states into a new family of spin networks with even larger graphical changes relative to the input spin network. If Eq.~\eqref{scalar_constr} is recursively applied many times, the number of spin networks with graphs that depart from the starting spin network structure increases drastically. 
It is therefore unclear whether such changes can be effectively absorbed into a coarse-grained spin network with graph-preserving dynamics.

To investigate the validity of this graph-preserving approximation, in the rest of this section we are going to discuss the transformation properties of some fiducial spin networks up to a certain order in perturbation theory, which corresponds to applying Eq.~\eqref{scalar_constr} recursively up to a fixed number of times. More concretely, we are going to study how the expectation value of the volume operator varies when comparing graph-changing and graph-preserving dynamics. The choice of the volume operator as the figure of merit is based on the central role that this operator plays in the dynamics (since it is present in the Hamiltonian) and in the conceptual foundations of LQG (it is one of the key operators of LQG and also leads to rather drastic and distinct quantum consequences, such as the discretization of spacetime geometry). 

The Hamiltonian constraint obtained after quantization of the Ashtekar-Barbero variables should in principle be integrated over the volume of a 3D manifold, smeared by a distribution corresponding to the (time) lapse. When a regularization protocol is adopted to allow for a description of the constraint in terms of operators acting locally at nodes of the spin network, as in Eq.~\eqref{scalar_constr}, the lapse distribution is reduced to a set of parameters $N_\boxtimes$, each of which is related to one of the nodes of the spin network, appearing as a summand in Eq.~\eqref{scalar_constr}. When considering only a node [so that the sum in Eq.~\eqref{scalar_constr} disappears], something that is justified by the independent action of the Hamiltonian on each node of the spin network, the single parameter $N=N_\boxtimes$ plays a role similar to time in the standard unitary description of quantum mechanics (in the absence of time ordering). Following this similarity, we choose to treat $N$ as our perturbation parameter. Our objective is to construct the unitary operator generated by the Hamiltonian constraint, $\hat{U}=\exp (-i\hat{C}_s[N])$~\cite{Reisenberger}, and expand it as a series in our perturbation parameter up to a specific desired order, acting with it on a given spin network and then estimating the expectation value of the volume with respect to the resulting transformed state. 

For the consideration of the action of the graph-changing constraint on all legs of an NLSN, we are going to compute terms only up to 3rd order in perturbation theory. The reason is that, in this case, the number of considered possibilities is rather large, and therefore also the computation times (cf. Table~\ref{func_cost}). In contrast, for the case in which the constraint is restricted to ignore the NLSN links along $p_4$, we are going to include terms up to 4th order. On the other hand, the considered graph-preserving constraint is going to act on spin networks that have three different structures [see Figs.~\ref{plot2}(a), 4(b) and 4(c)] by inserting extended loops solely between neighboring intertwiners, and the perturbative expansion of the unitary transformation in these cases is going to be truncated at 4th order. 

It is worth noting that 3rd- and general odd-order contributions to the expectation values are absent in all calculations, but the reasons for this differ between graph-preserving and graph-changing dynamics. Since, under the action of the graph-changing constraint, any graph can only be recovered after applying the constraint an even number of times, while the volume operator does not change the graph, but rather shuffles spin assignments, any term of the form $\langle  \hat{C}^n_s \hat{V}^l \hat{C}^m_s\rangle $ in which $l, n, m\in \mathbb{N}$, $n$ is even and $m$ is odd, or vice versa, gives zero whenever one starts from a single spin network. This is also true for matrix elements of the Hamiltonian itself, with $\langle  \hat{C}^m_s\rangle =0$ for $m$ odd (since the structures generated by $\hat{C}^m_s$ starting from any spin network differ graphically from that of the starting spin network). For the graph-preserving case, however, a certain spin network can in general be recovered also after an odd number of applications of the Hamiltonian constraint, as is the case for ladder-like spin networks [cf. Fig.~\ref{plot2}(c)] if one couples loops not only between each pair or intertwiners, but also from the ``sides'' of the ladder structure (which is closed via contraction of the lower legs of the bottommost intertwiner to the upper legs of the uppermost intertwiner, making the ladder into a possibly twisted ring). If such spin network is composed of an odd number $m$ of intertwiners, $m$ different loops coupled between intertwiners with two additional side loops suffice to recover the initial spin network. Although still $\langle  \hat{C}^{m+2}_s\rangle = 0$ (because the constraint generates a Hermitian and purely imaginary matrix~\cite{QSD, QSD2}), the terms of odd order in the expectation value of the transformed volume, e.g., $\langle  \hat{C}^{m+2}_s\hat{V}\rangle$, can fail to cancel out in general, leading to asymmetries in the volume dependence on $N$ (and hence on the proper time $T=\int \mathrm{d}t N(t)$~\cite{Reisenberger}). But, since ``side'' loops are neglected in our discussion (the expressions for their couplings depend on the specific number of intertwiners in the entire spin network), we observe no such behavior in our graph-preserving plots. 

The volume expectation value, as a function of the (perturbative) lapse parameter for two fiducial NLSNs, is shown in Fig.~\ref{plot1} for the complete (up to 3rd order) and link-excluded (up to 4th order) cases, respectively. Note that, in the last case, since the link of spin $j_4$ is effectively disregarded in the transformation, inner loops can only appear in locations 1, 2 and 5 (nonetheless, the same volume operators are used as in the general graph-changing case). We choose NLSNs of the form~\eqref{4val_spinnet} with $j_1=j_2=j_3=j_4=1/2$, $\varepsilon =0$, and $i=0$ (red curves) or $i=1$ (black curves). The choice of spin assignments for the considered spin networks aims at minimizing the computational times, since higher spins also imply an increase in the time cost of computations, as shown in Table~\ref{spin_cost}.

\begin{table}[h!]
\centering
\begin{tabular}{cc}
\hline
{$\{j_1, j_2, j_3, j_4, i\}$} & {Time (seconds)} \\
\hline
$\{1/2, 1/2, 1/2, 1/2, 0\}$ & 168.8\\
$\{1/2, 1/2, 1/2, 1/2, 1\}$ & 187.5\\
$\{1, 1/2, 1/2, 1, 1/2\}$ & 668.8\\
$\{1, 1/2, 1/2, 1, 3/2\}$ & 687.9\\
$\{1/2, 1, 1/2, 1, 0\}$ & 822.4\\
$\{1/2, 1, 1/2, 1, 1\}$ & 826.0\\
$\{1, 1, 1, 1, 0\}$ & 6543.8\\
$\{1, 1, 1, 1, 1\}$ & 6744.7\\
$\{1, 1, 1, 1, 2\}$ & 6482.7\\
\hline
\end{tabular}
\caption{Computational times for the application of the Hamiltonian constraint, Eq.~\eqref{scalar_constr}, on node-like spin networks without inner loops [Eq.~\eqref{4val_spinnet} for $\varepsilon =0$, $\alpha = j_3$ and $\beta=j_2$] for several choices of link spins. As higher spins are chosen, the computational times rise considerably. We estimate that the time $T$ scales as $T\sim 2^4 j_1 j_2 j_3 j_4 \text{max}\{i\} T_0$, where $\text{max}\{i\}$ is the maximum value of the spin $i$ allowed by the Clebsch-Gordan conditions and $T_0$ is the time cost of the lowest spin choice, $\{1/2, 1/2, 1/2, 1/2, 0\}$. Times were recorded on a MacBook Pro with M1 chip.}
\label{spin_cost}
\end{table}

For comparison, we also show in Fig.~\ref{plot1} the volume for the (twisted) ladder-like spin networks transformed under graph-preserving dynamics [represented in Fig.~\ref{plot2}(c)], expanded up to 4th order in $N$. Since graph-preserving calculations depend on specificities of the choice of spin network, we provide in Fig.~\ref{plot2}(d) additional data to compare the behaviors of three spin network structures with different modular ``patches'', shown in Figs.~\ref{plot2}(a)-4(c). This additional comparison supports the choice of the spin network displayed in Fig.~\ref{plot2}(c) as a reference to study the deviations between graph-changing and graph-preserving dynamics. Indeed, the spin networks in Figs.~\ref{plot2}(b) and 4(c) seemingly have very close volume profiles, possibly indicating that other more complicated spin networks could have volumes not so far from those plotted in Fig.~\ref{plot2}(d) when graph-preserving dynamics is implemented. To that extent, we may regard this ladder-like spin network as a good representative for the study of graph-preserving dynamics. It is worth noting that the ``bubble-like'' spin networks shown in Fig.~\ref{plot2}(a) are eigenstates of the graph-preserving Hamiltonian, therefore their volume and volume variance remain equal to zero.

The results displayed in the figures indicate that the graph-preserving approximation leads to a misestimation of the geometric observables of the system by nearly one order of magnitude at moderate values of the lapse. As discussed previously, the fact that $\langle  \hat{C}^m_{\tilde{s}}\hat{V}\rangle$ can be different from zero for $m$ odd in the graph-preserving scenario (assuming one allows for all possible loop-coupling locations) also shows that the dynamics of the constraint is affected by this approximation. Although computational-time limitations have prevented us from completing the graph-changing calculations at 4th order, the results for graph-changing dynamics of NLSNs without acting on one of the links indicate that the volume tends to increase for $N\gsim 3/4$ under graph-changing dynamics, even somewhat higher than the volume increase for $N\gsim 1/2$ observed under graph-preserving dynamics \footnote{Calculations for such values of the lapse, already close to the unit, might be questionable in a perturbative study. Nonetheless, our results about the relative values of the 2nd-order and 4th-order contributions provide support to their validity, or at least are not in conflict with it.}. The behaviors of two (out of the three) investigated spin network graphs for the graph-preserving dynamics are qualitatively similar to the 4th-order graph-changing case.  
For the considered NLSNs, the 4th-order contributions to the volume under graph-changing dynamics only become comparable to the 2nd-order ones at $N\sim 10/9$, while this happens at $N\sim 4/5$ in the graph-preserving case. This fact supports the idea that the graph-changing perturbative results are more reliable than the graph-preserving ones. Furthermore, within the range of positivity of the variance, which provides an estimate of the maximum value of the lapse for which perturbation theory is acceptable, the two NLSNs transformed under graph-changing dynamics have the same volume profile, in contrast to what we see in the graph-preserving case. 

Although consideration of only two spin networks does not provide a proof that the graph-preserving approximation leads to a considerable departure from the graph-changing dynamics, the presented numerical data serves as the first evidence that this might indeed be the case. Further scrutinization of the different behavior of the volume expectation value between the graph-changing and graph-preserving scenarios is still needed, as well as consideration of other figures of merit beyond the volume operator. We note that, although the spin networks considered in the analysis of the volume dynamics are not solutions to the Hamiltonian, they can describe the gravitational part of physical states in the presence of a suitable scalar field or nonrotational dust serving as clock~\cite{Ilkka_GC, dust}. 
In this context, the different volume profiles in graph-changing and graph-preserving approaches can have great influence, for instance in cosmology~\cite{ashtekar_review, Ashtekar_cosmo,hybrid}, leading to different expansion rates or distinct behaviors, compatible or not with an inflationary regime. Similarly, for black hole geometries, they can modify the dynamical properties of a black-to-white hole transition and its characteristic time~\cite{WH1,WH2,WH3}.
Moreover, even though we have employed the Euclidean constraint, this Hamiltonian is known to become proportional to the Lorentzian one in flat cosmological scenarios~\cite{ABL}, and one could then expect that the perturbative evolution that we have considered can shed light on some distinctive dynamical features of these cosmological systems.

\begin{table*}[h!t!]
\begin{tabular}{lc@{\hskip 2mm}c@{\hskip 2.5mm}c@{\hskip 2mm}ccc}
\hline
& & \multicolumn{5}{c}{{Time (seconds)}} \\
\cline{3-7}
Functional & $\{j_1, j_2, j_3, j_4, i\}$ & 1 & $C_s$ & $C^2_s$  & $C^3_s$ & $C^4_s$ \\
\hline
   \multirow{4}{*}{Hamiltonian}  & $\{1/2, 1/2, 1/2, 1/2,0\}|_\text{gc}$  &  NA  &  168.8  & 10058.5  &  $10^6$ &  NA  \\
    & $\{1/2, 1/2, 1/2, 1/2,1\}|_\text{gc}$  & NA &  187.5  & 10438.1 & $10^6$ &  NA \\
    & $\{1/2, 1/2, 1/2, 1/2,0\}|^{e}_\text{gc}$  &  NA  &  44.8  & 1048.1  &  9793.3 &  85985.6  \\
    & $\{1/2, 1/2, 1/2, 1/2,1\}|^e_\text{gc}$  & NA &  42.0  & 1054.0 & 10230.4 &  85636.3 \\
     & $\{1/2, 1/2, 1/2, 1/2,0\}|^\circ_\text{gp}$  &  NA  & 70.1   &  1057.7 & 7282.4   &  21233.1 \\
      &  $\{1/2, 1/2, 1/2, 1/2,1\}|^\circ_\text{gp}$ &  NA  &  71.2  & 1054.0  &  6141.1 &  21196.5  \\
\hline
\end{tabular}
\caption{Computational times for the application of several functionals at different perturbative levels on node-like spin networks (NLSNs) given by Eq.~\eqref{4val_spinnet} for $\varepsilon =0$, $\alpha =\beta=j_1=j_2=j_3=j_4=1/2$ and $i=0,1$. Subscripts gc and gp denote NLSNs transformed under graph-changing or graph-preserving constraints, respectively. Superscripts $e$ refer to the exclusion of one link from the Hamiltonian action, while $\circ, \Delta, \square$ represent the spin network structures depicted respectively in Figs~\ref{plot2}(a), 4(b) and 4(c). The number of recursive applications of the Hamiltonian constraint is labeled by the exponent $n$ in $C^n_s$, with the unit $1$ representing no application. In the Hamiltonian rows, the entries correspond to times consumed when applying the constraint the n-th time. The other considered functionals are generally applied afterwards. However, in the case of the normalizer and volume functionals, they are also applied on the initial spin networks. Therefore, they possess entries at the column labeled by 1, corresponding to the level prior to the first application of the Hamiltonian. Normalization times marked with an asterisk were recorded without using the collector functional before, since for very large linear combinations of ghost functions the collector offers no time advantage relative to a direct application of the normalizer. Not every functional needs to be applied to every output NLSN superpostion to calculate the volume expectation value, therefore some entries are marked as ``non applicable'' (NA). Times were recorded on a MacBook Pro with M1 chip.}
\label{func_cost}
\end{table*}
\clearpage

\begin{figure}[t!h]\includegraphics[width=0.48\textwidth]{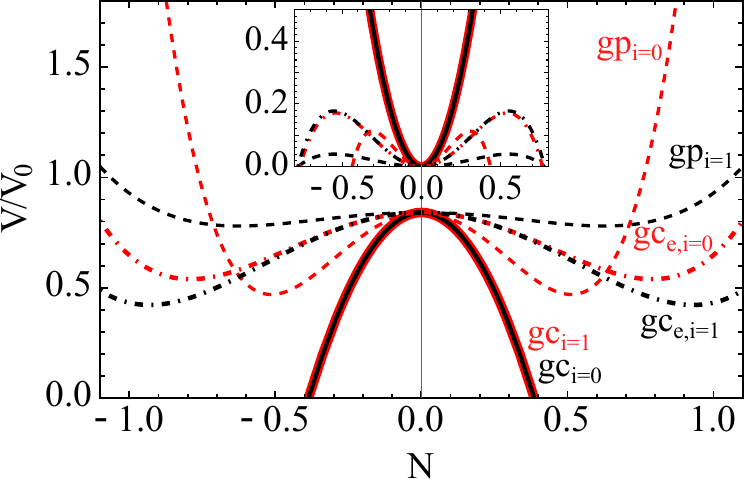}
	\caption{Variation of the dimensionless volume expectation value with respect to the lapse under different scenarios. The curves correspond to two fiducial spin networks with $j_1 = j_2 = j_3 = j_4 =1/2$, $\varepsilon = 0$ and $i=0$ (red) or $i=1$ (black). We compare graph-changing dynamics with (dot-dashed lines) and without (solid lines) allowing the Hamiltonian to act on the link along $p_4$, as well as a graph-preserving transformation of the spin network shown in Fig.~\ref{plot2}(c) (dashed lines). Unitary transformations are expanded up to 3rd order in $N$ for the full graph-changing case, and 4th order otherwise. Note that, as expected from our discussion in the main text, the 3rd-order contributions to the volume vanish. It is possible to see that the graph-preserving dynamics misestimates the volume expectation value both in absolute value and in the location of its minima. The graph-preserving dynamics also violates the equality between volume profiles observed for the two choices of inner virtual spins, $i=0$ and $i=1$. The input spin networks employed in the calculations are normalized. The inset gives the corresponding curves for the variance, $(\langle \hat{V}^2\rangle -\langle \hat{V}\rangle^2)/ V^2_0$. Recall that $V_0$ is the global constant factor introduced in our definition of the volume operator}.
	\label{plot1}
\end{figure}

\begin{figure}[t!h]\includegraphics[width=0.48\textwidth]{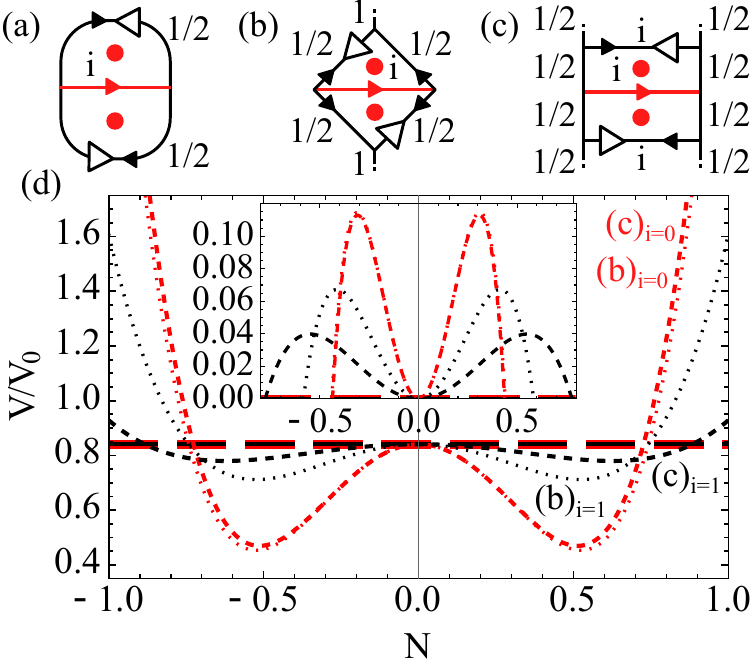}
	\caption{(a)-(c) Three different choices of spin networks, selected for the study of graph-preserving dynamics. Structures (b) and (c) are modular, as implied by the dotted links above and below. The red dots mark locations where loops can be coupled (note the absence of loops coupled from the sides of the spin networks, even though this is technically possible). The loops extend along
the entire perimeter of the regions marked by these red dots. The red link represents the sole intertwiner acted upon by the graph-preserving Hamiltonian, and its spin takes values $i=0$ or $i=1$, each corresponding to red or black volume profiles in (d), respectively. (d) Variation of the dimensionless volume expectation value with respect to the lapse $N$ for the three different choices of spin networks. The unitary transformation is expanded up to 4th order in $N$. Volumes of spin networks (a), (b) and (c) are represented by long-dashed, dotted and dashed lines, respectively. Note the similarity of the results for spin networks (b) and (c), particularly when $i=0$. The inset shows the corresponding curves for the variance, $(\langle \hat{V}^2\rangle -\langle \hat{V}\rangle^2)/ V^2_0$, where $V_0$ is the global constant factor introduced in our definition of the volume operator.}
	\label{plot2}
\end{figure}

\section{Conclusions and outlook}\label{conclusions}

In the first part of our work, we have made use of the modern conventions in recoupling theory to fully derive the action of the LQG Euclidean scalar constraint around $3$-valent and $4$-valent nodes of spin networks. These results represent an update, as well as an extension, of previous derivations~\cite{borissov, gaul, Ma, Alesci}. Our discussion shows how reversibility and self-adjointness can be directly visualized in the spin networks acted upon by the Hamiltonian constraint: inner loops can both be added or removed around the intertwiner. In fact, we show that, when acting on spin networks with inner loops, possible outcomes of the scalar constraint are spin networks with the same graph, but with different spin assignments on the bridging link of the inner loops. We derive this ``loop-coupling" mechanism using the tools from recoupling theory, something so far not yet presented in the literature. These calculations should serve as a reference for future studies of the full Euclidean Hamiltonian constraint in the graph-changing regime. 

We have then introduced a novel numerical approach that enables us to encode spin networks and implement the action of the Hamiltonian constraint on them. The code allows us to explore the effects of the graph-changing behavior of the scalar constraint on the expectation value of the volume operator and even compare them to the approximated, graph-preserving constraint. Our results show that the assumption that the graph-changing dynamics can be properly approximated by a graph-preserving Hamiltonian might not be firmly justified for generic spin networks states, at least if we restrict them to low spins. In addition to this analysis, we have also managed to determine with our numerical methods two families of solutions of the Euclidean Hamiltonian. It is reasonable to expect that solutions of the form~\eqref{eq_eigenstate} live in the algebraic dual of the dense set of spin network states on which we define our constraint. Their definitive physical interpretation should depend on the feasibility of introducing an inner product that endows them with a physical Hilbert-space structure. However, even though the set of such solutions is remarkably infinite (in contrast with the situation before our work), they do not seem to include yet enough degrees of freedom to capture the complete  phenomenology of general relativity (two per the discrete counterpart of a spatial point). The determination of a physical inner product seems still out of reach, and further scrutiny of the solutions and their properties is needed and planned for future works.

It is worth noting that, as we showed in our time-cost analysis, computations on a single computer are expectedly demanding and processing times increase rapidly both with the number of recursive applications of the Hamiltonian and with the spins involved. Therefore, it would be interesting to explore the potential to run these calculations in a computational cluster.

Our work is a thorough study of the graph-changing aspects of the Euclidean scalar constraint, both analytically and numerically, and introduces a new tool to further explore its action on spin networks. These contributions enable new analyses in LQG, allowing developments in areas previously assumed to be numerically unfeasible. More concretely, we expect future works on LQG to further use and build on our numerical approach and therewith extend our results to a wider domain of validity, potentially also unveiling new families of eigenstates and new phenomenology in LQG. In particular, it would be interesting to discuss how precise LQG formulations would affect relevant semi-classical results such as the black-to-white-hole tunneling~\cite{bounce}, the potential tiny-white-hole nature of dark matter~\cite{rovelli_white-hole_2018} or the black-hole halos potentially left from past universes through bounces~\cite{Cong}. Comparison of our results with compatible spin-foam numerical data might also shed light on the possible connection between canonical and covariant formalisms in LQG. Note, however, that unlike current numerical results in covariant LQG, our calculations include all possible superpositions of graphs generated by the Hamiltonian, up to a desired order in the lapse, a feature we believe is needed to fully embrace the graph-changing character of the theory. Lastly, in future studies we plan to focus our investigations on the numerical implementation of the Lorentzian constraint and on the search for its solutions.

\section*{Acknowledgments}
The authors are grateful for discussions with and suggestions by Ilkka M\"akinen, Jorge Pullin, Carlo Rovelli and Etera Livine. This work is supported by the ID\# 62312 grant from the John Templeton Foundation, as part of the project \href{https://www.templeton.org/grant/the-quantum-information-structure-ofspacetime-qiss-second-phase}``The Quantum Information Structure of Spacetime'' (QISS). 

TLMG and MM acknowledge support by the ERC Starting Grant QNets through Grant Number 804247, and the European Union’s Horizon Europe research and innovation program under Grant Agreement Number 101114305 (“MILLENION-SGA1” EU Project) and under Grant Agreement Number 101046968 (BRISQ). MM furthermore acknowledges support by the Deutsche Forschungsgemeinschaft (DFG, German Research Foundation) under Germany’s Excellence Strategy ‘Cluster of Excellence Matter and Light for Quantum Computing (ML4Q) EXC 2004/1’ 390534769. This research is also part of the Munich Quantum Valley (K-8), which is supported by the Bavarian state government with funds from the Hightech Agenda Bayern Plus.

FV's research at Western University is supported by the Canada Research Chairs Program and by the Natural Science and Engineering Council of Canada (NSERC) through the Discovery Grant ``Loop Quantum Gravity: from Computation to Phenomenology." FV acknowledges support from the Perimeter Institute for Theoretical
Physics through its affiliation program. Research
at Perimeter Institute is supported by the Government of Canada through Industry Canada and by the Province of Ontario through the Ministry
of Economic Development and Innovation.
FV acknowledges the Anishinaabek, Haudenosaunee, L\=unaap\'eewak, Attawandaron, and neutral people, on whose traditional lands Western University and the Perimeter Institute are located.

GAMM acknowledges support by MCIN/AEI/ 10.13039/\-501100011033/FEDER from Spain under the Grant Number PID2020-\-118159GB-\-C41 and also partly by the Grant Number PID2023-149018NB-C41, funded by MCIU/AEI/10.13019/501100011033 and by FSE+.

\appendix

\begin{widetext}

\section{Additional formulas from recoupling theory}\label{appendix_recoup}

One important point for the derivation of the action of the scalar constraint~\eqref{scalar_constr} on $4$-valent NLSNs is how to factor out diagonal inner links. The relation that allows this requires the use of several expressions from Sec.~\ref{recoupling_theory}. For this reason, we include its derivation here. The idea is that one uses a double braid operation (namely on the nodes of spins $\{a,j_3, 1/2\}$ and $\{j_3, j_1, p\}$) to convert the spin network into a more tractable form while picking up a phase from these braidings [cf. Eq.~\eqref{wigner3j} and the discussion thereafter]. One then uses Eq.~\eqref{wigner3j_invar} to introduce arrows on the four links connected to the intertwiner. This permits us to use Eq.~\eqref{pachner} on the links with spins $j_3$ and $q$. The double Pachner move combined with the removal of three arrows at the node $\{p, o, 1/2\}$ [which requires flipping $\epsilon^{(p)}$ and therefore contributes with a phase factor $(-1)^{(2p)}$], gives the expression in the second row of Eq.~\eqref{app1}. We are required to flip the  {cyclicity} of the node $\{p, o, 1/2\}$ to allow for the use of Eq.~\eqref{bubble}, which factors out the inner ``bubble" loop and removes a prefactor of $d_n$. Flipping both  $\epsilon^{(j_1)}$ and  $\epsilon^{(j_2)}$ leads to the expression in the third row of the equation. Finally, since $(-1)^{(2j_3)}=(-1)^{(2a+1)}$ and $(-1)^{(2n)}=(-1)^{(2j_2+2m)}$ (note the triangularity conditions imposed by the Wigner 6j symbols), we can rewrite the final expression in the form given in the last row of Eq.~\eqref{app1} (using the symmetry of the Wigner 6j symbols). 
\begin{equation}

\label{app1}
\end{equation}

Another important relation appears in the derivation of the action of the volume operator (or, more precisely, in the action of $\hat{W}$). We use Eq.~\eqref{theorem} to introduce the resolution of the identity in such a way that it isolates the spin-1 links, i.e., it ``splits" the links of spins $j_1$, $j_3$ and $m$. The factored term forms a hexagonal spin network that corresponds to the Wigner 9j symbol [cf. Eq.~\eqref{hexagon}],
\begin{equation}
%
\label{app2}
\end{equation}

We now show a simple example of how the arrows within a tetrahedral spin network of the form~\eqref{tetrahedron} can be rearranged using Eq.~\eqref{wigner3j_invar} to give similar structures, all of which correspond to the Wigner 6j symbol up to varying phase factors. In the concrete example below, we introduce three arrows on the node joining the spins $\{k_3, k_1, j_3\}$, and use Eqs.~\eqref{levi2} and \eqref{levi3} to remove doubled arrows, obtaining a phase, 
\begin{equation}
%
 
\label{app3}
\end{equation}
Expressions like the one on the right-hand side are often factored out in the equations of Secs.~\ref{sec_3_val} and \ref{sec_4_val}.

Finally, the 2-2 Pachner move can be derived by means of a braid, Eq.~\eqref{prepachner} and another braid, in the following sequence:
\begin{equation}
%
 
\label{app4}
\end{equation}
Note that, in the last row of this equation, the intertwiner has been rotated clockwise by $\pi/2$. Furthermore, the (Clebsch-Gordan) relation $(-1)^{2l+2j_2+2j_4}=1$ has been employed in the phase factor.

\end{widetext}

\section{Temperley-Lieb algebra}\label{Temperley-Lieb}

Instead of using the modern orthonormalized spin networks as quantum states in our description, we will use here the old-fashioned, yet more graphically intuitive description of such systems in terms of Temperley-Lieb tangles~\cite{temperley}. We introduce below the main working tools from recoupling theory  with Temperley-Lieb algebra, that will be needed to follow the following derivations (note the difference from what was introduced in {Sec.}~\ref{recoupling_theory}),

\begin{equation}
 
\label{recoup8}
\end{equation}

The labels {($a,b,c,d,i,j,m,n,r,s,t$)} used in Eqs. \eqref{recoup1}-\eqref{recoup8} are ``colors'', corresponding to twice the spins. Equation~\eqref{recoup1} represents a so-called 2-2 Pachner move~\cite{temperley}. The coefficient in the summand of Eq. \eqref{recoup1}, related through Eq.~\eqref{recoup2} to other commonly occurring symbols in recoupling theory, is  the 6j symbol in the Temperley-Lieb normalization. The tetrahedral net symbol with inputs ($a,b,c,d,i,j$) on the (numerator on the) right-hand side of Eq. \eqref{recoup2} will regularly appear throughout the following calculations. Its formula~\footnote{It is worth noting that the expression given in Ref.~\cite{temperley} fails to retrieve the correct answer, which is zero, in certain specific cases with violation of triangularity, such as, for example, when the central color [$j$ in Eq.~\eqref{recoup2}] alone violates it.} can be found in Sec. 9.11 of Ref.~\cite{temperley}, but we will often convert it to the most widely used 6j symbol with spin entries rather than colors~\footnote{Note that the original expression of the formula in Sec. 8.5 of Ref.~\cite{temperley}, also reproduced in Refs.~\cite{rovelli2004book, borissov, gaul}, has a typo (a minus sign turned into a plus) relative to the correct derivation presented in Sec. 9.11 of Ref.~\cite{temperley}, also reproduced in Refs.~\cite{rovelli2004book, borissov, gaul}. Furthermore, the derivation only holds when the triangularity condition is fulfilled, therefore numerical evaluations using this formula require the additional inclusion of triangularity-based selection rules.}, often encountered in mathematical softwares [e.g., the "SixJSymbol" function in Mathematica] and represented here in parentheses,
    \begin{equation}
\begin{tikzpicture}[baseline=(current  bounding  box.center)];
\node at (-0.7,0) {$ \begin{pmatrix}
a/2 & b/2 & i/2\\
c/2 & d/2 & j/2
\end{pmatrix} =$};
\draw (1,0)--(5,0);
\draw (1.4+0.2,-0.8) ellipse (0.4 and 0.3);
\draw (1+0.2,-0.8)--(1.8+0.2,-0.8);
\draw (2.4+0.2,-0.8)  ellipse (0.4 and 0.3);
\draw (2+0.2,-0.8)-- (2.8+0.2,-0.8);
\node at (2.4+0.2,-1.3) {$i$};
\node at (1.4+0.2,-1.3) {$i$};
\node at (2.4+0.2,-0.25) {$b$};
\node at (1.4+0.2,-0.25) {$a$};
\node at (1.4+0.2,-0.65) {$d$};
\node at (2.4+0.2,-0.65) {$c$};
\draw (3.4+0.2,-0.8) ellipse (0.4 and 0.3);
\draw (3+0.2,-0.8)--(3.8+0.2,-0.8);
\draw (4.4+0.2,-0.8)  ellipse (0.4 and 0.3);
\draw (4+0.2,-0.8)-- (4.8+0.2,-0.8);
\node at (4.4+0.2,-1.3) {$j$};
\node at (3.4+0.2,-1.3) {$j$};
\node at (4.4+0.2,-0.25) {$d$};
\node at (3.4+0.2,-0.25) {$a$};
\node at (3.4+0.2,-0.65) {$b$};
\node at (4.4+0.2,-0.65) {$c$};
\draw (2.2+0.4,0.2) -- (1.75+0.4,0.65) -- (2.2+0.4, 1.1)-- (2.65+0.4,0.65)--cycle;
\draw (1.75+0.4,0.65) -- node[above] {$j$} (2.65+0.4,0.65);
\draw (2.2+0.4,0.2) -- (3+0.4,0.2)-- node[right] {$i$} (3+0.4,1.1)-- (2.2+0.4, 1.1);
\node at (1.8+0.4,0.35) {$a$};
\node at (1.8+0.4,0.95) {$b$};
\node at (2.65+0.4,0.4) {$d$};
\node at (2.65+0.4,0.95) {$c$};
\draw (0.95,-0.15)-- (1.05,-1.0)--(1.2,-0.1)--(4.9,-0.1);
\end{tikzpicture}.
\label{recoup9}
\end{equation}
    
It is worth noting that the tetrahedral net  symbol is invariant with respect to the following permutations of arguments: $(a,b,i,c,d,j)$, $(b,a,i,d,c,j)$, $(a,i,b,c,j,d)$, $(a,d,j,c,b,i)$, $(c,d,i,a,b,j)$ and $(c,b,j,a,d,i)$. Another important property of this function is its {triangularity}, i.e., it only assumes nonzero values if the triples $(a,b,i)$, $(i,c,d)$, $(d, j, a)$ and $(c,b,j)$ simultaneously fulfil the triangle/Clebsch-Gordan conditions for all (permutations of) its entries. Equation \eqref{recoup5} is a symbolic representation of the Clebsch-Gordan spin coupling, with the colors $n$ and $m$ summing up to all allowed values $i$ such that $|m-n|\leq i \leq m+n$, with the additional (gauge-invariance) constraint $m+n+i = 2k \;(k\in \mathbb{N})$, referred to as the Clebsch-Gordan or triangle conditions. One interesting aspect of the Temperley-Lieb algebra is the fact that the geometric arrangement of colors in Eq.~\eqref{recoup9} differs from that in Eq.~\eqref{tetrahedron}, resulting in different predictions for the two approaches considered. The single loop in Eq.~\eqref{recoup3} represents, up to a possible $-1$ factor, the dimension of the color-$i$ representation (i.e., for $i=2j$, $d=2j+1$) and results from summing over all tangle permutations (the permutation is represented by a white square). The remaining equations are mostly used for the purpose of renormalization of virtual edges (edges added to the spin network through manipulations). Making use of those equations we will proceed with the derivation of the action of the scalar constraint on general spin networks. 

We start with a generic collection of three linearly independent edges attached to a common vertex of valence 3. We label their colors as $r$, $p$ and $q$. Following {Refs.}~\cite{borissov, gaul}, both the edges and the paths of the holonomies in Eq.~\eqref{scalar_constr} are oriented towards the vertex (inverse holonomies are therefore associated with segments oriented away from the vertex). Whenever necessary, the orientation of edges and holonomies will be indicated by an arrow. The orientation is important, since the consecutive application of the holonomies in Eq.~\eqref{scalar_constr} should follow a cyclic orientation closed by the trace (i.e., two holonomies connected by a virtual 2-valent vertex should not be simultaneously oriented towards this vertex). 

We proceed with the application of the first holonomy of the first term on the right-hand side of Eq.~\eqref{scalar_constr} to the three fiducial edges from the same vertex. At first, we consider the action of $\hat{h}^{-1}[p_k]$ only along the path $p_r$ parallel to the edge labelled by $r$, i.e.,
\begin{equation}
\begin{tikzpicture}[baseline=(current  bounding  box.center)];
\node at (-0.45,0.1) {$\hat{h}^{-1}[p_r]$};
\node at (1.6,0) {$\equiv$};
\draw (0+1.8,-0.7)  --(0.7+1.8,0.1)--(0.7+1.8,0.7);
\draw (0.7+1.8,0.1)--(1.4+1.8,-0.7);
\draw[thick,->] (0.7+1.9,0.1)--node[right] {$1$}(0.7+1.9,0.7);
\node at (1.85,-0.4) {$p$};
\node at (1.85+1.4,-0.4) {$q$};
\node at (2.35,0.75) {$r$};
\node at (3.5,0) {$=\sum_c$};
\draw (0,-0.7)  --(0.7,0.1)--(0.7,0.7);
\draw (0.7,0.1)--(1.4,-0.7);
\node at (0,-0.4) {$p$};
\node at (0.5,0.6) {$r$};
\node at (1.4,-0.4) {$q$};
\draw[rounded corners] (2+2.3, 0.2) rectangle  (2.4+2.3, 0.5) node[above left] {$c$} {};
\filldraw[fill=white, draw=black] (2.15+2.3,0.2-0.05) rectangle (2.25+2.3,0.25);
\draw (1.7+2.3,0)--(2.65+2.3,0);
\draw (2.2+2.3,-0.8) ellipse (0.4 and 0.3);
\draw (1.8+2.3,-0.8)--(2.6+2.3,-0.8);
\node at (2.2+2.3,-1.3) {$c$};
\node at (2.2+2.3,-0.25) {$1$};
\node at (2.2+2.3,-0.65) {$r$};
\draw (0+5.2,-0.7)  --(0.7+5.2,-0.1)--(0.7+5.2,0.7);
\draw (0.7+5.2,-0.1)--(1.4+5.2,-0.7);
\draw[<-] (0.7+5.2,0.15)--(6.2,0.1);
\draw[->] (0.7+5.2,0.55)--(6.2,0.6);
\node at (0+5.2,-0.4) {$p$};
\node at (0.5+5.2,0.7) {$r$};
\node at (0.5+5.2,0.37) {$c$};
\node at (0.5+5.2,0) {$r$};
\node at (1.4+5.2,-0.4) {$q$};
\node at (0.5+5.85,0.65) {$1$};
\node at (0.5+5.85,0.1) {$1$};
\end{tikzpicture} .
\label{Hstep1}
\end{equation}
The action of the holonomies $\hat{h}^{-1}[p_p]$ and $\hat{h}^{-1}[p_q]$ along the edges $p$ and $q$ follow analogous relations with permuted labels. Note that the holonomy $\hat{h}^{-1}[p_r]$ is represented as an arrow of color 1 [the fundamental representation of the SU(2) group] with orientation opposite to its adjacent edge, with label given by $r$. Using Eq.~\eqref{recoup5}, the two parallel segments with labels $r$ and $1$ (i.e., the edge and the holonomy) can be coupled, with their combined colors/spins assuming all values $c$ allowed by the Clebsch-Gordan conditions, namely $c=r+\epsilon$ with $\epsilon=\pm 1$. The use of Eq.~\eqref{recoup5} automatically results in a trivalent decomposition, with the small $r$-colored segment attached to the original vertex being actually virtual (i.e., it has no physical extension in the manifold), so that this vertex becomes effectively 4-valent with edges $p$, $q$, $c$ and 1. The increased valence of the original vertex, however, is not permanent, since the new inwards-oriented edge of color 1 is supposed to be tied to the other holonomies in Eq.~\eqref{scalar_constr}, in a similar way to how the indices of a product of matrices have to be contracted pairwise (and therefore no free indices are left after a trace is applied). The same holds for the oriented edge of color 1 created on the upper-most virtual vertex. 

The considered spin networks, and therefore also their corresponding tangles, are eigenstates of the volume operator. This operator acts on the (physical) vertices of the graph, giving zero contribution from vertices with valence below 4, while higher-valence vertices have a contribution defined by the colors of the edges attached to them. In practice, $\hat{V}\equiv \frac{l^3_0}{4}\sqrt{|i\hat{W}^{(4)}_{[p,q,c]}|}$ is defined in terms of the Planck length $l_0$ and the operator $\hat{W}^{(4)}_{[p,q,c]}$. Applying $\hat{W}^{(4)}_{[p,q,c]}$ on the right-hand side of Eq.~\eqref{Hstep1}, which contains an effective 4-valent vertex with edges of colors $p$, $q$, $c$ and 1 decomposed in a trivalent arrangement, leads to
\begin{equation}
\begin{tikzpicture}[baseline=(current  bounding  box.center)];
\node at (-0.6,0.1) {$\hat{W}^{(4)}_{[p,q,c]}$};
\node at (0.5+0.75,0.65) {$1$};
\node at (0.5+0.75,0.1) {$1$};
\node at (0.5,0.37) {$c$};
\node at (0.5,0) {$r$};
\draw[<-] (0.7,0.15)--(1.1,0.1);
\draw[->] (0.7,0.55)--(1.1,0.6);
\node at (3.3,0) {$=\sum_\beta {W}^{(4)}_{[p,q,c]}(p,q,1,c)^\beta_r$};
\draw (0,-0.7)  --(0.7,-0.1)--(0.7,0.7);
\draw (0.7,-0.1)--(1.4,-0.7);
\node at (0,-0.4) {$p$};
\node at (0.5,0.7) {$r$};
\node at (1.4,-0.4) {$q$};
\draw (0+5.2,-0.7)  --(0.7+5.2,-0.1)--(0.7+5.2,0.7);
\draw (0.7+5.2,-0.1)--(1.4+5.2,-0.7);
\draw[<-] (0.7+5.2,0.15)--(6.2,0.1);
\draw[->] (0.7+5.2,0.55)--(6.2,0.6);
\node at (0+5.2,-0.4) {$p$};
\node at (0.5+5.2,0.7) {$r$};
\node at (0.5+5.2,0.37) {$c$};
\node at (0.5+5.2,0) {$\beta$};
\node at (1.4+5.2,-0.4) {$q$};
\node at (0.5+5.85,0.65) {$1$};
\node at (0.5+5.85,0.1) {$1$};
\end{tikzpicture} .
\label{Hstep2}
\end{equation}
The matrix ${W}^{(4)}_{[p,q,c]}(p,q,1,c)^\beta_r$ in Eq.~\eqref{Hstep2} is skew-symmetric and, with the exception of two entries, is composed of zeros. The two nonzero entries depend on the value of $c$: if $c=r+1$, the entries with row $r-2$ and column $r$, and vice versa, are nonzero, while if $c=r-1$, the entries with row $r$ and column $r+2$, and vice versa, are nonzero instead~\cite{borissov, gaul, volume}. These matrix elements read
\begin{widetext}
\begin{equation}
\begin{aligned}
{W}^{(4)}_{[p,q,c]}(p,q,1,c)^{r\pm 2}_r =& \pm (-1)^{(p+q+r+1\pm 1)/2}\\
{\times}&\left[ \frac{1}{2^8}(p+q+r\pm 1+3)(1+p+q-r\mp 1)(1+p+r\pm 1 -q)(1+q+r\pm 1 -p)\right]^{1/2}.
\end{aligned}
\end{equation}
\end{widetext}
While ${W}^{(4)}_{[p,q,c]}(p,q,1,c)^\beta_\alpha$ is not diagonal, it can be easily diagonalized by a unitary matrix $U$. The two eigenvalues of ${W}^{(4, \text{diag})}_{[p,q,c]}(p,q,1,c)=U{W}^{(4)}_{[p,q,c]}(p,q,1,c) U^\dag$ are $ {W}^{(4)}_{[p,q,c]}(p,q,1,c)^{r\pm 2}_r$ and $-{W}^{(4)}_{[p,q,c]}(p,q,1,c)^{r\pm 2}_r$. The square root of the matrix $i {W}^{(4)}_{[p,q,c]}(p,q,1,c)$ can be expanded to show that 
\begin{eqnarray}
&&\sqrt{|i {W}^{(4)}_{[p,q,c]}(p,q,1,c)|}\nonumber \\&=&U^\dag \sqrt{|i {W}^{(4, \text{diag})}_{[p,q,c]}(p,q,1,c)|}U \nonumber \\
&=&U^\dag \mathbb{1} \sqrt{|{W}^{(4)}_{[p,q,c]}(p,q,1,c)^{r\pm 2}_r |} U \nonumber \\
&=& \mathbb{1} \sqrt{|{W}^{(4)}_{[p,q,c]}(p,q,1,c)^{r\pm 2}_r |}.
\end{eqnarray} 
The matrix representation of the volume operator is therefore diagonal. 

The next operator on the right-hand side of Eq.~\eqref{scalar_constr} is the holonomy $\hat{h}[p_k]$. For the specific case of the path $p_r$ along the $r$-colored edge, $\hat{h}[p_r]$ is graphically represented by an arrow parallel, but oppositely oriented, to the arrow representing $\hat{h}^{-1}[p_r]$ in Eq.~\eqref{Hstep1}. $\hat{h}[p_r]$ can be directly attached~\footnote{By convention, the attachment of a holonomy is always performed from the right side of an edge. This ordering will be important later, when Eq.~\eqref{recoup8} comes into play. If one consistently attaches holonomies from the same side, however, it does not matter which side is chosen as convention.} to the loose end of the upper edge of color 1 on the right-hand side of Eq.~\eqref{Hstep1} to give
\begin{equation}
\begin{tikzpicture}[baseline=(current  bounding  box.center)];
\node at (-0.4,0.1) {$\hat{h}(p_{r})$};
\node at (0.5+0.75,0.65) {$1$};
\node at (0.5+0.75,0.1) {$1$};
\node at (0.5,0.37) {$c$};
\node at (0.5,0) {$r$};
\draw[<-] (0.7,0.15)--(1.1,0.1);
\draw[->] (0.7,0.55)--(1.1,0.6);
\node at (1.8,0) {$\equiv$};
\draw (0,-0.7)  --(0.7,-0.1)--(0.7,0.7);
\draw (0.7,-0.1)--(1.4,-0.7);
\draw[<-] (2.8,0.15)--(3.2,0);
\draw[->] (2.8,0.55)--(3.2,0.6);
\draw[thick, ->] (2.7,0.7)--(2.7,0.1);
\node at (0.5+2.8,0.65) {$1$};
\node at (0.5+2.8,0) {$1$};
\node at (+2.55,0.4) {$1$};
\node at (+2.6,-0.05) {$r$};
\node at (+2.9,0.8) {$r$};
\node at (+2.96,0.35) {$c$};
\node at (0+2.1,-0.4) {$p$};
\node at (1.4+2.1,-0.4) {$q$};
\node at (0,-0.4) {$p$};
\node at (0.5,0.7) {$r$};
\node at (1.4,-0.4) {$q$};
\node at (3.9,0) {=};
\draw (2.1,-0.7)  --(2.8,-0.1)--(2.8,0.7);
\draw (2.8,-0.1)--(3.5,-0.7);
\draw (0+4.3,-0.7)  --(0.7+4.3,-0.1)--(0.7+4.3,0.7);
\draw (0.7+4.3,-0.1)--(1.4+4.3,-0.7);
\draw[<-] (0.7+4.3,0.15)--(5.3,0);
\draw[->] (0.7+4.3,0.55)--(5.15,0.6)--(5.15,0.2);
\node at (0+4.3,-0.4) {$p$};
\node at (0.5+4.3,0.7) {$r$};
\node at (0.5+4.3,0.37) {$c$};
\node at (0.5+4.3,0) {$r$};
\node at (1.4+4.3,-0.4) {$q$};
\node at (0.5+4.8,0.6) {$1$};
\node at (0.5+4.95,0) {$1$};
\end{tikzpicture} .
\label{Hstep3}
\end{equation}

The holonomies over the triangular loop, $\hat{h}[\alpha_{ij}]-\hat{h}[\alpha_{ji}]$, should be applied on the right-hand side of Eq.~\eqref{Hstep3} in such a way that the orientation of the sequentially coupled holonomies is preserved. Since $\alpha_{ij}$ and $\alpha_{ij}$ have opposite orientations, they are attached to the loose ends of the two virtual edges of color 1 (which are physically at the same point of the manifold) in different ways. The presence of the trace in Eq.~\eqref{scalar_constr} enforces that all virtual edges should be tied together, such that no loose virtual edges remain. As a result,
\begin{equation}
 .
\label{Hstep4}
\end{equation}
The effect of coupling holonomies (with color 1) to the edges of colors $p$ and $q$ can be accounted for by applying Eq.~\eqref{recoup5} to each of those edges. For the first subgraph (i.e., a region of a spin network around one of its vertices) on the right-hand side of Eq.~\eqref{Hstep4} this leads to
\begin{equation}
%
 ,
\label{Hstep5}
\end{equation}
where $a=p+\epsilon'$ and $b=q+\epsilon''$ (with $\epsilon',\epsilon''=\pm 1$ according to the Clebsch-Gordan conditions) and all virtual edges originating from holonomies have color 1 (this also applies for the following derivations). We can now use Eq.~\eqref{recoup8} to rearrange the crossings of edges~\footnote{Note that $\lambda^{a,b}_{c}=\lambda^{a,c}_{b}=\lambda^{b,c}_{a}${, because} $(-1)^{-1}=-1$.},
\begin{equation}
%
 .
\label{Hstep6}
\end{equation}
This operation is performed three times, once for each of the original edges that we are considering. 
The virtual loops that appear on the right-hand side of Eq.~\eqref{Hstep6} can be renormalized with the aid of  Eq.~\eqref{recoup6} to give 
\begin{equation}
%
 .
\label{Hstep7}
\end{equation}
Similar operations can be performed on the second graph on the right-hand side of Eq.~\eqref{Hstep4}.

After taking into account that the trace gives an additional $-1$ prefactor, Eqs.~\eqref{Hstep1}-\eqref{Hstep6} synthesize the action of the first term of the Hamiltonian~\eqref{scalar_constr} on the fiducial three edges attached to the same vertex. The second term of the Hamiltonian, arising from the alternate order of operators in the anticommutator, has a similar effect, but since the holonomies $\hat{h}[\alpha_{ij}]-\hat{h}[\alpha_{ji}]$ are applied before the volume operator, the latter has instead the matrix elements 
$$\frac{l^3_0}{4}\sqrt{|{W}^{(4)}_{[p+\epsilon',q+\epsilon'',c]}(p+\epsilon',q+\epsilon'',1,c)^{r\pm 2}_r |}.$$ Calling 
\begin{eqnarray}
K^{[a,b,c]}_{[p,q,c]}&=&\sqrt{|{W}^{(4)}_{[p,q,c]}(p,q,1,c)^{r\pm 2}_r |} \nonumber\\ &+&\sqrt{|{W}^{(4)}_{[a,b,c]}(a,b,1,c)^{r\pm 2}_r |}, 
\end{eqnarray} 
the action of the Hamiltonian~\eqref{scalar_constr} can thus be written as
\\

\begin{widetext}
\begin{equation}
 
\label{Hstep8}
\end{equation}

Since the Hamiltonian is Hermitian, if it takes an input spin network $|A\rangle$ into an output spin network $|B\rangle$, it should take $|B\rangle$ into $|A\rangle$ with equal probability, i.e., $\langle A| \hat{C}_s|B\rangle = \langle B| \hat{C}_s|A\rangle^*$. We apply our constraint operator on the first subgraph on the right-hand side of Eq.~\eqref{Hstep8} to check this condition. Similar arguments to those presented in the derivation of Eqs.~\eqref{Hstep1}-\eqref{Hstep8} show that one of the terms resulting from the application of Eq.~\eqref{scalar_constr} on this subgraph introduces a triangular loop inside the previously added loop, i.e.
\begin{equation}
 ,
\label{Hstep9}
\end{equation}
\end{widetext}
where the sum over $\xi', \xi'', \epsilon$ (with $d=a+\xi'$, $c=r+\epsilon$, $e=b+\xi''$ and $\xi',\xi'',\epsilon=\pm 1$), as well as the factor $\lambda^{1,r}_{c}\lambda^{1,a}_{d}\lambda^{1,b}_{e}K^{[d,e,c]}_{[a,b,c]}$, have been omitted. The two previously inexistent edges of color 1 in the subgraph on the right-hand side of Eq.~\eqref{Hstep9} can be combined with the help of relation~\eqref{recoup5}, which merges the two into a single edge that can take the colors $0$ or $2$,
\begin{equation}
%
 .
\label{Hstep10}
\end{equation}

When the merging of the edges of color 1 leads to an edge of color 2, its attachments to the two previously existent edges can be renormalized by means of Eq.~\eqref{recoup6} to remove the virtual triangles. When this merging process results in no edge ($l=0$), the coefficient on the right-hand side in Eq.~\eqref{Hstep10} becomes $-1/2$ and the remaining ``loop'' in two of the edges can be removed with the aid of Eq.~\eqref{recoup7}, recovering the original subgraph (with no added triangular loops) with an overall coefficient $\lambda^{1,r}_{c}\lambda^{1,a}_{p}\lambda^{1,b}_{q}K^{[p,q,c]}_{[a,b,c]}/2 (=\lambda^{1,c}_{c}\lambda^{1,p}_{a}\lambda^{1,q}_{b}K^{[a,b,c]}_{[p,q,c]}/2)$ multiplied by
\begin{widetext}
\begin{equation}
 
\label{Hstep11}
\end{equation}
\end{widetext}
This differs from the first coefficient in Eq.~\eqref{Hstep8} by
\begin{equation}
%
 .
\label{Hstep12}
\end{equation}
Equation~\eqref{Hstep12} seems in conflict with the self-adjointness of the Hamiltonian. This illusory tension is a by-product of the usage of Temperley-Lieb tangles rather than normalized spin networks. As show in Ref.~\cite{volume} (cf. {Sec.} VIII), tangles have to be normalized by the square root of the product of loops of the form~\eqref{recoup3}, one for each edge of the tangle, divided by the product of theta symbols of the form~\eqref{recoup4}, one for each vertex of the tangle (the indices of the symbols are the colors of the corresponding edges and vertices), i.e.
\begin{equation}
    |\text{spin network}\,\{j_i\}\rangle = \sqrt{\prod_v\prod_e {\frac{\Delta_e} { \theta_v} } }  | \text{tangle} \,\{2j_i\}\rangle ,
    \label{normalize_tangle}
\end{equation}
where $\Delta_e$ is given by Eq.~\eqref{recoup3} with the same color as edge $e$, $\theta_v$ by Eq.\eqref{recoup4} for the vertex $v$ (the arguments being the three colors of the edges connected to that vertex) and products run over edges $e$ and vertices $v$. Note that the orthonormalized spin network state in Eq.~\eqref{normalize_tangle} has a set $\{j_i\}$ of spins attached to its vertices, while the tangle has a corresponding set of colors $\{2j_i\}$. Hence, if the subgraph on the left-hand side of Eq.~\eqref{Hstep9} is associated with the spin network $|A\rangle$ and the subgraph on the left-hand side of Eq.~\eqref{Hstep8} (before the application of the Hamiltonian) is associated with $|B\rangle$ (the rest of the two graphs being identical), the ratio $\langle A|A\rangle /\langle B|B\rangle$ is exactly given by Eq.~\eqref{Hstep12} (up to a minus sign). Therefore $\langle A| \hat{C}_s|B\rangle /\langle B| \hat{C}_s|A\rangle =-1$ and the scalar Hamiltonian constraint is self-adjoint.

Notwithstanding the apparently cumbersome form of Eq.~\eqref{Hstep8}, a few properties can be extracted from its coefficients. The tetrahedral net symbol, for example, can be recast using Eq.~\eqref{recoup9} as a spin-normalized 6j symbol, with well known properties. It is therefore clear that, if $p=q$ and $a=b$, the pairs of tetrahedral net symbols (related by $p \leftrightarrow q $, $a \leftrightarrow b$ up to symmetry) will cancel each other in the first coefficient of the right-hand side of Eq.~\eqref{Hstep8}. Similarly, its second  coefficient vanishes when $r=p$ and $c=a$ (owing to the swap $p \leftrightarrow r $, $a \leftrightarrow c$ between pairs of symbols), while the third coefficient vanishes for $r=q$ and $c=b$ (owing to the argument swap $q \leftrightarrow r $, $b \leftrightarrow c$).

Triangularity in Eq.~\eqref{recoup9} requires that $(a,1,c)$, $(c,r,q)$, $(q, p, a)$, $(r,1,p)$, $(a,c,r)$, $(r,1,b)$, $(b, q, a)$ and $(1,c,q)$ all fulfill the triangular condition in order for the first pair of tetrahedral net symbols on the right-hand side of Eq.~\eqref{Hstep9} to be nonzero~\cite{turaev}. If $p=q=r=1$, for example, the aforementioned term will give zero whenever $a=b=0$, $c=a=0$, $c=b=0$, $a\neq b$, $c\neq a$ or $c\neq b$ (because, in this case, $a$, $b$ and $c$ 
can only assume the values 0 or 2). Additionally, if $a=b=2$, the two pairs of  tetrahedral net symbols subtracted from each other will be equal and therefore cancel out (since they differ by a $a \leftrightarrow b$ argument swap). As a result, a vertex with all edges of color 1 is annihilated by the action of the Hamiltonian [in other words, it is a zero-eigenvalue eigenstate of Eq.~\eqref{scalar_constr}]. 

Similarly, if $p=q=1$ and $r\geq 3$, the triples $(r,1,p)$ and $(r,1,q)$ are not triangular. Permutation of these labels reveals that no combination of the labels $1$, $1$ and $n$ (with $n\geq 3$) can simultaneously fulfill all the triangularity conditions. It is worth noting that these vertices violate the gauge constraint, therefore they are not contained in the physical Hilbert subspace. Nonetheless, 3-valent vertices with edges of colors $1$, $1+n$ and $1+n$  with $n\in \mathbb{N}^*$ fulfill the gauge constraint, but when acted upon by the Hamiltonian constraint cannot satisfy the triangularity conditions of the 6j symbols in Eq.~\eqref{Hstep8} (giving a zero outcome), and are therefore zero-eigenvalue eigenstates of the Hamiltonian constraint when the Temperley-Lieb algebra is adopted.

\bibliography{lit_LQGNew}

\end{document}